\documentclass[aps,prd,reprint,preprintnumbers,showpacs,floatfix,nofootinbib,superscript address]{revtex4-2}
\usepackage[utf8]{inputenc}
\usepackage{parskip}
\usepackage{amssymb}
\usepackage{stix}
\usepackage{hhline}
\usepackage{amsmath}
\usepackage{amsfonts}
\usepackage{mathtools}
\usepackage[dvipsnames]{xcolor}
\usepackage{xspace}
\usepackage{multirow,tabularx}
\usepackage{siunitx}
\usepackage{multirow}
\usepackage{graphicx}
\usepackage{xstring}
\usepackage{etoolbox}
\usepackage{notoccite}
\usepackage{tocbasic}
\usepackage{textgreek}
\usepackage{upgreek}
\usepackage{natbib}
\usepackage{mathrsfs}
\usepackage{lineno}
\usepackage{tensor}
\usepackage{accents}
\usepackage{tikz}
\usepackage{listings}
\usepackage{eso-pic}
\usepackage{atbegshi}
\usepackage{pdflscape}
\usepackage{pdfpages}
\usepackage{bm}
\allowdisplaybreaks
\newcommand{\beq} {\begin{equation}}
\newcommand{\eeq} {\end{equation}}
\usepackage{xspace}

\newrobustcmd{\PSALTer}{\textit{PSALTer}\xspace}
\newrobustcmd{\xAct}{\textit{xAct}\xspace}
\newrobustcmd{\xTensor}{\textit{xTensor}\xspace}
\newrobustcmd{\xCoba}{\textit{xCoba}\xspace}
\newrobustcmd{\xPerm}{\textit{xPerm}\xspace}
\newrobustcmd{\xCore}{\textit{xCore}\xspace}
\newrobustcmd{\xTras}{\textit{xTras}\xspace}
\newrobustcmd{\SymManipulator}{\textit{SymManipulator}\xspace}
\newrobustcmd{\SymPy}{\textit{SymPy}\xspace}
\newrobustcmd{\Maple}{\textit{Maple}\xspace}
\newrobustcmd{\RectanglePacking}{\textit{RectanglePacking}\xspace}
\newrobustcmd{\Inkscape}{\textit{Inkscape}\xspace}
\newrobustcmd{\Mathematica}{\textit{Mathematica}\xspace}
\newrobustcmd{\Macaulay}{\textit{Macaulay2}\xspace}
\newrobustcmd{\xPert}{\textit{xPert}\xspace}
\newrobustcmd{\MathGR}{\textit{MathGR}\xspace}
\newrobustcmd{\HiGGS}{\textit{HiGGS}\xspace}
\newrobustcmd{\Windows}{\textit{Microsoft Windows}\xspace}
\newrobustcmd{\Mac}{\textit{macOS}\xspace}
\newrobustcmd{\Linux}{\textit{Linux}\xspace}
\newrobustcmd{\GitHub}{\textit{GitHub}\xspace}
\newrobustcmd{\Bash}{\textit{bash}\xspace}
\newrobustcmd{\WolframLanguage}{\textit{Wolfram Language}\xspace}
\newrobustcmd{\CPP}{\textit{C++}\xspace}

\usepackage{titlesec}
\newrobustcmd{\pea}[1]{%
	\emph{#1}\textbf{\ \ \ ---}
}
\titleformat{\paragraph}[runin]{\normalfont\normalsize\bfseries}{\emph\theparagraph}{1em}{\pea}
\DeclareTOCStyleEntry[numwidth=20pt,linefill=\bfseries\TOCLineLeaderFill]{tocline}{section}
\DeclareTOCStyleEntry[entryformat=\textit,numwidth=10pt,linefill=\TOCLineLeaderFill]{tocline}{subsection}
\DeclareTOCStyleEntry[entryformat=\textit,numwidth=10pt,linefill=\TOCLineLeaderFill]{tocline}{subsubsection}
\makeatletter
\AtBeginDocument{\let\LS@rot\@undefined}
\makeatother
\usepackage{hyperref}
\hypersetup{%
     colorlinks = true,%
     linkcolor = Blue,%
     citecolor = Blue,%
     filecolor = Blue,%
     urlcolor = Blue%
     }%

\usepackage[capitalize]{cleveref}
\allowdisplaybreaks
\newrobustcmd{\CInvar}[1]{%
	{\tensor{\mathscr{C}}{_{#1}}}%
}
\newrobustcmd{\AInvar}[1]{%
	{\tensor{\mathscr{A}}{_{#1}}}%
}
\newrobustcmd{\XInvar}[1]{%
	{\tensor{\mathsf{x}}{_{#1}}}%
}
\newrobustcmd{\FieldH}[1]{%
  \tensor{h}{#1}
}
\newrobustcmd{\FieldG}[1]{%
	\tensor{g}{#1}
}
\newrobustcmd{\FieldHyp}[1]{%
	\tensor{\Delta}{#1}
}
\newrobustcmd{\FieldEta}[1]{%
	\tensor{\eta}{#1}
}
\newrobustcmd{\MAGA}[1]{%
  \tensor{\Gamma}{#1}
}

\begin{document}

\title{Complete background cosmology of parity-even quadratic metric-affine gravity}

\author{T. Dyer}
\email{td455@cantab.ac.uk}
\affiliation{Astrophysics Group, Cavendish Laboratory, JJ Thomson Avenue, Cambridge CB3 0HE, UK}
\affiliation{Kavli Institute for Cosmology, Madingley Road, Cambridge CB3 0HA, UK}

\author{W. Barker}
\affiliation{Central European Institute for Cosmolgy and Fundamental Physics, Institute of Physics of the Czech Academy of Sciences, Na Slovance 1999/2, 182 00 Prague 8, Czechia}

\author{D. Iosifidis}
\email{damianos.iosifidis@ut.ee}
\affiliation{Laboratory of Theoretical Physics, Institute of Physics, University of Tartu, W. Ostwaldi 1, 50411 Tartu, Estonia.}

\begin{abstract}
	The cosmology of metric-affine gravity is studied for the general, parity preserving action quadratic in curvature, torsion and non-metricity. The unconstrained model has too many parameters to be predictive, so this work serves to emulate and extend the well-known zero-non-metricity analysis of Goenner and M\"uller-Hoissen by charting the general field equations and phenomenology: these products will be useful for the future study of motivated special cases. We also make some progress towards identifying such cases, through combined constraints on the background phenomenology, and the unitarity of the quantum field theory on Minkowski spacetime. Linear and higher order relations between the quadratic operators in a Friedmann--Lema{\^i}tre--Robertson--Walker spacetime are obtained, along with the modified Friedmann, torsion and non-metricity equations. Extra parameter constraints lead to two special branches of the model. Firstly, a branch is found in which the Riemannian spatial curvature (thought to be slightly closed or flat in the \textLambda CDM model of our Universe) is entirely screened from all the field equations, regardless of its true value. Secondly, an integrable branch is found which yields (anti) de Sitter expansion at late times. The particle spectra of these two branches are studied, and the need to eliminate higher-spin particles as well as ghosts and tachyons motivates further parameter constraints in each case. The most general model is also found which reproduces the exact Friedmann equations of general relativity. The full set of equations describing closed, open or flat cosmologies, for general parity-even quadratic metric-affine gravity, is made available for \SymPy, \Mathematica and \Maple platforms.
\end{abstract}

\maketitle

\tableofcontents

\section{Introduction}\label{Introduction}

\begin{figure}
\includegraphics[width=1\linewidth]{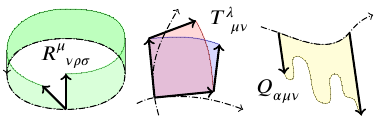}
	\caption{Left: the effect of the curvature in~\cref{riemann} -- the rotation of a vector after parallel transport in a closed loop. Centre: the effect of torsion in~\cref{TorsionDef} -- non-closure of infinitesimal parallelograms formed from parallely-transported vectors. Right: the effect of non-metricity in~\cref{NonMetDef} -- the change in vector norm under parallel transport.}
\label{Fig1}
\end{figure}

\paragraph*{Reformulation of gravity} Metric-affine gravity (MAG)~\cite{Hehl_1995,Iosifidis:2019dua,Damos_main_2020,Percacci_2020,JimenezCano:2021rlu,Belarbi:2021qiw,Baldazzi:2021kaf,Blagojevic:2012bc} is a natural generalisation of Einstein's general relativity (GR)~\cite{Einstein:1915}. MAG assumes that the connection, uniquely defined in GR as the Levi--Civita connection, is an independent dynamical variable in addition to the metric. A completely general affine connection endows the spacetime with \emph{non-Riemannian} geometric properties beyond the Riemannian curvature of GR, namely torsion and non-metricity, see~\cref{Fig1} for an illustration. MAG can be motivated as a natural extension of the theory of elastic continua in three dimensions (which describes deformations beyond curvature) to a four-dimensional spacetime~\cite{Hehl:1976one,Hehl:1976two,Hehl:1976three,Hehl_1995}. The conjugate source of the affine connection, the hypermomentum, is equivalently the source for torsion and non-metricity~\cite{Hehl:1976one,Hehl:1976three,Damos_main_2020,Iosifidis:2020upr,Iosifidis:2023kyf}. In this framework, not only does gravitation represent a `metrical elasticity' of space~\cite{Andrei_D_Sakharov_1991}, but it also possesses torsional and non-metric degrees of freedom (d.o.f) generated by the microstructure of matter, for which GR has no equivalent. Since its inception~\cite{Hehl:1976inc}, MAG has had many applications and has received interest in a range of areas (see reviews~\cite{Hehl_1995,Blagojevic:2012bc}). In particular today, the phenomenology of torsion and non-metricity couplings in the standard model of particle physics is an area of active focus~\cite{Rigouzzo:2023sbb,barker2024particle,Barker:2024dhb}.

\paragraph*{Motivation for extending GR} While GR is an extremely successful physical theory of gravitation and has passed many experimental and observational tests to high precision, in isolation it fails to account for certain phenomena. Chief among these are the strongly accelerated expansion of the early Universe~\cite{Starobinsky:1979ty,Guth:1980zm}, and the enhancement of gravity across a range of astrophysical and cosmological scales~\cite{Zwicky:1933gu}. Accordingly, a separate inflationary mechanism is sought after, along with a feebly-interacting dust-like ingredient in the concordance model of cosmology: the dark-energy-cold-dark-matter (\textLambda CDM) paradigm~\cite{Planck:2018vyg,Scott:2018adl}. In the last decade, high-precision data have revealed pronounced tensions in the~\textLambda CDM framework~\cite{Handley:2019tkm,DiValentino:2021izs,Perivolaropoulos:2021jda}. Planck 2018 CMB data under base-\textLambda CDM imply~$H_0=(67.4\pm0.5)\,\si{km.s^{-1}.Mpc^{-1}}$,~$\Omega_m=0.315\pm0.007$ and~$S_8=0.831\pm0.013$~\cite{Planck2018}, whereas the SH0ES distance ladder finds~$H_0=(73.04\pm1.04)\,\si{km.s^{-1}.Mpc^{-1}}$~\cite{Riess2022}, constituting a~$\sim5\sigma$ discrepancy. Meanwhile, cosmic-shear surveys report low clustering: KiDS-1000 yields~$S_8=0.759^{+0.024}_{-0.021}$~\cite{Asgari2021} and DES Year-3 (3$\times$2pt) finds~$S_8=0.776\pm0.017$ with~$\Omega_m\simeq0.339^{+0.032}_{-0.031}$, both lower than Planck’s predictions. However, DES Year-3 (3$\times$2pt) combined with available baryon acoustic oscillation (BAO), redshift-space distortion, type Ia supernovae, and Planck CMB data yields~$S_8=0.812\pm0.008$ with~$\Omega_m\simeq0.306$,~$h\simeq0.680$~\cite{DESY3_Abbott2022}. In parallel, combined ACT/SPT CMB analyses show~$\sim2$–$3\sigma$ shifts compared to Planck within~\textLambda CDM~\cite{Handley2021}. These inconsistencies have been tested against systematic error (e.g.\ alternative distance ladders) with no resolution, so Bayesian model comparison quantifies the tension as highly significant. Indeed, typical extensions have been tightly constrained and, as conventional fixes have failed, the community attention has turned to dark energy.\footnote{For example, Planck+BAO give~$N_{\rm eff}=2.99\pm0.17$ and~$\sum m_\nu<0.12\,$eV~\cite{Planck2018}, with only a mild~$A_{\rm lens}$ anomaly ($\sim2\sigma$) remaining~\cite{Planck2018}.} In this context, DESI’s recent results are pivotal: the DR2 BAO data plus Planck and updated SNe suggest deviation from~$w=-1$. Both parametric ($w_0w_a$) and non-parametric reconstructions prefer a phantom-crossing~$w(z)$ at late times~\cite{Lodha2025}.\footnote{Quantitatively, introducing~$w_0$ and~$w_a$ increases the Bayesian evidence, and this preference is robust under different reconstructions (binning, Gaussian processes)~\cite{Lodha2025}.} Indeed, on the Jeffreys scale one finds moderate-to-strong support for dynamical dark energy, with simple quintessence (no phantom crossing) disfavoured~\cite{Lodha2025}. In summary, while the~$H_0$ and~$S_8$ tensions persist across Planck, SH0ES, DES, KiDS, ACT and SPT data, DESI’s 2025 analysis provides robust Bayesian evidence that the low-redshift acceleration may deviate from a pure cosmological constant~\cite{Lodha2025}. Underpinning the \textLambda CDM model are the Friedmann equations, which describe the background expansion of the Universe when GR is coupled to matter-like sources, along with the theory of perturbations around this background. By modifying GR, so the Friedmann and perturbation equations are themselves modified: there is some hope that the resulting phenomenology will lead to a less ad hoc cosmological model than \textLambda CDM, or one with fewer tensions.

\paragraph*{A very general geometry} Non-Riemannian models offer a natural alternative to GR. For two very special cases, GR itself can be equivalently formulated with vanishing curvature using either only torsion, or only non-metricity (all other quantities in~\cref{Fig1} vanishing). These are called the teleparallel and symmetric-teleparallel equivalents of GR (TEGR~\cite{Bahamonde:2021gfp} and STEGR).\footnote{There is also the generalized teleparallel gravity with both torsion and non-metricity~\cite{BeltranJimenez:2019odq}.} Together with Riemannian GR, these three formulations are frequently referred to as the `\emph{geometrical trinity}' of gravity~\cite{jimenez2019geometrical}. However, the geometry does not have to be thus restricted. An admixture of geometries is found in Weyl spacetime~\cite{Weyl:1918,Weyl:1922}, which admits part of the general non-metricity tensor in addition to curvature. A more substantial extension is to Riemann--Cartan spacetime~\cite{Cartan:1923,Eddington:1923,Cartan:1924,Cartan:1925,Einstein:1925,Einstein:1928,Einstein:19282}, where the full torsion is instead admitted: models with this geometry may be thought of as examples of Poincar\'e gauge theory (PGT), where the gauge-theoretic structure is motivated by graivtational coupling to spinorial matter~\cite{Utiyama:1956sy,Sciama:1962,Kibble:1961ba}. By blending the Weyl and Riemann--Cartan spacetimes, models of dynamical Weyl--Cartan spacetime may be thought of as examples of Weyl gauge theory (WGT)~\cite{Blagojevic:2002du}. At the most general level, however, MAG encompasses all these non-Riemannian geometries, and specialises to them depending on properties of the connection.

\paragraph*{A very general Lagrangian} The choice of geometry is not the only consideration when constructing a non-Riemannian model: one must also select the operators which appear in the action. Quadratic gravity (QG) is the class of theories in Riemannian geometry that model gravity with an action up to quadratic order in curvature~\cite{Salvio:2018crh}. It is closely related to quadratic MAG (QMAG), in which operators quadratic in torsion and non-metricity are also included~\cite{Percacci_2020}. Classically, QMAG is a straightforward extension of QG, being indistinguishable from Riemannian quadratic gravity coupled (albeit in a highly non-minimal way) to torsion and non-metricity tensors, which can be thought of as extra matter fields. QG is a generic expectation of GR as a low-energy effective field theory~\cite{Salam:1978,Stelle:1977,Einhorn:2017icw,Donoghue:1994dn}: the same may possibly be true of QMAG, though investigations into the non-Riemannian quantum theory are still relatively nascent~\cite{Pagani:2015ema,Percacci:2020bzf,Baldazzi:2021kaf,Melichev:2023lwj,Melichev:2024hih}.\footnote{An immediate objection to QMAG being quantum in origin is that the low-energy effective theory of a dynamical connection should be a priori spanned by a much larger basis of operators than those which can be formed from the squares of curvature, torsion and non-metricity.} Famously, unitarity is not a general feature of the QG theory-space, according to the standard definition of ghost modes~\cite{Salvio:2018crh}. The particle spectrum of QMAG is much richer than that of QG: without careful tuning the linear spectrum is also blighted by ghosts and tachyons~\cite{Sezgin:1981xs,Blagojevic:1983zz,Blagojevic:1986dm,Kuhfuss:1986rb,Yo:1999ex,Yo:2001sy,Blagojevic:2002,Puetzfeld:2004yg,Yo:2006qs,Shie:2008ms,Nair:2008yh,Nikiforova:2009qr,Chen:2009at,Ni:2009fg,Baekler:2010fr,Ho:2011qn,Ho:2011xf,Ong:2013qja,Puetzfeld:2014sja,Karananas:2014pxa,Ni:2015poa,Ho:2015ulu,Karananas:2016ltn,Obukhov:2017pxa,Blagojevic:2017ssv,Blagojevic:2018dpz,Tseng:2018feo,Lin:2018awc,BeltranJimenez:2019acz,Zhang:2019mhd,Aoki:2019rvi,Zhang:2019xek,Jimenez:2019qjc,Lin:2019ugq,Percacci:2019hxn,Barker:2020gcp,BeltranJimenez:2020sqf,MaldonadoTorralba:2020mbh,Barker:2021oez,Marzo:2021esg,Marzo:2021iok,delaCruzDombriz:2021nrg,Baldazzi:2021kaf,Annala:2022gtl,Mikura:2023ruz,Mikura:2024mji,Barker:2024ydb}, though these are not generally the same species which spoil QG. Moreover, there is a particular danger in QMAG that the linear spectrum is not a faithful representation of the general degrees of freedom that are propagating: some modes may become strongly coupled near Minkowski spacetime~\cite{Moller:1961,Pellegrini:1963,Hayashi:1967se,Cho:1975dh, Hayashi:1979qx,Hayashi:1979qx,Dimakis:1989az,Dimakis:1989ba,Lemke:1990su,Hecht:1990wn,Hecht:1991jh,Yo:2001sy,Afshordi:2006ad,Magueijo:2008sx,Charmousis:2008ce,Charmousis:2009tc,Papazoglou:2009fj,Baumann:2011dt,Baumann:2011dt,DAmico:2011eto,Gumrukcuoglu:2012aa,Wang:2017brl,Mazuet:2017rgq,BeltranJimenez:2020lee,JimenezCano:2021rlu,Barker:2022kdk,Delhom:2022vae,Annala:2022gtl,Barker:2022kdk,Barker:2023fem}.

\paragraph*{In this work} Our objective is not to formulate from QMAG a comprehensive alternative to the \textLambda CDM model, suitable for empirical testing. Indeed, QMAG is insufficiently developed at the theoretical level for such a task. However, we do make some important steps towards this objective. We obtain background cosmology of QMAG, with the arbitrary but pragmatic restriction that parity-violating operators are excluded. Indeed, parity is not a good quantum number, but given a particular non-Riemannian geometry it is common to first explore the quadratic theory in the parity-preserving case, following up the investigation with the parity-violating extension. Cosmology has been extensively studied in various kinds of non-Riemannian geometry, see~\cite{Damos_main_2020,iosifidis2024magverse,Shimada:2018lnm,Gialamas:2022xtt,minkevich1998isotropic,obukhovQuad,Iosifidis_2022} (also~\cite{Puetzfeld:2004yg} and references therein). One of the best known among these analyses is that of Goenner and M\"uller-Hoissen~\cite{Gonner:1984rw}, who performed a comprehensive study of the background cosmology in full PGT. Recent investigations into the full MAG geometry show a diverse collection of phenomena~\cite{IosifidisQuad,Iosifidis_2022}, particularly when including the hypermomentum as a source~\cite{iosifidis2024magverse}. Despite these advances, the full QMAG action remains unexplored. This is true even at the level of background cosmology, where the basic set of equations has not actually been derived. In the long term, of course, such equations should be studied in tandem with the metric-affine perturbation theory, however cosmological perturbations of MAG offer a far more substantial challenge than the background cosmology, and the field it as-yet just beginning to be developed~\cite{HeisenbergHohmann2023,CastilloFelisola2024,Aoki2024}. The rest of this work is structured as follows. In~\cref{Theory} we set out the MAG formalism and discuss our methods. In~\cref{Quadratic} we obtain the cosmological equations and discuss their structure. In~\cref{Specials} we motivate two special cases of QMAG by their cosmology, and use linearised particle spectroscopy near Minkowski spacetime to further constrain them. Conclusions follow in~\cref{ConcludingRemarks}, along with technical appendices. We work in natural units~$c\equiv\hbar\equiv 1$. Further conventions are introduced as needed.

\section{Theoretical development}\label{Theory}

Having motivated MAG in~\cref{Introduction}, we will define our conventions for the 64 extra kinematic d.o.f implied by metric-affine geometry (relative to the ten d.o.f present in the Riemannian case) in~\cref{Geometry}. In~\cref{CosmologySetup} will then understand how the restriction to an isotropic and homogeneous cosmology eliminates all but five of these.

\subsection{Metric-affine geometry}\label{Geometry}

Our notation mainly follows~\cite{Damos_main_2020}. A non-Riemannian spacetime is a differentiable manifold equipped with a symmetric metric~$\tensor{g}{_{\mu\nu}}$ of signature~$\left(-,+,+,+\right)$ and an independent affine connection~$\tensor{\Gamma}{^\lambda_{\mu\nu}}$, where the covariant derivative~$\tensor{\nabla}{_\mu}$ of some tensor field~$\tensor{X}{^\mu_\nu}$ is
\begin{equation}\label{CovariantDerivative}
    \nabla{_{\alpha}}\tensor{X}{^{\mu}_{\nu}} \equiv \partial{_\alpha}\tensor{X}{^{\mu}_{\nu}}-\tensor{\Gamma}{^{\lambda}_{\nu\alpha}}\tensor{X}{^{\mu}_{\lambda}}+\tensor{\Gamma}{^{\mu}_{\lambda\alpha}}\tensor{X}{^{\lambda}_{\nu}}.
\end{equation}
Since the difference of two connections is always a tensor, the deviation of the affine connection from the unique, symmetric and metric-compatible Levi--Civita connection~$\tensor{\tilde{\Gamma}}{^{\lambda}_{\mu\nu}}$ gives the~$4\times 4\times 4=64$ new d.o.f (i.e. the components of an asymmetric rank-three four-tensor) of the distortion tensor
\begin{equation}  \label{PRD}
    \tensor{N}{^\lambda_{\mu\nu}} \equiv \tensor{\Gamma}{^\lambda_{\mu\nu}}-\tensor{\tilde{\Gamma}} {^{\lambda}_{\mu\nu}},
\end{equation}
where~$\tensor{\tilde{\Gamma}} {^{\lambda}_{\mu\nu}}$ is explicitly given by the Christoffel formula
\begin{equation}  \label{LCD}
    \tensor{\tilde{\Gamma}}{^\lambda_{\mu\nu}} \equiv \frac{1}{2}\tensor{g}{^{\alpha\lambda}}\left(\tensor{\partial}{_{\mu}}\tensor{g}{_{\mu\nu}}+\tensor{\partial}{_{\nu}}\tensor{g}{_{\alpha\mu}}-\tensor{\partial}{_{\alpha}}\tensor{g}{_{\mu\nu}}\right).
\end{equation}
From this point onwards, quantities with a tilde over them will denote Riemannian parts derived from~$\tensor{\tilde{\Gamma}}{^{\lambda}_{\mu\nu}}$ unless stated otherwise. It is possible to derive two non-Riemannian quantities from the affine connection: the torsion and the non-metricity. The torsion tensor is defined to be the antisymmetric part of the connection, with~$4\times 6=24$ d.o.f
\begin{equation} \label{TorsionDef}
    \tensor{T}{^{\lambda}_{\mu\nu}}\equiv\tensor{\Gamma}{^{\lambda}_{[\mu\nu]}}\equiv\tensor{N}{^{\lambda}_{[\mu\nu]}}.
\end{equation}
The torsion in~\cref{TorsionDef} can be said to measure the non-closure of infinitesimal parallelograms formed from parallel transport of tangents along two intersecting paths (see the central diagram in~\cref{Fig1}). The non-metricity tensor with~$4\times 10=40$ d.o.f is defined as the degree to which the metricity condition is violated by the independent connection (see rightmost diagram in~\cref{Fig1})
\begin{equation} \label{NonMetDef}
    \tensor{Q}{_{\alpha\mu\nu}}\equiv-\tensor{\nabla}{_{\alpha}}\tensor{g}{_{\mu\nu}}\equiv 2\tensor{N}{_{(\mu\nu)\alpha}}.
\end{equation}
The second equality of~\cref{NonMetDef} follows from expanding the covariant derivative into the connection using~\cref{CovariantDerivative}, and then expanding the connection using~\cref{PRD}. The Levi--Civita connection in~\cref{LCD} uniquely cancels all but the non-Riemannian contributions. This post-Riemannian decomposition (PRD) --- whereby the original Palatini variables~$\left\{\tensor{g}{_{\mu\nu}},\tensor{\Gamma}{^{\lambda}_{\mu\nu}}\right\}$ are superseded by~$\left\{\tensor{g}{_{\mu\nu}},\tensor{N}{^{\lambda}_{\mu\nu}}\right\}$ or~$\left\{\tensor{g}{_{\mu\nu}},\tensor{T}{^{\lambda}_{\mu\nu}},\tensor{Q}{_\lambda_\mu_\nu}\right\}$ through a simple field reparameterisation --- will be used throughout our investigations.

Curvature is defined as the failure of covariant derivatives to commute (see leftmost diagram in~\cref{Fig1}). The non-Riemannian curvature tensor is
\begin{equation}\label{riemann}
    \tensor{R}{^{\mu}_{\nu\rho\sigma}}\equiv\partial_{\rho}\tensor{\Gamma}{^{\mu}_{\nu\sigma}}-\partial_{\sigma}\tensor{\Gamma}{^{\mu}_{\nu\rho}}+\tensor{\Gamma}{^{\tau}_{\nu\sigma}}\tensor{\Gamma}{^{\mu}_{\tau\rho}}-\tensor{\Gamma}{^{\tau}_{\nu\rho}}\tensor{\Gamma}{^{\mu}_{\tau\sigma}}.
\end{equation}
Note that, unlike in the Riemannian case, the non-Riemannian curvature is only antisymmetric in its last two indices. By plugging~\cref{PRD} into~\cref{riemann} the post-Riemannian decomposition of the curvature tensor is found to be
\begin{equation} \label{PRDRiemann}
	\begin{aligned}
		\tensor{R}{^{\mu}_{\nu\rho\sigma}}&\equiv \tensor{\tilde{R}}{^{\mu}_{\nu\rho\sigma}}+\tilde{\nabla}_{\rho}\tensor{N}{^{\mu}_{\nu\sigma}}-\tilde{\nabla}_{\sigma}\tensor{N}
            {^{\mu}_{\nu\rho}}
            \\& \ \ \ 
            +\tensor{N}{^{\tau}_{\nu\sigma}}\tensor{N}{^{\mu}_{\tau\rho}}-\tensor{N}{^{\tau}_{\nu\rho}}\tensor{N}{^{\mu}_{\tau\sigma}}.
	\end{aligned}
\end{equation}
Without using the metric, two independent contractions of the curvature can be defined
\begin{equation}\label{ricciDef1}
	\tensor{R}{_{\mu\nu}}\equiv \tensor{R}{^{\alpha}_{\mu\alpha\nu}}, \quad \tensor{\Hat{R}}{_{\mu\nu}}\equiv \tensor{R}{^{\alpha}_{\alpha\mu\nu}}.
\end{equation}
The former is the Ricci tensor and the latter is the homothetic curvature. A third tensor --- the co-Ricci --- is defined as~$\tensor{\check{R}}{^\mu_{\rho}}\equiv\tensor{R}{^\mu_{\nu\sigma\rho}}\tensor{g}{^{\nu\sigma}}$. The Ricci scalar remains uniquely defined, since
\begin{equation} \label{scalar}
	R\equiv\tensor{R}{_{\mu\nu}}\tensor{g}{^{\mu\nu}} \equiv -\tensor{\check{R}}{_{\mu\nu}}\tensor{g}{^{\mu\nu}}\equiv-\tensor{R}{_{\mu}^{\alpha}_{\alpha\nu}}\tensor{g}{^{\mu\nu}},
\end{equation}
meanwhile the homothetic curvature in~\cref{ricciDef1} is traceless due to its antisymmetry~$\tensor{\Hat{R}}{_{\mu\nu}}\tensor{g}{^{\mu\nu}}\equiv 0$. Note that the notation and index conventions here are different to~\cite{Percacci_2020}. The latter two traces of the Riemann tensor do not exist in Riemannian geometry due to the index symmetry~$\tensor{\tilde{R}}{_{\alpha\beta\mu\nu}}\equiv-\tensor{\tilde{R}}{_{\beta\alpha\mu\nu}}$. Consequently, we have~$\tensor{\Hat{R}}{_{\mu\nu}}\equiv\tensor{\tilde{R}}{^{\alpha}_{\alpha\mu\nu}}\equiv0$ and~$\tensor{\Check{R}}{_{\mu\nu}}\equiv\tensor{\tilde{R}}{_{\mu}^{\alpha}_{\alpha\nu}}\equiv-\tensor{\tilde{R}}{_{\mu\nu}}$ in the Riemannian limit. 

It is also useful to define non-Riemannian four-vectors using the torsion and non-metricity. Based on~\cref{TorsionDef,NonMetDef} the torsion forms one vector and the non-metricity forms two vectors:
\begin{equation}\label{VecDef}
    \tensor{t}{_{\mu}}\equiv\tensor{T}{^{\lambda}_{\mu\lambda}}, \quad \tensor{Q}{_{\alpha}}\equiv\tensor{Q}{_{\alpha\mu\nu}}\tensor{g}{^{\mu\nu}}, \quad \tensor{q}{_{\alpha}}\equiv\tensor{Q}{_{\mu\nu\alpha}}\tensor{g}{^{\mu\nu}}.
\end{equation}
Note that~\cref{VecDef} omits the axial vector~$\tensor{\epsilon}{_{\mu\alpha\beta\gamma}}\tensor{T}{^{\alpha\beta\gamma}}$, which will not need a separate notation in this work. The non-metricity does not form an axial vector because of its index symmetries. Having defined the basic geometric ingredients, their explicit forms in a cosmological setting can be introduced.

\subsection{Background cosmology}\label{CosmologySetup}

We will now introduce the Friedmann--Lema{\^i}tre--Robertson--Walker (FLRW) metric in the context of MAG, giving the forms of non-Riemannian ingredients from~\cref{Geometry} in an FLRW coordinate basis derived from the assumed symmetries of the spacetime. The line element in a spatially flat FLRW spacetime is given by
\begin{equation}\label{FLRWBasic}
	\mathrm{d}s^{2}= -\mathrm{d}t^2 + a^2\left[\mathrm{d}r^2+r^2\left(\mathrm{d}\uptheta^2+\sin^2(\uptheta)\mathrm{d}\upphi^2\right)\right],
\end{equation}
in spatially co-moving spherical polar coordinates, where the dimensionless scale factor~$a\equiv a(t)$ is normalised to unity at the current epoch. In~\cref{Linear,Quadratic}, a minisuperspace method will be used to extract the background dynamics, and for this a `lapse-like' function~$b\equiv b(t)$ will be introduced, alongside~$a$. This is done so that the action can then be varied with respect to~$a$ and~$b$ separately, setting~$b$ to unity afterwards as a choice of time gauge. With the inclusion of the `lapse-like' function,~\cref{FLRWBasic} becomes
\begin{equation}\label{FLRWelement}
	\mathrm{d}s^{2}= -b^2\mathrm{d}t^2 + a^2\left[\mathrm{d}r^2+r^2\left(\mathrm{d}\uptheta^2+\sin^2(\uptheta)\mathrm{d}\upphi^2\right)\right].
\end{equation}
The co-moving observer in a homogeneous, isotropic Universe has four-velocity~$\tensor{u}{_{\mu}}\tensor{u}{^{\mu}}\equiv-1$. Based on this velocity, a projection tensor can be defined as
\begin{equation}\label{projdef}
    \tensor{P}{_{\mu\nu}}\equiv\tensor{g}{_{\mu\nu}}+\tensor{u}{_{\mu}}\tensor{u}{_{\nu}}.
\end{equation}
The conditions of homogeneity and isotropy imply that, of the 64 d.o.f in~\cref{PRD}, only five remain. We parameterise these using the `scalars'~$\left\{X,Y,Z,V,W\right\}$, which depend only on cosmic time~$t$ so that~$X\equiv X(t)$, etc. Using~\cref{projdef} these scalars are distributed according to 
\begin{equation}\label{distFLRW}
	\begin{aligned}
		\tensor{N}{_{\lambda\mu\nu}}& \equiv X~ \tensor{P}{_{\mu\nu}}\tensor{u}{_{\lambda}} + Y~ \tensor{P}{_{\nu\lambda}}\tensor{u}{_{\mu}} 
		\\ &\ \ \ 
		+ Z~\tensor{P}{_{\mu\lambda}}\tensor{u}{_{\nu}} + V~\tensor{u}{_{\lambda}}\tensor{u}{_{\mu}}\tensor{u}{_{\nu}} + W~ \tensor{\epsilon}{_{\lambda\mu\nu\kappa}}\tensor{u}{^\kappa}.
	\end{aligned}
\end{equation}
Equally, with~\cref{TorsionDef,NonMetDef} we see that~\cref{distFLRW} becomes
\begin{subequations}
\begin{align}
	\tensor{Q}{_{\alpha\mu\nu}}&\equiv 2Z\tensor{u}{_{\alpha}}\tensor{P}{_{\mu\nu}}+2\left(X+Y\right)\tensor{P}{_{\alpha(\mu}}\tensor{u}{_{\nu)}}+2V\tensor{u}{_{\alpha}}\tensor{u}{_{\mu}}\tensor{u}{_{\nu}},\label{hyperdist}
	\\
	\tensor{T}{_{\alpha\mu\nu}}&\equiv \left(Y+Z\right)\tensor{u}{_{[\mu}}\tensor{P}{_{\nu]\alpha}}+W\tensor{\epsilon}{_{\mu\nu\alpha\rho}}\tensor{u}{^{\rho}}.\label{hypodist}
\end{align}
\end{subequations}
The restrictions in~\cref{distFLRW,hyperdist,hypodist} may be formally derived from the vanishing of the Lie derivative --- see e.g.~\cite{Tsamparlis_1979,minkevich1998isotropic,Damos_main_2020} --- however they are also self-evident. Indeed, it follows in cosmology that only the spin-zero parts of the tensor field~$\tensor{N}{^\lambda_{\mu\nu}}$ may contribute to the dynamics. As is well known, an asymmetric tensor of third rank contains five spin-zero modes, one of negative parity ($W$ in this notation) and all the others of positive parity. Thus,~\cref{distFLRW,hyperdist,hypodist} constitute the spin-zero part of the spin-parity decomposition of the distortion tensor. This decomposition is used for, example, in the canonical analysis, where the analogue of~$\tensor{u}{^\mu}$ is provided by the normal to the space-like hypersurfaces. As will be shown in~\cref{ParticleSpectra}, this decomposition is also used in the study of the particle spectrum, where the analogue of~$\tensor{u}{^\mu}$ is provided by~$\tensor{n}{^\mu}\equiv\tensor{k}{^\mu}/k$ for~$k^2\equiv\tensor{k}{_\mu}\tensor{k}{^\mu}$ and~$\tensor{k}{^\mu}$ the four-momentum of the (massive) particles. The same principle applies to the metric: it is well known that a symmetric tensor of second rank contains two scalars, both of positive parity. This is the reason why we need the two variables~$\left\{a,b\right\}$ in~\cref{FLRWelement} for a lossless minisuperspace reduction. As a final comment before using these definitions, it will be useful for us to define the following combination of distortion scalars

\begin{equation}
    U \equiv V+Z,
\end{equation}
for various results in~\cref{AppendixFried}.

\subsection{The Einstein--Hilbert case}\label{Linear}

Having restricted to cosmology in~\cref{CosmologySetup} we will compute the background cosmological equations for the minimal case of MAG, in which the Lagrangian density comprises only the Einstein--Hilbert term
\begin{equation}\label{LinearAction}
    S=\frac{1}{2 \kappa} \int \mathrm{d}^{4}x \sqrt{-g}~R.
\end{equation}
The first method to obtain the Friedmann equations is covariantly, via field equations. We will work with the PRD and separately vary~\cref{LinearAction} with respect to the distortion and the metric
    \begin{align}
	    \delta S= \frac{1}{2 \kappa} \int & \mathrm{d}^{4}x \sqrt{-g}~\tensor{\delta g}{^{\mu\nu}} \bigg[ \Big. \tensor{\tilde{R}}{_{\mu\nu}}-\frac{1}{2}\tensor{g}{_{\mu\nu}}\tilde{R} + \big( \big. \tensor{N}{^{\sigma}_{\lambda\sigma}}\tensor{N}{^{\lambda}_{\mu\nu}} \nonumber\\& 
	    -\tensor{N}{^{\sigma}_{\lambda\nu}}\tensor{N}{^{\lambda}_{\mu\sigma}} \big)\big. -\frac{1}{2} \tensor{g}{_{\mu\nu}} \tensor{g}{^{\alpha\beta}} \big(\big. \tensor{N}{^{\sigma}_{\lambda\sigma}}\tensor{N}{^{\lambda}_{\alpha\beta}} 
	    \nonumber\\&
	    -\tensor{N}{^{\sigma}_{\lambda\beta}}\tensor{N}{^{\lambda}_{\alpha\sigma}}\big)\big. \bigg] \Big.
         + \sqrt{-g}~\tensor{\delta N}{^{\sigma}_{\kappa\gamma}} \bigg[ \tensor{\delta}{^{\gamma}_{\sigma}}\tensor{N}{^{\kappa\alpha}_{\alpha}}
	    \nonumber\\&
	    +\tensor{g}{^{\kappa\gamma}}\tensor{N}{^{\alpha}_{\sigma\alpha}}-\tensor{N}{^{\kappa\gamma}_{\sigma}}-\tensor{N}{^{\gamma}_{\sigma}^{\kappa}}\bigg].\label{TotalVariation}
    \end{align}
On shell we take~$\delta S = 0$ so that~\cref{TotalVariation} yields the field equations
\begin{subequations}
\begin{align}
	&\tensor{\tilde{R}}{_{\mu\nu}}-\frac{1}{2}\tensor{g}{_{\mu\nu}}\tilde{R} + \tensor{D}{_{\mu\nu}} -\frac{1}{2} \tensor{g}{_{\mu\nu}} \tensor{D}{^{\alpha}_{\alpha}} = 0,
    \\
	&\tensor{\delta}{^{\gamma}_{\sigma}}\tensor{N}{^{\kappa\alpha}_{\alpha}}+\tensor{g}{^{\kappa\gamma}}\tensor{N}{^{\alpha}_{\sigma\alpha}}-\tensor{N}{^{\kappa\gamma}_{\sigma}}-\tensor{N}{^{\gamma}_{\sigma}^{\kappa}} = 0,
\end{align}
\end{subequations}
where~$\tensor{D}{_{\mu\nu}}\equiv 2~\tensor{N}{^{\lambda}_{\mu [ \nu |}}\tensor{N}{^{\sigma}_{\lambda | \sigma ] }}$. Referring back to the line element in~\cref{FLRWBasic} and the cosmological form of the distortion~\cref{distFLRW}, the FLRW component forms of these equations simplify to give
\begin{equation} \label{LinearFriedmann}
        \begin{aligned}
            H = 
            W= X= Y= V+Z=0,
        \end{aligned}
\end{equation}
where~$H \equiv H(t)$ is the Hubble parameter defined as~$H \equiv \dot{a}/a$, and an over-dot will always denote differentiation with respect to the cosmic time~$t$, so that~$\dot{a}\equiv\mathrm{d}a/\mathrm{d}t$. Recall that the `lapse-like' function introduced in~\cref{FLRWelement} is set to unity throughout. Note that~\cref{LinearFriedmann} agrees with a well known result in MAG: an action linear in the Ricci scalar should return the same dynamics as normal GR in a vacuum. Note however that there is one d.o.f between~$V$ and~$Z$ (namely~$U$) which is left unconstrained by~\cref{LinearFriedmann}. We will show in~\cref{case2} that this is a result of the invariance of~\cref{LinearAction} under the projective transformation
\begin{equation}\label{Projective}
    \tensor{\Gamma}{^{\lambda}_{\mu\nu}} \mapsto \tensor{\Gamma}{^{\lambda}_{\mu\nu}} + \delta_{\mu}^{\lambda} \tensor{\xi}{_{\nu}},
\end{equation}
where~$\tensor{\xi}{_{\nu}}\equiv\tensor{\xi}{_{\nu}}(x)$ is an arbitrary local four-vector gauge generator~\cite{Iosifidis:2019fsh,Percacci_2020}.

The second method of calculating the Friedmann equations is to use the minisuperspace\footnote{The term `\textit{minisuperspace}' comes from quantum cosmology, when considering the full gravitational path integral. Instead of integrating over all components of the metric~$\tensor{g}{_{\mu\nu}}\equiv\tensor{g}{_{\mu\nu}}(x)$, an approximation can be made by assuming the off-diagonal components to vanish, and and the remaining components to be functions only of~$t$, thereby reducing the number of fields to two -- the `lapse-like' and `shift-like' functions. Including the `lapse-like' function is what allows for time reparameterisation invariance, and its variation yields a constraint equation.} method (or Weyl method). This method can be adapted to classical cosmology~\cite{Remmen:2013eja,Achour:2021lqq}, it involves imposing~\cref{FLRWelement,distFLRW,hyperdist,hypodist} first, before taking variational derivatives. The three-dimensional space-like hypersurfaces are integrated out of the original action with an appropriate cut-off to give a fiducial volume. In general, the imposition of constraints on the fields prior to variations is not guaranteed to lead to the correct equations of motion, since the two operations do not commute. However, for Bianchi class~A cosmological models, the method holds as guaranteed by Palais' principle of symmetric criticality~\cite{Deser:2003up,Maccallum:1972er,Fels:2001,Hawking:1969} and since the metric here fulfills the requirements\footnote{The metric ansatz in~\cref{FLRWelement} meets the conditions in~\cite{Maccallum:1972er,Fels:2001}.} we can proceed to obtain the correct equations of motion. The Ricci scalar in a metric-affine FLRW spacetime is
\begin{equation} \label{RicciFLRW}
\begin{aligned}
    R&=3 \Bigg[ \Bigg. 2\left(\frac{\ddot{a}}{ab^2} + \left(\frac{\dot{a}}{ab}\right)^2-\frac{\dot{a}\dot{b}}{ab^3}\right) +3\frac{\dot{a}}{ab}(X-Y)\\&+\frac{1}{b}(\dot{X}-\dot{Y}) +(X+Y)(V+Z)-2XY-2W^2 \Bigg] \Bigg. .
\end{aligned}
\end{equation}
Now, using also~$\sqrt{-g}=a^3b$, we can use~\cref{RicciFLRW} to write the Lagrangian density in~\cref{LinearAction} as  
\begin{align}
	S\propto\int \mathrm{d}t \bigg[& a^{3}b \big( 
	(X+Y)(Z+V)-2 X Y-2 W^{2} \big) 
	\nonumber \\
	&
	-2\frac{a \dot{a}^{2}}{b}
	+\frac{d}{dt}\Big[  2 \frac{a^{2}\dot{a}}{b}+ a^{3}(X-Y)\Big]\bigg], \label{Rwithb}
\end{align}
where the proportionality holds up to constants and the fiducial volume. Note that in~\cref{Rwithb} the last term can be dropped, since this is a total derivative. Variation with respect to the minisuperspace field variables~$\left\{a,b,X,Y,Z,V,W\right\}$ gives us seven (or six independent) equations, respectively
    \begin{equation}\label{LinearMinisuperspaceEqns}
    \begin{gathered}
	    6 H^2 + 4\dot{H}  = 6 W^2 + 6 X Y - 3 \left(X + Y\right)\left(V + Z\right),
        \\
        2 H^2  = 2 W^2 + 2 X Y - \left(X + Y\right) \left(V + Z\right),
        \\
        V + Z  = 2Y,
        \quad
        V + Z  = 2X,
        \quad
        X + Y = 0,
        \\
        X + Y = 0,
        \quad
        W  = 0.
    \end{gathered}
    \end{equation}
Collectively,~\cref{LinearMinisuperspaceEqns} simplify to yield precisely~\cref{LinearFriedmann}, showing that the covariant and minisuperspace methods are consistent.

\section{Structure of the field equations}\label{Quadratic}
As discussed in~\cref{Introduction}, it is interesting to consider the most general action quadratic in curvature, torsion, and non-metricity, i.e. QMAG. The QMAG action is given by\footnote{Note that the couplings~$\left\{c_{1},\ldots,c_{16}\right\}$ and~$\left\{a_{0},\ldots,a_{11}\right\}$ are carefully tuned so as to be identical to those in~\cite{Percacci_2020}. Since our conventions for the geometry itself, as set out in~\cref{Geometry}, follow~\cite{Damos_main_2020} instead of~\cite{Percacci_2020}, there are apparent sign changes in front two terms.}
\begin{align}\label{fullAction}
    &S= -\frac{1}{2} \int \mathrm{d}^4 x \sqrt{-g}~ \Bigg[ \Big. -a_{0} R + R^{\rho\sigma\mu\nu} \Big(
  c_{1} R_{\rho\sigma\mu\nu}
+ c_{2} R_{\sigma\rho\mu\nu}
\nonumber\\&
+ c_{3} R_{\mu\nu\rho\sigma}
+ c_{4} R_{\mu\sigma\rho\nu}
+ c_{5} R_{\mu\rho\sigma\nu}
+ c_{6} R_{\rho\mu\sigma\nu} \Big)
+ c_{7} \tensor{R}{^{\mu\nu}} \tensor{R}{_{\mu\nu}}
\nonumber\\&
+ c_{8} \tensor{R}{^{\mu\nu}} \tensor{R}{_{\nu\mu}}
+ c_{9}  \tensor{\check{R}}{^{\mu\nu}} \tensor{\check{R}}{_{\mu\nu}}
+ c_{10} \tensor{\check{R}}{^{\mu\nu}} \tensor{\check{R}}{_{\nu\mu}}
+ c_{11} \tensor{R}{^{\mu\nu}}\tensor{\check{R}}{_{\mu\nu}}
\nonumber\\&
+ c_{12} \tensor{R}{^{\mu\nu}} \tensor{\check{R}}{_{\nu\mu}}
+ c_{13} \tensor{\hat{R}}{^{\mu\nu}}\tensor{\hat{R}}{_{\mu\nu}}
- c_{14} \tensor{R}{^{\mu\nu}}\tensor{\hat{R}}{_{\mu\nu}}
- c_{15} \tensor{\hat{R}}{^{\mu\nu}}\tensor{\check{R}}{_{\mu\nu}}\nonumber\\&
+ c_{16} R^{2}
+ 4 a_{1} \tensor{T}{^{\mu\nu\rho}} \tensor{T}{_{\mu\nu\rho}}
+ 4 a_{2} \tensor{T}{^{\mu\nu\rho}} \tensor{T}{_{\nu\mu\rho}}
+ 4 a_{3} \tensor{t}{^{\mu}}\tensor{t}{_{\mu}}
\nonumber\\&
+ a_{4} \tensor{Q}{^{\mu\nu\rho}}\tensor{Q}{_{\mu\nu\rho}}
+ a_{5} \tensor{Q}{^{\mu\nu\rho}}\tensor{Q}{_{\nu\mu\rho}}
+ a_{6} \tensor{Q}{^{\mu}}Q_{\mu}
+ a_{7} \tensor{q}{^{\mu}}\tensor{q}{_{\mu}}
\nonumber\\&
+ a_{8} \tensor{Q}{^{\mu}}\tensor{q}{_{\mu}}
+ 2 a_{9} Q^{\mu\nu\rho} T_{\nu\rho\mu}
+2\Big( a_{10} \tensor{Q}{^{\mu}}  
+a_{11} \tensor{q}{^{\mu}}\Big) \tensor{t}{_{\mu}} \Big. \Bigg].
\end{align}
This action has many parameters and is very cumbersome; therefore, the goal of~\cref{Invariants} will be to make it as hygienic as possible to work with.

\subsection{Reduced couplings for cosmology}\label{Invariants}

As in~\cref{Linear}, the action in~\cref{fullAction} can be reduced by using the FLRW forms of tensors which were introduced in~\cref{CosmologySetup}. The large symmetry of the cosmological ansatz restricts the forms of the operators in the action, causing some to vanish and others to be related to each other. Accordingly, we will next obtain the \emph{reduced} couplings in~\cref{mycoefficients}. Before quoting these results, however, we will discuss the formal procedure for obtaining them.

We define the operators of dimensions four and two as being respectively~$\CInvar{i}\equiv \delta S/\delta c_{i}$ and~$\AInvar{i}\equiv \delta S/ a_{i}$ for all relevant~$i$. The operators are hexic polynomials in~$\left\{\XInvar{1},\ldots,\XInvar{13}\right\}\equiv \left\{X,Y,Z,V,W,\dot{X},\dot{Y},\dot{W},a^{-1},\dot{a},\ddot{a},b^{-1},\dot{b}\right\}$ with rational coefficients. For QMAG, the problem of extracting from the~$\left\{c_{1},\ldots,c_{16}\right\}$ and~$\left\{a_{0},\ldots,a_{11}\right\}$ some minimal set of couplings relevant to the background cosmology is evidently related to finding all non-trivial solutions for~$\left\{\chi_{1},\ldots,\chi_{16}\right\}$ and~$\left\{\alpha_{0},\ldots,\alpha_{11}\right\}$ to the equations
\begin{equation}\label{InvarEqns}
    \sum_{i=1}^{16} \chi_{i} \CInvar{i} \equiv 0, \quad \sum_{i=0}^{11} \alpha_{i} \AInvar{i} \equiv 0.
\end{equation}
The problem in~\cref{InvarEqns} is linear: its non-linear generalisation in commutative algebra is the so-called \emph{ideal of relations} for each of the algebraically dependent sets~$\{\CInvar{1},\ldots,\CInvar{16}\}$ and~$\{\AInvar{0},\ldots,\AInvar{11}\}$~\cite{CommAlg}. This latter problem is not relevant to QMAG, having instead applications to quartic or hexic MAG and beyond.\footnote{Moreover, quartic and hexic MAG are additionally spanned by many more operators which are not complete squares of the QMAG operators, in addition to cubic operators and quintic operators.} Nonetheless, the full ideal of relations can be obtained with surprisingly little effort using computer algebra techniques, and so we use two such techniques to solve~\cref{InvarEqns}.

A Gr{\"o}bner basis technique was used for the~$\left\{\AInvar{0},\ldots,\AInvar{11}\right\}$. We introduce new indeterminates~$\left\{\mathsf{a}_{0},\ldots,\mathsf{a}{_{11}}\right\}$, and ascribe a monomial elimination order such that for any~$i$ and~$j$ we have~$\XInvar{i}\succ\mathsf{a}_{j}$. With this order, we compute the Gr{\"o}bner basis for the ideal~$\left\langle\mathsf{a}_{0}-\AInvar{0},\ldots,\mathsf{a}_{11}-\AInvar{11}\right\rangle$, which is a subset of the polynomial ring~$\mathbb{Q}\left[\XInvar{1},\ldots, \XInvar{13},\mathsf{a}_{0},\ldots, \mathsf{a}_{11}\right]$ over the rationals. By dropping all elements of this basis which refer explicitly to the~$\left\{\XInvar{1},\ldots,\XInvar{13}\right\}$, so the ideal of relations is obtained. Those elements of the ideal of relations which are linear in the~$\left\{\mathsf{a}_{0},\ldots,\mathsf{a}{_{11}}\right\}$ evidently solve for the~$\left\{\alpha_{0},\ldots\alpha_{11}\right\}$ in~\cref{InvarEqns}, once the identifications~$\mathsf{a}_i\equiv\AInvar{i}$ are made for all~$i$. Using the \Mathematica{} function \texttt{GroebnerBasis}, we find the three relations
\begin{equation}\label{MD2Invar}
    \begin{gathered}
        \AInvar{1}+
        \AInvar{2}-\AInvar{3}\equiv 0,\quad  3\AInvar{9} + \AInvar{10} - \AInvar{11}\equiv 0,
        \\3\AInvar{4} - 3\AInvar{5} - \AInvar{6} - \AInvar{7}+ 2\AInvar{8}\equiv 0.
    \end{gathered}
\end{equation}
Consequently, of the 11 original non-Riemannian operators, it follows from~\cref{MD2Invar} that only eight are independent. These relations are consistent with the results in~\cite{Iosifidis_2022}. The non-linear extension also reveals a fourth, quartic relation: this is shown in~\cref{NonRiemRel3}.

The \texttt{GroebnerBasis} implementation was not powerful enough for the~$\left\{\CInvar{1},\ldots,\CInvar{16}\right\}$. An alternative method for finding the ideal of relations is to find the kernel of the ring map~$F: \mathbb{Q}[\CInvar{1},\ldots, \CInvar{16}] \mapsto \mathbb{Q}[\XInvar{1},\ldots, \XInvar{13}]$, i.e. those elements which map to zero, since these are the polynomial functions of the~$\left\{\CInvar{1},\ldots,\CInvar{16}\right\}$ which identically vanish. An algorithm implemented in \Macaulay~\cite{M2} finds
\begin{gather}
\CInvar{16} - \CInvar{1} - 2\CInvar{2} - \CInvar{6} - 4\CInvar{5} -\CInvar{4} + 6\CInvar{12} + \CInvar{3} \equiv0,
\nonumber\\
\CInvar{8} - \CInvar{1} - \CInvar{2} - \CInvar{6} - 2\CInvar{5} - \CInvar{4} + 2\CInvar{12} + \CInvar{10} \equiv0,
\nonumber\\
\CInvar{7} - \CInvar{8} \equiv \CInvar{9} - \CInvar{10} \equiv \CInvar{11} - \CInvar{12} \equiv0,
\nonumber\\
\CInvar{13}\equiv \CInvar{14}\equiv \CInvar{15}\equiv 0.\label{curvatureRelations}
\end{gather}
This implementation finds in addition to~\cref{curvatureRelations} two extra relations in~\cref{CurvRel1,CurvRel3} which apply to quartic MAG, and one invariant in~\cref{CurvRel2} which applies to hexic MAG. The final three relations in~\cref{curvatureRelations} follow by construction because the homothetic curvature~$\tensor{\Hat{R}}{_{\mu\nu}}\equiv 0$ is defined to be the Maxwell operator for the Weyl vector, and the Maxwell tensor always vanishes in cosmology. Similarly, the third, fourth and fifth relations in~\cref{curvatureRelations} are equivalent to
\begin{equation}\label{MD4Invar}
    \begin{gathered}
        \tensor{R}{^{[\bar{\mu}\bar{\nu}]}} \tensor{R}{_{[\bar{\mu}\bar{\nu}]}}\equiv 
        \tensor{\check{R}}{^{[\bar{\mu}\bar{\nu}]}} \tensor{\check{R}}{_{[\bar{\mu}\bar{\nu}]}}\equiv 
        \tensor{R}{^{[\bar{\mu}\bar{\nu}]}}\tensor{\check{R}}{_{[\bar{\mu}\bar{\nu}]}}\equiv 0,
    \end{gathered}
\end{equation}
which is to be expected for the same reason. The conditions in~\cref{curvatureRelations} imply that of the original 16 curvature operators, only eight are independent. This is also consistent with the results in~\cite{Iosifidis_2022}. In summary, of the total 28 operators entering~\cref{fullAction}, we expect only 17 to be linearly independent. The couplings entering in~\cref{fullAction} are variables that form the basis of a 28-dimensional space, and their linear combinations in front of the monomials are vectors in that space. This linear system can be reduced via standard Gaussian elimination to yield a set of 17 linearly independent coefficients: we relabel these as~$\left\{a_{0},d_{1},\ldots,d_{16}\right\}$. Thus the Einstein--Hilbert parameter~$a_{0}$ already plays a unique role in the cosmology; the relevant formulae for the other new couplings are:
\begin{widetext}
\begin{equation}\label{mycoefficients}
    \begin{gathered}
    d_{1} \equiv -6 \left(2 c_{1}+2 c_{10}-2 c_{11}-2 c_{12}+6 c_{16}-2 c_{2}+2 c_{3}+c_{4}-c_{5}+c_{6}+2 \left(c_{7}+c_{8}+c_{9}\right)\right),
    \\
    d_{2} \equiv 12 \left(c_{10}-c_{11}-c_{12}+6 c_{16}+c_{7}+c_{8}+c_{9}\right),
    \\
    d_{3} \equiv -6 \left(4 c_{1}+4 c_{10}-5 c_{11}-5 c_{12}+2 \left(9 c_{16}-2 c_{2}+2 c_{3}+c_{4}-c_{5}+c_{6}+3 \left(c_{7}+c_{8}\right)+2 c_{9}\right)\right),
    \\
    d_{4} \equiv -3 \left(6 c_{1}+7 c_{10}-6 c_{11}-6 c_{12}+27 c_{16}-4 c_{2}+5 c_{3}+3 c_{4}-2 c_{5}+3 c_{6}+9 \left(c_{7}+c_{8}\right)+7 c_{9}\right),
    \\
    d_{5} \equiv 6 \left(14 c_{1}+4 c_{10}-4 c_{11}-4 c_{12}+12 c_{16}-14 c_{2}+12 c_{3}-7 c_{4}+7 c_{5}-7 c_{6}+4 \left(c_{7}+c_{8}+c_{9}\right)\right),
    \\
    d_{6} \equiv 6 \left(12 c_{1}+4 c_{10}-5 c_{11}-5 c_{12}+18 c_{16}-12 c_{2}+8 c_{3}-2 c_{4}+4 c_{5}-6 c_{6}+6 \left(c_{7}+c_{8}\right)+4 c_{9}\right),
    \\
    d_{7} \equiv -6 \left(12 c_{1}+6 c_{10}-5 c_{11}-5 c_{12}+18 c_{16}-12 c_{2}+8 c_{3}-6 c_{4}+4 c_{5}-2 c_{6}+4 \left(c_{7}+c_{8}\right)+6 c_{9}\right), 
    \\
    d_{8} \equiv 24 \left(c_{1}-c_{6}\right), \quad
    d_{9} \equiv 3 \left(2 a_{4} + a_{5} + 3 a_{7}\right), \quad
    d_{10} \equiv 3 \left(3 a_{11} + 4 a_{4} + 2 a_{5} + 6 a_{7} + a_{9}\right),
    \\
    d_{11} \equiv 3 \left(2 a_{1} + 
   3 a_{11} + a_{2} + 3 a_{3} + 2 a_{4} + a_{5} + 3 a_{7} + a_{9}\right),
    d_{12} \equiv -6 \left(2 a_{7} + 
   a_{8}\right), \quad
    d_{13} \equiv -6 \left(a_{10} + a_{11} + 2 a_{7} + a_{8}\right),
    \\
    d_{14} \equiv4 \left(a_{4} + a_{5} + a_{6} + a_{7} + 
   a_{8}\right), \quad
    d_{15} \equiv -3 \left(3 a_{11} - 4 a_{5} - 6 a_{8} + a_{9}\right), \quad
    d_{16} \equiv -24 \left(a_{1} - a_{2}\right).
     \end{gathered}
\end{equation}
\end{widetext}
Note in~\cref{mycoefficients} that the couplings~$\left\{d_{1},\ldots,d_{8}\right\}$ are of mass dimension zero whilst the~$\left\{d_{9},\ldots,d_{16}\right\}$ are of mass dimension two.

\subsection{Minisuperspace computation}\label{MSS}

The minisuperspace method outlined in~\cref{Linear} is now used to calculate the modified Friedmann equations corresponding to the full model in~\cref{fullAction}. As found in~\cref{Invariants}, the minisuperspace Lagrangian density corresponding to~\cref{fullAction} can be expressed entirely in terms of the reduced couplings in~\cref{mycoefficients}. This action is then varied with respect to the seven fields~$\left\{a,b,X,Y,Z,V,W\right\}$, subsequently setting the `lapse-like' function to unity and so restricting to FLRW coordinates by passing from~\cref{FLRWelement} to~\cref{FLRWBasic}. To our knowledge, the full equations were not yet obtained elsewhere. They are, however, very cumbersome, and we defer them to~\cref{AppendixFried}.

The~$b$-equation is a constraint which we take to be the first modified Friedmann equation, in the sense that it always allows one to solve for the term proportional to~$a_0 H^2$. The first Friedmann equation in~\cref{Friedmann1} has mass dimension four, and contains besides the~$a_0 H^2$ term a whole host of other terms. Terms linear in~$W$ or its velocities are generally absent, since these are pseudoscalar in nature, whilst~\cref{fullAction} is parity-preserving.

We prefer that the second modified Friedmann equation should allow one to solve for the term~$a_0 \dot{H}$; it is useful to construct this equation as a combination of the other equations.\footnote{This process is purely for mathematical convenience and, since we are not discarding any on-shell relations, it does not constrain the theory in any way.} Thus, the second modified Friedmann equation is constructed from the equations of motion for~$a$,~$b$,~$X$ and~$Y$ according to
\begin{equation}\label{F2fromABXY}
	\frac{\delta S}{\delta a} - 3\frac{\delta S}{\delta b}+\frac{\mathrm{d}}{\mathrm{d}t}\left(\frac{\delta S}{\delta X}-\frac{\delta S}{\delta Y} \right) = 0.
\end{equation}
The reason for the final term in~\cref{F2fromABXY} is to remove from the~$a$-equation in~\cref{Friedmann2} the triple-derivative structure
\begin{equation}\label{TripleDeriv}
	\frac{\delta S}{\delta a}\supset 2 d_1 \dddot{H}+\left(3 d_1-d_3-\tfrac{d_2}{2}\right) \dddot{X}+\left(d_1-d_3-\tfrac{d_2}{2}\right)\dddot{Y}.
\end{equation}
As an aside, an alternative route to removing the higher-derivative terms in~\cref{TripleDeriv} is to set the coefficients in~\cref{TripleDeriv} to zero, resulting in a special case of QMAG. Indeed, the resulting conditions~$d_1 = 0$ and~$d_2 + 2d_3 = 0$ greatly simplify the whole system, though we do not assume them moving forward. In GR, the second Friedmann equation can be derived from the first by taking its derivative and subtracting from it the product of itself with the Hubble parameter. In the metric-affine case, this procedure generalises to the identity
\begin{equation}\label{Bianchi?}
    \begin{aligned}
	    H\frac{\delta S}{\delta a} & \equiv \left(\frac{\mathrm{d}}{\mathrm{d}t}+3H\right)\frac{\delta S}{\delta b} 
	-\dot{X}\frac{\delta S}{\delta X}
	    \\ &\ \ \ 
	-\dot{Y}\frac{\delta S}{\delta Y}
	-\dot{Z}\frac{\delta S}{\delta Z}
	-\dot{V}\frac{\delta S}{\delta V}
	-\dot{W}\frac{\delta S}{\delta W},
    \end{aligned}
\end{equation}
which holds both on- and off-shell. The reasoning behind~\cref{Bianchi?} is as follows. In the case of GR,~\cref{Bianchi?} holds because of the projection of the Riemannian Bianchi identity~$2\tensor{u}{^\nu}\tensor{\tilde{\nabla}}{^{\mu}}\tensor{\tilde{R}}{_{\mu\nu}}\equiv\tensor{u}{^\nu}\tensor{\tilde{\nabla}}{_{\nu}}\tensor{\tilde{R}}{}$ along the time direction. When the theory is extended from GR to~\cref{fullAction}, the Lagrangian density acquires (i) terms quadratic in~$\tensor{\tilde{R}}{^\mu_{\nu\rho\sigma}}$, (ii) a dynamical theory of~$\tensor{N}{^\lambda_{\mu\nu}}$ minimally coupled to the metric, and (iii) non-minimal couplings between~$\tensor{N}{^\lambda_{\mu\nu}}$ and~$\tensor{\tilde{R}}{^\mu_{\nu\rho\sigma}}$ (i.e., the cross-terms from the PRD of operators quadratic in~$\tensor{R}{^\mu_{\nu\rho\sigma}}$). Focussing just on the type-(i) terms, it is well known that the Bianchi identity extends to QG~\cite{Kolekar:2022vkp} --- a result which is consistent with the LHS and the first term on the RHS of~\cref{Bianchi?}. To understand the origin of the remaining terms on the RHS of~\cref{Bianchi?}, we consider the type-(ii) operators. These extend QG with an effective matter sector, such that the Bianchi identity must be extended by the four-divergence of the effective stress-energy tensor. As expected, when~$\tensor{N}{^\lambda_{\mu\nu}}$ is restricted as in~\cref{distFLRW}, it can be shown that the extra terms in~\cref{Bianchi?} derive from applying the operator~$\tensor{u}{^\nu}\tensor{\tilde{\nabla}}{^{\mu}}\delta /\delta\tensor{g}{^{\mu\nu}}$ to the action of a \emph{general} theory of~$\tensor{N}{^\lambda_{\mu\nu}}$ minimally coupled to the metric, and then assuming FLRW-type background values for all fields. This also explains why~\cref{Bianchi?} is invariant under linear redfinitions of the distortion scalars, i.e. it doesn't matter how we normalise the~$\left\{X,Y,Z,V,W\right\}$. The type-(iii) terms are similarly consistent with~\cref{Bianchi?}.

As a calibration of the full system of equations we compared against~\cite{Iosifidis_2022,obukhovQuad}, where an action quadratic in torsion and non-metricity was considered and the vacuum dynamics found to reduce to GR. In terms of the reduced couplings in~\cref{mycoefficients}, this model corresponds to setting all the~$\left\{d_{1},\ldots,d_{8}\right\}$ to zero. After simplifying, the result is
\begin{equation}\label{reductionofQMAG}
    H = W = X = Y = V = Z = 0,
\end{equation}
which differs from~\cref{LinearFriedmann} by the requirement that~$V$ and~$Z$ both vanish. This is an expected consequence of the broken projective invariance of the model in~\cite{Iosifidis_2022,obukhovQuad}.

\subsection{Conditions for projective invariance}\label{case2}

It remains to explain the claim, regarding~\cref{LinearFriedmann}, that projective-invariant models exclusively propagate the linear combination~$V+Z$, rather than~$V$ and~$Z$ separately. The transformation of the affine connection in~\cref{Projective} can be equivalently viewed as a transformation of the distortion
\begin{equation}\label{ProjectiveDist}
    \tensor{N}{^{\lambda}_{\mu\nu}} \mapsto \tensor{N}{^{\lambda}_{\mu\nu}} + \delta_{\mu}^{\lambda} \tensor{\xi}{_{\nu}},
\end{equation}
where~$\tensor{\xi}{_{\nu}}\equiv \tensor{\xi}{_{\nu}}(x)$ is an arbitrary local vector gauge generator. By combining~\cref{distFLRW,projdef}, the distortion in~\cref{PRD} is restricted by cosmology to
\begin{equation}\label{ExpandedDist}
\begin{aligned}
\tensor{N}{^{\lambda}_{\mu\nu}} &= (X+V+Y+Z)~\tensor{u}{^{\lambda}}\tensor{u}{_{\mu}}\tensor{u}{_{\nu}} + X \tensor{g}{_{\mu\nu}}\tensor{u}{^{\lambda}} \\&\ \ \  + Z ~\delta_{\mu}^{\lambda}\tensor{u}{_{\nu}} +Y ~\delta_{\nu}^{\lambda}\tensor{u}{_{\mu}} + W \tensor{\epsilon}{^{\lambda}_{\mu\nu\kappa}}\tensor{u}{^\kappa}.
\end{aligned}
\end{equation}
The only projective transformation which is consistent with isotropy and homogeneity is~$\tensor{\xi}{_{\nu}}=\lambda~\tensor{u}{_{\nu}}$, where~$\lambda\equiv\lambda(t)$. By combining~\cref{ProjectiveDist,ExpandedDist} it follows that the scalars transform according to
\begin{equation}\label{ProjectiveScalars}
Z \mapsto Z - \lambda, \quad V \mapsto V + \lambda,
\end{equation}
and from~\cref{ProjectiveScalars} the invariance of~$V+Z$ follows immediately. As a consequence, projective invariance in cosmology requires the minisuperspace Lagrangian just to depend on the combination~$V+Z$. Equivalently, the condition
\begin{equation}\label{ProjDiff}
    \frac{\delta S}{\delta V} - \frac{\delta S}{\delta Z} = 0,
\end{equation}
must hold also off-shell, not just on-shell, for a model invariant under the projective transformation. Calculating~\cref{ProjDiff} explicitly for the QMAG model and couplings in~\cref{fullAction,mycoefficients} yields
    \begin{align}
        &\left(d_{10} + 2 d_{12} - d_{13} - 8 d_{14} + d_{15} + 2 d_{9}\right) V  + \left(-d_{12} + d_{15}\right) X 
	   \nonumber \\&
	    + \left(-2 d_{11} + 3 d_{12} - 4 d_{13} + d_{15} + 2 d_{9}\right) Y  + \big(-3 d_{10} + 2 d_{11}
\nonumber  \\&
	    - 8 d_{12} + 7 d_{13} + 24 d_{14} - 5 d_{15} - 8 d_{9}\big) Z= 0.\label{ProjDiff2}
    \end{align}
If~\cref{ProjDiff2} holds off-shell as well as on-shell, then the coefficients of each scalar must vanish separately. When reduced as far as possible, and translated back into the couplings in~\cref{fullAction} using~\cref{mycoefficients}, the conditions are
\begin{equation}\label{ProjectiveEqn}
\begin{gathered}
2a_{1} - 4 a_{10} - a_{11} + a_{2} + 3 a_{3} + a_{9} = 0,\\
 4 a_{4} - 3 a_{10} + 16 a_{6} + 2 a_{8} + a_{9} = 0,\\
 4 a_{5} - 3 a_{11} + 4 a_{7} + 8 a_{8} - a_{9} = 0.
\end{gathered}
\end{equation}
It is noted that~\cref{ProjectiveEqn} are precisely the constraints on the mass-dimension-two parameters obtained in~\cite{Percacci_2020,Iosifidis:2019fsh}. In~\cite{Percacci_2020}, three constraints on the mass-dimension-zero couplings are obtained, which beyond~\cref{ProjectiveEqn} are additionally needed for full projective invariance of~\cref{fullAction} in completely general spacetimes. By contrast, the mass-dimension-zero couplings are entirely unconstrained by invariance under our cosmological projective transformation in~\cref{ProjectiveScalars}: all the quadratic curvature operators automatically have this restricted version of the symmetry. In fact, this is to be expected. The restricted transformation is a special case of the scenario~$\tensor{\xi}{_\mu}=\tensor{\partial}{_\mu}\phi$ where a conservative field --- equivalently the local scalar~$\phi\equiv\phi(x)$ --- acts as the vector generator. It has been shown already in~\cite{Iosifidis:2018zwo} that any QMAG constructed entirely from curvature operators supports this restricted symmetry. In the case of cosmology, the scalar generator `rolls' in the direction of time according to~$\phi\equiv\phi(t)$, and in this sense it may be termed a \emph{khronon} field~\cite{Koivisto:2019ejt}.

\section{Special models and their particles}\label{Specials}
The field equations obtained in~\cref{Quadratic} still lack any predictive power, due to the abundance of unconstrained parameters. We will now identify some special cases of the model in~\cref{fullAction} whose cosmological dynamics have some interesting features. Regardless of how compelling these features might be, this is evidently a very superficial analysis, not extending, for example, to the perturbation theory. This is a severe limitation, because almost all metric-affine models one can write down are expected to be inconsistent as physical theories, regardless of how well they compare against observations. The most pernicious inconsistencies cannot be detected at the level of the background cosmology, having instead to do with unitarity violation, or strong coupling.

As a first step towards consistency, we compute for each special case the linear particle spectra near Minkowski spacetime. If the models admit a perturbative, weak gravity regime at all, then the elimination of higher-spin particles along with ghosts and tachyons is a necessary (but not sufficient) condition for consistency. The computations are performed with the \PSALTer{} software~\cite{PSALTer}, and the resulting particle spectrographs are presented in~\cref{ParticleSpectra}.

\subsection{Gravity with~$K$-screening}\label{spectra}

As mentioned already in~\cref{CosmologySetup}, generally accepted bounds on the \textLambda CDM model~\cite{Planck:2018vyg} are not strong enough to guarantee that the spatially flat line element in~\cref{FLRWelement} describes the background metric of our Universe. In general, some curvature scale~$K$ should be assumed, according to
\begin{equation}\label{FLRWCurved}
	\mathrm{d}s^{2}= -\mathrm{d}t^2 + a^2\left[\frac{\mathrm{d}r^2}{1-Kr^2}+r^2\left(\mathrm{d}\uptheta^2+\sin^2(\uptheta)\mathrm{d}\upphi^2\right)\right].
\end{equation}
All the methods and conclusions of~\cref{Quadratic} apply equally well to the spatially curved line element in~\cref{FLRWCurved}, though the expressions are more cumbersome and the full field equations are confined to the supplemental materials~\cite{Supplement}. An obvious question to ask, given this full set of equations, is whether it is possible to completely screen~$K$ from the background dynamics. This so-called `$K$-screening' effect certainly does not apply in GR, so that~$K$ is in principle an important observable. Recent observational probes of \textLambda CDM are found to be in moderate tension over~$K$, and their exclusion suggests~$K$ to be a small positive scale, or an effectively negative energy density weighing in at a few percent of the combined baryonic and dark-sector energy budget~\cite{Handley:2019tkm}. If this is true, then the Riemannian geometry of the observable Universe is very slightly closed. In any case, the smallness of this curvature scale (the base model of \textLambda CDM actually \emph{assumes}~$K=0$) is a key motivation for inflation in the early Universe. Just at the background level, therefore, the possibility of a~$K$-screening effect seems like a promising starting point for modifying GR so as to alleviate cosmological tensions, or formulate a more holistic replacement for \textLambda CDM.

There is, in fact, a clear precedent for~$K$-screening leading to a compelling non-Riemannian model. Thorough analyses of the particle spectra of parity-preserving, quadratic PGT (i.e. the~$\tensor{Q}{_{\mu\nu\sigma}}\equiv 0$ limit of MAG) identified certain special cases which (i) are free of ghosts and tachyons, (ii) propagate light d.o.f which might plausibly correspond to graviton polarisations, and (iii) are power-counting renormalisable~\cite{Lin:2018awc,Lin:2019ugq}. The presence of a graviton-like particle is very significant in these models, because the PGT~$K$-screening conditions eliminate the Einstein--Hilbert term.\footnote{Since the Riemannian Einstein--Hilbert term may emerge dynamically after post-Riemannian expansion, this feature does not automatically disqualify such theories as `lacking an Einstein limit'. Such analyses should be performed on a case-by-case basis, which goes beyond the scope of the current work.} A study of the background cosmology later revealed that these promising models are precisely those which are~$K$-screened~\cite{Barker:2020gcp}. Moreover, these models affect the background dynamics in other ways, suggesting a possible mechanism for alleviating yet another tension within \textLambda CDM --- the so-called \emph{Hubble tension} --- regarding the current value of~$H$~\cite{Barker:2020gcp,Barker:2020elg,Rew:2023zxy}. 

For the full MAG model in~\cref{fullAction}, and referring to the reduced couplings in~\cref{mycoefficients}, we find that~$K$-screening occurs (i.e.~$K$ is eliminated from all the equations) when the following conditions are satisfied:
\begin{equation}\label{kscreened}
    a_{0} = d_{1} = d_{2} = d_{3} = 0.
\end{equation}
Thus, in MAG as with PGT, the Einstein--Hilbert term is excluded by~\cref{kscreened}. We will make no attempt here to connect~\cref{kscreened} with the exact solutions proposed in~\cite{Barker:2020gcp}, but instead proceed directly to the particle analysis. The constraints in~\cref{kscreened} are so-far quite minimal, and even when~$a_{1}$ and~$a_{4}$ are the only mass-dimension-two parameters present, the preliminary analysis in~\cref{ParticleSpectra} suggests a rich spectrum of massive modes. By imposing further, minimal constraints in an algorithmic manner, it is possible to ensure that the squares of the masses of all the particles are given by rational functions of the remaining Lagrangian couplings. The resulting model propagates only a massive~$0^-$ particle (see unitarity conditions in~\cref{UnitarityConditionsKScreened}), without any massleess d.o.f. This is somehow disappointing, but hardly surprising. Indeed, the conditions~\cref{KScreeningRules,NoA,Rule2m,Rule1p,Rule1mA,Rule1mB}, when imposed on~\cref{fullAction}, reduce the quadratic curvature sector to the square of the Holst pseudoscalar~$\tensor{\epsilon}{_{\mu}^{\nu\rho\sigma}}\tensor{R}{^\mu_{\nu\rho\sigma}}$. It is well known (see e.g.~\cite{Barker:2024dhb}) that this operator propagates a single massive pseudoscalar particle (without strong coupling). Moreover, it does not involve the Riemannian curvature at all: this is consistent with the expectation of~$K$-sceening. Our analysis shows only that~$K$-screening in QMAG \emph{can} lead to wholly self-consistent theoretical models, it does not (yet) point to a phenomenologically viable model. In particlular, a replacement for the graviton is still needed. This should presumably be derived from the~$2^+$ or~$2^-$ sectors of the theory, which still contain a priori many massive modes once the~$K$-screening condition in~\cref{kscreened} is imposed. A thorough investigation of this matter is left to future work, as is an analysis of the rather large array of source constraints in the minimal Holst-squared model in~\cref{ParticleSpectrographKScreeningNoA2m1p1mA1mB}.

\subsection{Integrable gravity and its Maxwell limit}\label{EvenHubble}
Finally, we provide one alternative branch of MAG to that proposed in~\cref{spectra}, motivated by an exact solution to the cosmological equations of motion. Our starting point will be to remove odd powers of~$H$ from the equations, wherever they are not coupled to other fields. This leads to the conditions
\begin{equation}\label{evenhubble}
    d_{6} = d_{7}, \quad d_{5} = d_{4} = d_{3} = d_{2} = d_{1} = 0.
\end{equation}
Whilst~\cref{evenhubble} are somewhat arbitrary, they greatly simplify the field equations, which can be written down compactly as:
\begin{subequations}
    \begin{align}
            &6 a_{0} H^2 + d_{14} V^2 - (6 a_{0} - d_{16}) W^2 + (3 a_{0} + d_{12}) V X + d_{9} X^2 
	    \nonumber\\&\ \ \
	    + d_{8} W^2 X^2 + (3 a_{0} + d_{13}) V Y - (6 a_{0} - d_{10}) X Y
	    \nonumber\\&\ \ \
	    - 2 ( d_{7} -  d_{8}) W^2 X Y + d_{11} Y^2 - (2 d_{7} - d_{8}) W^2 Y^2 
	    \nonumber\\&\ \ \
	    + (d_{10} + 2 d_{12} - d_{13} - 6 d_{14} + d_{15} + 2 d_{9}) V Z 
	    \nonumber\\&\ \ \
	    + (3 a_{0} + d_{15}) X Z +(3 a_{0} - 2 d_{11} + 3 d_{12} - 3 d_{13} 
	    \nonumber\\&\ \ \
	    + d_{15} + 2 d_{9}) Y Z - (d_{10} - d_{11} + 3 d_{12} - 3 d_{13} - 9 d_{14} 
	    \nonumber\\&\ \ \
	    + 2 d_{15} + 3 d_{9}) Z^2 = 0,
	    \label{C1}
\\\nonumber\\
	    &12 a_{0} \dot{H}- d_{7} W \left(5 \dot{W} \left(X+Y\right)+W \left(\dot{X}+\dot{Y}\right)\right)= 0,
	    \label{C2}
\\\nonumber\\
	    &(3 a_{0} + d_{12}) V + d_{7} H W^2 + 2 d_{9} X + 2 d_{8} W^2 X 
	    \nonumber\\&\ \ \
	    + (d_{10} - 6a_{0}) Y + 2 (d_{8} - d_{7}) W^2 Y + (3 a_{0} + d_{15}) Z 
	    \nonumber\\&\ \ \
	    - d_{7} W \dot{W} = 0,
	    \label{C3}
\\\nonumber\\
        &(3 a_{0} + d_{13}) V + d_{7} H W^2 + (d_{10} - 6a_{0}) X 
	    \nonumber\\&\ \ \
	    + 2 (d_{8} - d_{7}) W^2 X + 2 d_{11} Y + 2(d_{8} - 2 d_{7}) W^2 Y 
	    \nonumber\\&\ \ \
	    + (3 a_{0} - 2 d_{11} + 3 d_{12} - 3 d_{13} + d_{15} + 2 d_{9}) Z 
	    \nonumber\\&\ \ \
	    - d_{7} W \dot{W} = 0,
	    \label{C4}
\\\nonumber\\
        &(d_{10} + 2 d_{12} - d_{13} - 6 d_{14} + d_{15} + 2 d_{9}) V 
	    \nonumber\\&\ \ \
	    + (3 a_{0} + d_{15}) X + (3 a_{0} - 2 d_{11} + 3 d_{12} - 3 d_{13} 
	    \nonumber\\&\ \ \
	    + d_{15} + 2 d_{9}) Y - 2 (d_{10} - d_{11} + 3 d_{12} - 3 d_{13} - 9 d_{14} 
	    \nonumber\\&\ \ \
	    + 2 d_{15} + 3 d_{9}) Z = 0,
	    \label{C5}
\\\nonumber\\
	    &2 d_{14} V + (3 a_{0} + d_{12}) X + (3 a_{0} + d_{13}) Y + (d_{10} + 2 d_{12} 
	    \nonumber\\&\ \ \
	    - d_{13} - 6 d_{14} + d_{15} + 2 d_{9}) Z = 0,
	    \label{C6}
\\\nonumber\\
	    &2 (d_{16} - 6a_{0}) W + 5 d_{7} H W \left(X+Y\right) + 2 d_{8} W X^2 
	    \nonumber\\&\ \ \
	    + 4 (d_{8} - d_{7}) W X Y + 2(d_{8} - 2 d_{7}) W Y^2 
	    \nonumber\\&\ \ \ 
	    + d_{7} W \left(\dot{X} + \dot{Y}\right)= 0.
	    \label{C7}
    \end{align}
\end{subequations}
It is important to note that if~$W= 0$ then everything trivializes as, in this case, it follows that~$X= Y= Z= V= H=0$ as can be easily verified. We will therefore look for solutions with~$W \not= 0$, so that~\cref{C7} becomes
    \begin{align}
        & 2 (d_{16} - 6a_{0})  + 5 d_{7} H  \left(X+Y\right) + 2 d_{8}  X^2 + 4 (d_{8} - d_{7})  X Y \nonumber \\&
	    + 2(d_{8} - 2 d_{7})  Y^2 + d_{7}  \left(\dot{X} + \dot{Y}\right)= 0. \label{Weq}
    \end{align}
\paragraph*{The case~$\bm{d_7=0}$} The parameter~$d_{7}$ heavily influences the behaviour of the solutions. For~$d_{7}=0$, and after some simple algebra, it follows that
\begin{equation}
\begin{gathered}
H= H_{0}\equiv \mathrm{const.}, \quad
Y\propto V \propto Z\propto X= X_{0}\equiv \mathrm{const.},\\
W= W_{0}\equiv \pm \sqrt{\frac{\sigma_{1}X_{0}^{2}-6 a_{0}H_{0}^{2}}{\sigma_{2}X_{0}^{2}+\sigma_{3}}},
\end{gathered}
\end{equation}
where the proportionality factors as well as the~$\left\{\sigma_{i}\right\}$ are some long expressions of the coefficients~$a_{0}$ and~$\left\{d_{i}\right\}$. Therefore, for~$d_{7}=0$ one can only have de Sitter expansion and constant distortion.

\paragraph*{The case~$\bm{d_7\neq0}$}Now let us focus on the non-trivial~$d_{7}\neq 0$ case. Subtracting~\cref{C3} and~\cref{C4} we get
\begin{align}
&(d_{12}-d_{13})V+(2 d_{9}-d_{10}+6 a_{0})X
    +(d_{10}-6 a_{0}
    -2 d_{11})Y
    \nonumber\\&
    +(2 d_{11}-3 d_{12}+3 d_{13}-2 d_{9})Z+2 d_{7}W^{2}\left(X+Y\right)= 0. \label{C12}
\end{align}
When the system in~\cref{C5,C6} is not degenerate we can solve for~$V$ and~$Z$ in terms of~$X$ and~$Y$, viz.
\begin{equation}
V= k_{1}X+k_{2}Y \quad Z= k_{3}X+k_{4}Y, \label{C13}
\end{equation}
where again the~$\left\{k_{i}\right\}$ are some linear combinations of the parameters of the theory. We may then substitute~\cref{C13} back into~\cref{C12} and solve for~$W^{2}$, to yield
\beq
W^{2}= \frac{k_{5}X+k_{6}Y}{2d_{7}(X+Y)}\label{W2} = \frac{k_6}{2d_7}+\frac{(k_5-k_6)X}{2d_7(X+Y)}. 
\eeq
\paragraph*{The case~$\bm{k_5\neq k_6}$}Note that choosing the theory parameters such that~$k_{6}=k_{5}$ implies that~$W= \mathrm{const.}$. Let us firstly consider the case~$k_{6}\neq k_{5}$. On substituting~\cref{C13} into~\cref{C1} we obtain
\begin{equation}\label{Intermediate}
    \begin{aligned}
&6a_{0}H^{2}+k_{7}X^{2}+k_{8}XY+k_{9}Y^{2}+W^{2}\Big[d_{16}-6a_{0} \\&+ d_{8}(X+Y)^{2}-2 d_{7}Y(X+Y) \Big]= 0.
    \end{aligned}
\end{equation}
Then, using~\cref{W2} we may eliminate~$W^{2}$ from~\cref{Intermediate} and solve for~$H$ in terms of~$X$ and~$Y$ according to
\begin{align}
	&6 a_{0} H^{2} - \frac{6a_0-d_{16}}{2 d_7}\left[k_6+(k_5-k_6)\frac{X}{X+Y}\right]  
	\nonumber \\&-\left[\frac{d_{8}}{2 d_{7}}k_{5}-k_{7}\right]X^{2}
	-\left[ \frac{d_{8}}{2 d_{7}}k_{6}-k_6-k_9\right]Y^{2}
	\nonumber \\ &
	-\left[\frac{d_{8}}{2 d_{7}}( k_{5}+k_{6}) -k_5-k_8\right]X Y=0.
    \label{HXY1}
\end{align}
We have therefore expressed everything in terms of~$X$ and~$Y$. These can be inverted in terms of~$H$ and~$W$ as the dynamical variables or any other pick of two out of four. Then, by eliminating the rest of the variables,~\cref{C2} combined with either of~\cref{C3} or~\cref{C4} gives us a system of coupled second order ordinary differential equations for, say, the variables~$H$ and~$W$. The resulting system is, in general, somewhat complicated. However, we do get some quite simple expressions if~$d_{16}=6 a_{0}$ and in~\cref{HXY1} remaining terms on the RHS form a perfect square~$(X+Y)^{2}$. For such a configuration, taking the square root of the aforementioned equation, yields
\beq
H=\lambda_{0}(X+Y),
\eeq
provided the following conditions hold:
\beq
k_5+k_8=k_6+k_7+k_9, \quad \frac{d_8 (k_7-k_8+k_9)}{k_6-k_7+k_9}+2 d_7=0.
\eeq
Combining this with~\cref{W2} we deduce
\beq
X=\frac{W^{2}-\lambda_{1}}{\lambda_{0}\lambda_{2}}H.
\eeq
Then eliminating~$X$ and~$Y$, equations~\cref{C2} and~\cref{C3} read
\beq
12 a_{0}\lambda_{0}\dot{H}=d_{7}\left(5 W \dot{W}H+W^{2}\dot{H}\right), \label{HWW}
\eeq
\beq
\lambda_{3}H+\lambda_{4}H W^{2}+\lambda_{5}W^{4}H=\lambda_{0}d_{7} W \dot{W}. \label{HWW2}
\eeq
Now,~\cref{HWW} directly integrates to
\beq
H=C_{1}\left(12 a_{0}\lambda_{0}-d_7 W^{2}\right)^{-\frac{5}{2}}.
\eeq
Substituting this in~\cref{HWW2} and changing variables as 
\beq
u=12 a_{0}\lambda_{0}-d_7 W^{2},
\eeq
we obtain the differential equation
\beq
\dot{u}=-\frac{2 C_{1}}{\lambda_{0}}\frac{(\tilde{\lambda}_{1}+\tilde{\lambda}_{4}u+\tilde{\lambda}_{5}u^{2})}{ u^{\frac{5}{2}}}.
\eeq
This is then integrated straightforwardly but its exact form depends crucially on the values of the parameters. For instance, if~$\tilde{\lambda}_{1}=0=\tilde{\lambda}_{5}$, it follows that~$u\propto t^{2/5}$ and consequently~$H\propto \frac{1}{t}$, yielding a power-law behaviour for the scale factor.

\paragraph*{The case~$\bm{k_5 = k_6}$}Let us now focus on the~$k_{6}=k_{5}$ case. From~\cref{W2} we find 
\beq\label{C14}
W^{2}= W_{0}^{2}\equiv \frac{k_{5}}{2 d_{7}}.
\eeq
First, note that under the conditions
\beq
6 a_0=d_{16}, \quad k_6=k_5, \quad k_7+k_9=k_8, \quad k_5+2 k_9=k_8,
\eeq
then the RHS of~\cref{HXY1} is once again a perfect square and combining~\cref{HXY1} with~\cref{C14,C2} we get an~$H=\text{const.}$ de-Sitter solution (even with~$d_7\neq0$).

Returning to the general case of~$k_6=k_5$, with~\cref{C14} we find that~\cref{C2} directly integrates to
\beq
H= \frac{k_{5}}{24 a_{0}}(X+Y)+\CInvar{0} \label{HXY},
\eeq
with~$\CInvar{0}$ the integration constant. Then, for~$W= \mathrm{const.}$ and taking also into account~\cref{C13}, we find that~\cref{C3} implies
\beq\label{C15}
k_{10}X+k_{11}Y= k_{5}H.
\eeq
Combining~\cref{C15} with~\cref{HXY}, we get
\beq\label{C16}
X=\lambda_{1}H+X_{0}, \quad 
Y=\lambda_{2}H+Y_{0}.
\eeq
From~\cref{C16} it follows that~\cref{C13} can also be expressed as
\beq
V=\lambda_{3}H+V_{0}, \quad 
Z=\lambda_{4}H+Z_{0}.
\eeq
Geometrically it follows that in the configuration space of the cosmology~$\left\{H,X,Y,V,Z,W\right\}$ the~$k_{6}=0$ solutions are represented by lines lying on the~$W= \mathrm{const.}$ hyperplane. Of course this is not the end of the story, if we plug the above results back to~\cref{Weq} we obtain the first order differential equation\footnote{\cref{C17} also known as a Riccati equation.}
\beq\label{C17}
\dot{H}= \mu_{2}H^{2}+\mu_{1}H+\mu_{0}.
\eeq
Obviously,~\cref{C17} is separable and can be readily integrated. However, the form of the solutions depends on the values~$\left\{\mu_{i}\right\}$ which are given in terms of the parameters of the theory and the integration constant. For~$\mu_{2}=0$ and~$\mu_{1}\neq 0$ the solution reads
\beq
H= \frac{1}{\mu_{1}}\Big( C_{1}e^{\mu_{1}t}-\mu_{0}\Big),
\quad
a= C_{2}e^{\frac{1}{\mu_{1}}\Big( C_{1}e^{\mu_{1}t}-\mu_{0}t\Big)},
\eeq
i.e. a super-exponential expansion. For~$\mu_{2}\neq 0$ we can define~$\gamma\equiv\mu_{1}^{2}/4 \mu_{2}-\mu_{0}$ and shift~$\tilde{H}\equiv H+\mu_{1}/2\mu_{2}$, then~\cref{C17} is written as
\beq\label{C18}
\frac{\mathrm{d} \tilde{H}}{\tilde{H}^{2}+\gamma/\mu_{2}}= \mu_{2}\mathrm{d}t,
\eeq
and of course the form of the solution to~\cref{C18} depends on the ratio~$\gamma/\mu_{2}$. For~$\gamma/\mu_{2}<0$ and defining~$\omega_{0}^{2}\equiv|\gamma/\mu_{2}|$ we have
\beq
H= \omega_{0}\frac{1+C_{1}e^{2\mu_{2}\omega_{0}t}}{1-C_{1}e^{2\mu_{2}\omega_{0}t}},
\quad\label{C19}
a= \frac{C_{2}e^{\omega_{0}t}}{\left(1-C_{1}e^{2\mu_{2}\omega_{0}t}\right)^{1/\mu_{2}}}.
\eeq
Notice that for late times~\cref{C19} always implies de Sitter or anti-de Sitter expansion depending on whether~$\mu_{2}\omega_{0}$ is positive or negative. When~$\gamma/\mu_{2}>0$ the solution to~\cref{C18} is
\beq
H= -\frac{\mu_{1}}{2 \mu_{2}}+\omega_{0}\tan{\left(\mu_{2}\omega_{0}t+C_{1}\right)}.
\eeq

The integrable model contains too many unconstrained parameters for a thorough analysis of its particle spectrum. A severe restriction on the model set out in~\cref{evenhubble} is the case where all of the~$\left\{d_{1},\ldots,d_{8}\right\}$ are set to zero. Whilst the Lagrangian density may still contain quadratic curvature operators in this case, they will not affect the cosmology at the background level. Since quadratic curvature operators endow the distortion with kinetic terms, it is interesting to consider the particle spectrum of such a model. The analogue in Riemannian cosmology would be Einstein--Maxwell theory: the Maxwell operator does not affect cosmology at the background level\footnote{Of course, we know that radiative d.o.f actually dominate the background dynamics in the early Universe, but this is an average effect of locally anisotropic and inhomogeneous field configurations.}, yet it still contributes two massless photon polarizations to the spectrum. For simplicity, we take the case wheere the~$\left\{a_{1},\ldots,a_{11}\right\}$ are all taken to be zero, so that~$a_{0}$ is the only non-vanishing coupling of mass dimension two. As shown in~\cref{ParticleSpectra}, this leads to a unitary model when some further constraints are imposed (concretely, sequential imposition of~\cref{AntiCosmologicalRules,AllAZero,FurtherRules,InfiniteMassVector} on~\cref{fullAction}). According to the resulting particle spectrograph in~\cref{ParticleSpectrographMaxwellNoHigherSpinNo1m}, this unitarity is guaranteed by~\cref{MaxwellUnitary}, which suggests that~$c_{13}>0$ is the no-ghost condition on the new species. By comparing~\cref{MaxwellUnitary} with~\cref{fullAction} we see that~$c_{13}$ modulates the square of the Homothetic curvature. We may therefore conjecture that, despite the presence of many other kinetic operators, the propagating part of the model is effectively Einstein--Maxwell theory, where the Weyl vector is massless. We note also in~\cref{ParticleSpectrographMaxwellNoHigherSpinNo1m} the presence of \emph{five} source constraints. This is expected: there are four gauge generators associated with diffeomeorphisms and one associated with the integrable (i.e. conservative) Weyl symmetry discussed in~\cref{case2}, which is common to all QMAG models built from the curvature alone~\cite{Iosifidis:2018zwo}.

\section{Concluding remarks}\label{ConcludingRemarks}

\paragraph*{Main findings} The results of this work are as follows:
\begin{itemize}
	\item Based on the general (i.e. spatially curved) Friedmann--Lema\^itre--Robertson--Walker line element in~\cref{FLRWCurved}, we have derived the equations describing the background dynamics of the general parity-preserving quadratic metric-affine gravity model in~\cref{fullAction}. The full field equations are confined to the supplemental materials~\cite{Supplement}, where versions have been prepared for direct import into \SymPy, \Mathematica and \Maple platforms. The simplified equations for the spatially flat case are on the edge of becoming human-readable, and so are formatted in~\cref{Quadratic,AppendixFried}.
	\item Only the reduced set of 16 parameters in~\cref{mycoefficients} are relevant to the background dynamics, down from the 28 Lagrangian couplings which parameterise~\cref{fullAction}.
	\item In terms of the reduced parameters in~\cref{mycoefficients}, the conditions in~\cref{kscreened} completely eliminate the spatial curvature~$K$ from the field equations: this phenomenon is termed `\emph{$K$-screening}'.
	\item As an alternative to~$K$-screening, the conditions in~\cref{evenhubble} lead to simple equations which may be integrated analytically. This model always transitions to a de Sitter or anti-de Sitter epoch in the late Universe.	
\end{itemize}

\paragraph*{Further work} The following should be investigated:
\begin{itemize}
	\item A straightforward extension of this analysis to parity-violating quadratic metric-affine gravity.
	\item The use of~$K$-screening to construct models which alleviate current tensions in the dark-energy-cold-dark-matter (\textLambda CDM) model. Any serious attempt at solving these problems requires a complete understanding of the cosmological perturbation theory.
	\item The degree --- if any --- to which the parameter constraints identified in this work may lead to consistent models, i.e. those which are unitary and weakly coupled. The relevant tools in this endeavour include non-linear Hamiltonian analysis and functional renormalisation group methods.
\end{itemize}

\paragraph*{Final note of caution} We reiterate that the last point of further work is especially crucial: the analyses conducted here are superficial when compared to those which are needed to demonstrate consistency. Despite the colourful dynamics of the background cosmology, metric-affine gravity in general has a poor record in this regard, and to date only a very few special cases (mostly minimal extensions to general relativity itself) were shown to be consistent. It is perhaps telling that, starting with constraints motivated by the background cosmology, a systematic search for consistent weak-field particle spectra leads back to two such theories: the square-Holst theory (see~\cref{ParticleSpectrographKScreeningNoA2m1p1mA1mB}) and the square-homothetic theory (see~\cref{ParticleSpectrographMaxwellNoHigherSpinNo1m}). It is conjectured in~\cite{Barker:2024dhb} that these \emph{are} the only self-consistent kinetic operators in QMAG, though the point remains to be proven.

\begin{acknowledgements}
We are grateful for correspondence with Will Handley, Anthony Lasenby and Tobias Mistele. This work is based in part on a Master's thesis by TD, which was itself improved by feedback from the assessor, Paul Rimmer.

This work used the DiRAC Data Intensive service (CSD3 \href{www.csd3.cam.ac.uk}{www.csd3.cam.ac.uk}) at the University of Cambridge, managed by the University of Cambridge University Information Services on behalf of the STFC DiRAC HPC Facility (\href{www.dirac.ac.uk}{www.dirac.ac.uk}). The DiRAC component of CSD3 at Cambridge was funded by BEIS, UKRI and STFC capital funding and STFC operations grants. DiRAC is part of the UKRI Digital Research Infrastructure.

This work also used the Newton server, access to which was provisioned by Will Handley using an ERC grant.

WB is grateful for the support of Marie Skłodowska-Curie Actions and the Institute of Physics of the Czech Academy of Sciences.

The work of DI was supported by the Estonian Research Council via the Center of Excellence `\emph{Foundations of the Universe}' (TK202U4).
Co-funded by the European Union (Physics for Future -- Grant Agreement No.~101081515).
Views and opinions expressed are however those of the author(s) only and do not necessarily reflect those of the European Union or European Research Executive Agency.
Neither the European Union nor the granting authority can be held responsible for them.
\end{acknowledgements}

\bibliography{NotINSPIRE,Manuscript}

\begin{thebibliography}{159}%
\makeatletter
\providecommand \@ifxundefined [1]{%
 \@ifx{#1\undefined}
}%
\providecommand \@ifnum [1]{%
 \ifnum #1\expandafter \@firstoftwo
 \else \expandafter \@secondoftwo
 \fi
}%
\providecommand \@ifx [1]{%
 \ifx #1\expandafter \@firstoftwo
 \else \expandafter \@secondoftwo
 \fi
}%
\providecommand \natexlab [1]{#1}%
\providecommand \enquote  [1]{``#1''}%
\providecommand \bibnamefont  [1]{#1}%
\providecommand \bibfnamefont [1]{#1}%
\providecommand \citenamefont [1]{#1}%
\providecommand \href@noop [0]{\@secondoftwo}%
\providecommand \href [0]{\begingroup \@sanitize@url \@href}%
\providecommand \@href[1]{\@@startlink{#1}\@@href}%
\providecommand \@@href[1]{\endgroup#1\@@endlink}%
\providecommand \@sanitize@url [0]{\catcode `\\12\catcode `\$12\catcode
  `\&12\catcode `\#12\catcode `\^12\catcode `\_12\catcode `\%12\relax}%
\providecommand \@@startlink[1]{}%
\providecommand \@@endlink[0]{}%
\providecommand \url  [0]{\begingroup\@sanitize@url \@url }%
\providecommand \@url [1]{\endgroup\@href {#1}{\urlprefix }}%
\providecommand \urlprefix  [0]{URL }%
\providecommand \Eprint [0]{\href }%
\providecommand \doibase [0]{https://doi.org/}%
\providecommand \selectlanguage [0]{\@gobble}%
\providecommand \bibinfo  [0]{\@secondoftwo}%
\providecommand \bibfield  [0]{\@secondoftwo}%
\providecommand \translation [1]{[#1]}%
\providecommand \BibitemOpen [0]{}%
\providecommand \bibitemStop [0]{}%
\providecommand \bibitemNoStop [0]{.\EOS\space}%
\providecommand \EOS [0]{\spacefactor3000\relax}%
\providecommand \BibitemShut  [1]{\csname bibitem#1\endcsname}%
\let\auto@bib@innerbib\@empty
\bibitem [{\citenamefont {Hehl}\ \emph {et~al.}(1995)\citenamefont {Hehl},
  \citenamefont {McCrea}, \citenamefont {Mielke},\ and\ \citenamefont
  {Ne’eman}}]{Hehl_1995}%
  \BibitemOpen
  \bibfield  {author} {\bibinfo {author} {\bibfnamefont {F.~W.}\ \bibnamefont
  {Hehl}}, \bibinfo {author} {\bibfnamefont {J.}~\bibnamefont {McCrea}},
  \bibinfo {author} {\bibfnamefont {E.~W.}\ \bibnamefont {Mielke}},\ and\
  \bibinfo {author} {\bibfnamefont {Y.}~\bibnamefont {Ne’eman}},\ }\bibfield
  {title} {\bibinfo {title} {Metric-affine gauge theory of gravity: field
  equations, noether identities, world spinors, and breaking of dilation
  invariance},\ }\bibfield  {journal} {\bibinfo  {journal} {Physics Reports}\
  }\textbf {\bibinfo {volume} {258}},\ \href
  {https://doi.org/10.1016/0370-1573(94)00111-f} {10.1016/0370-1573(94)00111-f}
  (\bibinfo {year} {1995})\BibitemShut {NoStop}%
\bibitem [{\citenamefont {Iosifidis}(2019)}]{Iosifidis:2019dua}%
  \BibitemOpen
  \bibfield  {author} {\bibinfo {author} {\bibfnamefont {D.}~\bibnamefont
  {Iosifidis}},\ }\emph {\bibinfo {title} {{Metric-Affine Gravity and
  Cosmology/Aspects of Torsion and non-Metricity in Gravity Theories}}},\
  \href@noop {} {Ph.D. thesis} (\bibinfo {year} {2019}),\ \Eprint
  {https://arxiv.org/abs/1902.09643} {arXiv:1902.09643 [gr-qc]} \BibitemShut
  {NoStop}%
\bibitem [{\citenamefont {Iosifidis}(2020{\natexlab{a}})}]{Damos_main_2020}%
  \BibitemOpen
  \bibfield  {author} {\bibinfo {author} {\bibfnamefont {D.}~\bibnamefont
  {Iosifidis}},\ }\bibfield  {title} {\bibinfo {title} {Cosmological
  hyperfluids, torsion and non-metricity},\ }\bibfield  {journal} {\bibinfo
  {journal} {The European Physical Journal C}\ }\textbf {\bibinfo {volume}
  {80}},\ \href {https://doi.org/10.1140/epjc/s10052-020-08634-z}
  {10.1140/epjc/s10052-020-08634-z} (\bibinfo {year}
  {2020}{\natexlab{a}})\BibitemShut {NoStop}%
\bibitem [{\citenamefont {Percacci}\ and\ \citenamefont
  {Sezgin}(2020{\natexlab{a}})}]{Percacci_2020}%
  \BibitemOpen
  \bibfield  {author} {\bibinfo {author} {\bibfnamefont {R.}~\bibnamefont
  {Percacci}}\ and\ \bibinfo {author} {\bibfnamefont {E.}~\bibnamefont
  {Sezgin}},\ }\bibfield  {title} {\bibinfo {title} {New class of ghost- and
  tachyon-free metric affine gravities},\ }\bibfield  {journal} {\bibinfo
  {journal} {Physical Review D}\ }\textbf {\bibinfo {volume} {101}},\ \href
  {https://doi.org/10.1103/physrevd.101.084040} {10.1103/physrevd.101.084040}
  (\bibinfo {year} {2020}{\natexlab{a}})\BibitemShut {NoStop}%
\bibitem [{\citenamefont {Jim\'enez~Cano}(2021)}]{JimenezCano:2021rlu}%
  \BibitemOpen
  \bibfield  {author} {\bibinfo {author} {\bibfnamefont {A.}~\bibnamefont
  {Jim\'enez~Cano}},\ }\emph {\bibinfo {title} {{Metric-affine Gauge theories
  of gravity. Foundations and new insights}}},\ \href@noop {} {Ph.D. thesis},\
  \bibinfo  {school} {Granada U., Theor. Phys. Astrophys.} (\bibinfo {year}
  {2021}),\ \Eprint {https://arxiv.org/abs/2201.12847} {arXiv:2201.12847
  [gr-qc]} \BibitemShut {NoStop}%
\bibitem [{\citenamefont {Belarbi}\ and\ \citenamefont
  {Meziane}(2021)}]{Belarbi:2021qiw}%
  \BibitemOpen
  \bibfield  {author} {\bibinfo {author} {\bibfnamefont {O.~A.}\ \bibnamefont
  {Belarbi}}\ and\ \bibinfo {author} {\bibfnamefont {A.}~\bibnamefont
  {Meziane}},\ }\bibfield  {title} {\bibinfo {title} {{Overview and
  perspectives on metric-affine gravity}},\ }\href
  {https://doi.org/10.1088/1742-6596/1766/1/012007} {\bibfield  {journal}
  {\bibinfo  {journal} {J. Phys. Conf. Ser.}\ }\textbf {\bibinfo {volume}
  {1766}},\ \bibinfo {pages} {012007} (\bibinfo {year} {2021})}\BibitemShut
  {NoStop}%
\bibitem [{\citenamefont {Baldazzi}\ \emph {et~al.}(2022)\citenamefont
  {Baldazzi}, \citenamefont {Melichev},\ and\ \citenamefont
  {Percacci}}]{Baldazzi:2021kaf}%
  \BibitemOpen
  \bibfield  {author} {\bibinfo {author} {\bibfnamefont {A.}~\bibnamefont
  {Baldazzi}}, \bibinfo {author} {\bibfnamefont {O.}~\bibnamefont {Melichev}},\
  and\ \bibinfo {author} {\bibfnamefont {R.}~\bibnamefont {Percacci}},\
  }\bibfield  {title} {\bibinfo {title} {{Metric-Affine Gravity as an effective
  field theory}},\ }\href {https://doi.org/10.1016/j.aop.2022.168757}
  {\bibfield  {journal} {\bibinfo  {journal} {Annals Phys.}\ }\textbf {\bibinfo
  {volume} {438}},\ \bibinfo {pages} {168757} (\bibinfo {year} {2022})},\
  \Eprint {https://arxiv.org/abs/2112.10193} {arXiv:2112.10193 [gr-qc]}
  \BibitemShut {NoStop}%
\bibitem [{\citenamefont {Blagojevic}\ and\ \citenamefont
  {Hehl}(2012)}]{Blagojevic:2012bc}%
  \BibitemOpen
  \bibfield  {author} {\bibinfo {author} {\bibfnamefont {M.}~\bibnamefont
  {Blagojevic}}\ and\ \bibinfo {author} {\bibfnamefont {F.~W.}\ \bibnamefont
  {Hehl}},\ }\bibfield  {title} {\bibinfo {title} {{Gauge Theories of
  Gravitation}},\ }\href@noop {} {\  (\bibinfo {year} {2012})},\ \Eprint
  {https://arxiv.org/abs/1210.3775} {arXiv:1210.3775 [gr-qc]} \BibitemShut
  {NoStop}%
\bibitem [{\citenamefont {Einstein}(1915)}]{Einstein:1915}%
  \BibitemOpen
  \bibfield  {author} {\bibinfo {author} {\bibfnamefont {A.}~\bibnamefont
  {Einstein}},\ }\bibfield  {title} {\bibinfo {title} {{Die Feldgleichungen der
  Gravitation}},\ }\href@noop {} {\bibfield  {journal} {\bibinfo  {journal}
  {Sitzungsber. Preuss. Akad. Wiss}\ }\textbf {\bibinfo {volume} {18}},\
  \bibinfo {pages} {844} (\bibinfo {year} {1915})}\BibitemShut {NoStop}%
\bibitem [{\citenamefont {Hehl}\ \emph
  {et~al.}(1976{\natexlab{a}})\citenamefont {Hehl}, \citenamefont {Kerlick},\
  and\ \citenamefont {von~der Heyde}}]{Hehl:1976one}%
  \BibitemOpen
  \bibfield  {author} {\bibinfo {author} {\bibfnamefont {F.~W.}\ \bibnamefont
  {Hehl}}, \bibinfo {author} {\bibfnamefont {G.~D.}\ \bibnamefont {Kerlick}},\
  and\ \bibinfo {author} {\bibfnamefont {P.}~\bibnamefont {von~der Heyde}},\
  }\bibfield  {title} {\bibinfo {title} {On hypermomentum in general relativity
  i. the notion of hypermomentum},\ }\href
  {https://doi.org/doi:10.1515/zna-1976-0201} {\bibfield  {journal} {\bibinfo
  {journal} {Zeitschrift für Naturforschung A}\ }\textbf {\bibinfo {volume}
  {31}},\ \bibinfo {pages} {111} (\bibinfo {year}
  {1976}{\natexlab{a}})}\BibitemShut {NoStop}%
\bibitem [{\citenamefont {Hehl}\ \emph
  {et~al.}(1976{\natexlab{b}})\citenamefont {Hehl}, \citenamefont {Kerlick},\
  and\ \citenamefont {von~der Heyde}}]{Hehl:1976two}%
  \BibitemOpen
  \bibfield  {author} {\bibinfo {author} {\bibfnamefont {F.~W.}\ \bibnamefont
  {Hehl}}, \bibinfo {author} {\bibfnamefont {G.~D.}\ \bibnamefont {Kerlick}},\
  and\ \bibinfo {author} {\bibfnamefont {P.}~\bibnamefont {von~der Heyde}},\
  }\bibfield  {title} {\bibinfo {title} {On hypermomentum in general relativity
  ii. the geometry of spacetime},\ }\href
  {https://doi.org/doi:10.1515/zna-1976-0602} {\bibfield  {journal} {\bibinfo
  {journal} {Zeitschrift für Naturforschung A}\ }\textbf {\bibinfo {volume}
  {31}},\ \bibinfo {pages} {524} (\bibinfo {year}
  {1976}{\natexlab{b}})}\BibitemShut {NoStop}%
\bibitem [{\citenamefont {Hehl}\ \emph
  {et~al.}(1976{\natexlab{c}})\citenamefont {Hehl}, \citenamefont {Kerlick},\
  and\ \citenamefont {von~der Heyde}}]{Hehl:1976three}%
  \BibitemOpen
  \bibfield  {author} {\bibinfo {author} {\bibfnamefont {F.~W.}\ \bibnamefont
  {Hehl}}, \bibinfo {author} {\bibfnamefont {G.~D.}\ \bibnamefont {Kerlick}},\
  and\ \bibinfo {author} {\bibfnamefont {P.}~\bibnamefont {von~der Heyde}},\
  }\bibfield  {title} {\bibinfo {title} {On hypermomentum in general relativity
  iii. coupling hypermomentum to geometry},\ }\href
  {https://doi.org/doi:10.1515/zna-1976-0724} {\bibfield  {journal} {\bibinfo
  {journal} {Zeitschrift für Naturforschung A}\ }\textbf {\bibinfo {volume}
  {31}},\ \bibinfo {pages} {823} (\bibinfo {year}
  {1976}{\natexlab{c}})}\BibitemShut {NoStop}%
\bibitem [{\citenamefont {Iosifidis}(2021)}]{Iosifidis:2020upr}%
  \BibitemOpen
  \bibfield  {author} {\bibinfo {author} {\bibfnamefont {D.}~\bibnamefont
  {Iosifidis}},\ }\bibfield  {title} {\bibinfo {title} {{Non-Riemannian
  cosmology: The role of shear hypermomentum}},\ }\href
  {https://doi.org/10.1142/S0219887821501292} {\bibfield  {journal} {\bibinfo
  {journal} {Int. J. Geom. Meth. Mod. Phys.}\ }\textbf {\bibinfo {volume}
  {18}},\ \bibinfo {pages} {2150129} (\bibinfo {year} {2021})},\ \Eprint
  {https://arxiv.org/abs/2010.00875} {arXiv:2010.00875 [gr-qc]} \BibitemShut
  {NoStop}%
\bibitem [{\citenamefont {Iosifidis}\ and\ \citenamefont
  {Koivisto}(2024)}]{Iosifidis:2023kyf}%
  \BibitemOpen
  \bibfield  {author} {\bibinfo {author} {\bibfnamefont {D.}~\bibnamefont
  {Iosifidis}}\ and\ \bibinfo {author} {\bibfnamefont {T.~S.}\ \bibnamefont
  {Koivisto}},\ }\bibfield  {title} {\bibinfo {title} {{Hyperhydrodynamics:
  relativistic viscous fluids from hypermomentum}},\ }\href
  {https://doi.org/10.1088/1475-7516/2024/05/001} {\bibfield  {journal}
  {\bibinfo  {journal} {JCAP}\ }\textbf {\bibinfo {volume} {05}},\ \bibinfo
  {pages} {001}},\ \Eprint {https://arxiv.org/abs/2312.06780} {arXiv:2312.06780
  [gr-qc]} \BibitemShut {NoStop}%
\bibitem [{\citenamefont {Sakharov}(1991)}]{Andrei_D_Sakharov_1991}%
  \BibitemOpen
  \bibfield  {author} {\bibinfo {author} {\bibfnamefont {A.~D.}\ \bibnamefont
  {Sakharov}},\ }\bibfield  {title} {\bibinfo {title} {Vacuum quantum
  fluctuations in curved space and the theory of gravitation},\ }\href
  {https://doi.org/10.1070/PU1991v034n05ABEH002498} {\bibfield  {journal}
  {\bibinfo  {journal} {Soviet Physics Uspekhi}\ }\textbf {\bibinfo {volume}
  {34}},\ \bibinfo {pages} {394} (\bibinfo {year} {1991})}\BibitemShut
  {NoStop}%
\bibitem [{\citenamefont {Hehl}\ \emph
  {et~al.}(1976{\natexlab{d}})\citenamefont {Hehl}, \citenamefont {Kerlick},\
  and\ \citenamefont {{Von der Heyde}}}]{Hehl:1976inc}%
  \BibitemOpen
  \bibfield  {author} {\bibinfo {author} {\bibfnamefont {F.}~\bibnamefont
  {Hehl}}, \bibinfo {author} {\bibfnamefont {G.}~\bibnamefont {Kerlick}},\ and\
  \bibinfo {author} {\bibfnamefont {P.}~\bibnamefont {{Von der Heyde}}},\
  }\bibfield  {title} {\bibinfo {title} {On a new metric affine theory of
  gravitation},\ }\href
  {https://doi.org/https://doi.org/10.1016/0370-2693(76)90393-2} {\bibfield
  {journal} {\bibinfo  {journal} {Physics Letters B}\ }\textbf {\bibinfo
  {volume} {63}},\ \bibinfo {pages} {446} (\bibinfo {year}
  {1976}{\natexlab{d}})}\BibitemShut {NoStop}%
\bibitem [{\citenamefont {Rigouzzo}\ and\ \citenamefont
  {Zell}(2023)}]{Rigouzzo:2023sbb}%
  \BibitemOpen
  \bibfield  {author} {\bibinfo {author} {\bibfnamefont {C.}~\bibnamefont
  {Rigouzzo}}\ and\ \bibinfo {author} {\bibfnamefont {S.}~\bibnamefont
  {Zell}},\ }\bibfield  {title} {\bibinfo {title} {{Coupling metric-affine
  gravity to the standard model and dark matter fermions}},\ }\href
  {https://doi.org/10.1103/PhysRevD.108.124067} {\bibfield  {journal} {\bibinfo
   {journal} {Phys. Rev. D}\ }\textbf {\bibinfo {volume} {108}},\ \bibinfo
  {pages} {124067} (\bibinfo {year} {2023})},\ \Eprint
  {https://arxiv.org/abs/2306.13134} {arXiv:2306.13134 [gr-qc]} \BibitemShut
  {NoStop}%
\bibitem [{\citenamefont {Barker}\ and\ \citenamefont
  {Marzo}(2024{\natexlab{a}})}]{barker2024particle}%
  \BibitemOpen
  \bibfield  {author} {\bibinfo {author} {\bibfnamefont {W.}~\bibnamefont
  {Barker}}\ and\ \bibinfo {author} {\bibfnamefont {C.}~\bibnamefont {Marzo}},\
  }\href@noop {} {\bibinfo {title} {Particle spectra of general ricci-type
  palatini or metric-affine theories}} (\bibinfo {year} {2024}{\natexlab{a}}),\
  \Eprint {https://arxiv.org/abs/2402.07641} {arXiv:2402.07641 [hep-th]}
  \BibitemShut {NoStop}%
\bibitem [{\citenamefont {Barker}\ and\ \citenamefont
  {Zell}(2024{\natexlab{a}})}]{Barker:2024dhb}%
  \BibitemOpen
  \bibfield  {author} {\bibinfo {author} {\bibfnamefont {W.}~\bibnamefont
  {Barker}}\ and\ \bibinfo {author} {\bibfnamefont {S.}~\bibnamefont {Zell}},\
  }\bibfield  {title} {\bibinfo {title} {{Consistent particle physics in
  metric-affine gravity from extended projective symmetry}},\ }\href@noop {} {\
   (\bibinfo {year} {2024}{\natexlab{a}})},\ \Eprint
  {https://arxiv.org/abs/2402.14917} {arXiv:2402.14917 [hep-th]} \BibitemShut
  {NoStop}%
\bibitem [{\citenamefont {Starobinsky}(1979)}]{Starobinsky:1979ty}%
  \BibitemOpen
  \bibfield  {author} {\bibinfo {author} {\bibfnamefont {A.~A.}\ \bibnamefont
  {Starobinsky}},\ }\bibfield  {title} {\bibinfo {title} {{Spectrum of relict
  gravitational radiation and the early state of the universe}},\ }\href
  {https://ui.adsabs.harvard.edu/abs/1979JETPL..30..682S} {\bibfield  {journal}
  {\bibinfo  {journal} {JETP Lett.}\ }\textbf {\bibinfo {volume} {30}},\
  \bibinfo {pages} {682} (\bibinfo {year} {1979})}\BibitemShut {NoStop}%
\bibitem [{\citenamefont {Guth}(1981)}]{Guth:1980zm}%
  \BibitemOpen
  \bibfield  {author} {\bibinfo {author} {\bibfnamefont {A.~H.}\ \bibnamefont
  {Guth}},\ }\bibfield  {title} {\bibinfo {title} {{The Inflationary Universe:
  A Possible Solution to the Horizon and Flatness Problems}},\ }\href
  {https://doi.org/10.1103/PhysRevD.23.347} {\bibfield  {journal} {\bibinfo
  {journal} {Phys. Rev. D}\ }\textbf {\bibinfo {volume} {23}},\ \bibinfo
  {pages} {347} (\bibinfo {year} {1981})}\BibitemShut {NoStop}%
\bibitem [{\citenamefont {Zwicky}(1933)}]{Zwicky:1933gu}%
  \BibitemOpen
  \bibfield  {author} {\bibinfo {author} {\bibfnamefont {F.}~\bibnamefont
  {Zwicky}},\ }\bibfield  {title} {\bibinfo {title} {{Die Rotverschiebung von
  extragalaktischen Nebeln}},\ }\href
  {https://doi.org/10.1007/s10714-008-0707-4} {\bibfield  {journal} {\bibinfo
  {journal} {Helv. Phys. Acta}\ }\textbf {\bibinfo {volume} {6}},\ \bibinfo
  {pages} {110} (\bibinfo {year} {1933})}\BibitemShut {NoStop}%
\bibitem [{\citenamefont {Aghanim}\ \emph {et~al.}(2020)\citenamefont {Aghanim}
  \emph {et~al.}}]{Planck:2018vyg}%
  \BibitemOpen
  \bibfield  {author} {\bibinfo {author} {\bibfnamefont {N.}~\bibnamefont
  {Aghanim}} \emph {et~al.} (\bibinfo {collaboration} {Planck}),\ }\bibfield
  {title} {\bibinfo {title} {{Planck 2018 results. VI. Cosmological
  parameters}},\ }\href {https://doi.org/10.1051/0004-6361/201833910}
  {\bibfield  {journal} {\bibinfo  {journal} {Astron. Astrophys.}\ }\textbf
  {\bibinfo {volume} {641}},\ \bibinfo {pages} {A6} (\bibinfo {year} {2020})},\
  \bibinfo {note} {[Erratum: Astron.Astrophys. 652, C4 (2021)]},\ \Eprint
  {https://arxiv.org/abs/1807.06209} {arXiv:1807.06209 [astro-ph.CO]}
  \BibitemShut {NoStop}%
\bibitem [{\citenamefont {Scott}(2020)}]{Scott:2018adl}%
  \BibitemOpen
  \bibfield  {author} {\bibinfo {author} {\bibfnamefont {D.}~\bibnamefont
  {Scott}},\ }\bibfield  {title} {\bibinfo {title} {{The standard model of
  cosmology: A skeptic\textquoteright{}s guide}},\ }\href
  {https://doi.org/10.3254/ENFI200007} {\bibfield  {journal} {\bibinfo
  {journal} {Proc. Int. Sch. Phys. Fermi}\ }\textbf {\bibinfo {volume} {200}},\
  \bibinfo {pages} {133} (\bibinfo {year} {2020})},\ \Eprint
  {https://arxiv.org/abs/1804.01318} {arXiv:1804.01318 [astro-ph.CO]}
  \BibitemShut {NoStop}%
\bibitem [{\citenamefont {Handley}(2021)}]{Handley:2019tkm}%
  \BibitemOpen
  \bibfield  {author} {\bibinfo {author} {\bibfnamefont {W.}~\bibnamefont
  {Handley}},\ }\bibfield  {title} {\bibinfo {title} {{Curvature tension:
  evidence for a closed universe}},\ }\href
  {https://doi.org/10.1103/PhysRevD.103.L041301} {\bibfield  {journal}
  {\bibinfo  {journal} {Phys. Rev. D}\ }\textbf {\bibinfo {volume} {103}},\
  \bibinfo {pages} {L041301} (\bibinfo {year} {2021})},\ \Eprint
  {https://arxiv.org/abs/1908.09139} {arXiv:1908.09139 [astro-ph.CO]}
  \BibitemShut {NoStop}%
\bibitem [{\citenamefont {Di~Valentino}\ \emph {et~al.}(2021)\citenamefont
  {Di~Valentino}, \citenamefont {Mena}, \citenamefont {Pan}, \citenamefont
  {Visinelli}, \citenamefont {Yang}, \citenamefont {Melchiorri}, \citenamefont
  {Mota}, \citenamefont {Riess},\ and\ \citenamefont
  {Silk}}]{DiValentino:2021izs}%
  \BibitemOpen
  \bibfield  {author} {\bibinfo {author} {\bibfnamefont {E.}~\bibnamefont
  {Di~Valentino}}, \bibinfo {author} {\bibfnamefont {O.}~\bibnamefont {Mena}},
  \bibinfo {author} {\bibfnamefont {S.}~\bibnamefont {Pan}}, \bibinfo {author}
  {\bibfnamefont {L.}~\bibnamefont {Visinelli}}, \bibinfo {author}
  {\bibfnamefont {W.}~\bibnamefont {Yang}}, \bibinfo {author} {\bibfnamefont
  {A.}~\bibnamefont {Melchiorri}}, \bibinfo {author} {\bibfnamefont {D.~F.}\
  \bibnamefont {Mota}}, \bibinfo {author} {\bibfnamefont {A.~G.}\ \bibnamefont
  {Riess}},\ and\ \bibinfo {author} {\bibfnamefont {J.}~\bibnamefont {Silk}},\
  }\bibfield  {title} {\bibinfo {title} {{In the realm of the Hubble
  tension\textemdash{}a review of solutions}},\ }\href
  {https://doi.org/10.1088/1361-6382/ac086d} {\bibfield  {journal} {\bibinfo
  {journal} {Class. Quant. Grav.}\ }\textbf {\bibinfo {volume} {38}},\ \bibinfo
  {pages} {153001} (\bibinfo {year} {2021})},\ \Eprint
  {https://arxiv.org/abs/2103.01183} {arXiv:2103.01183 [astro-ph.CO]}
  \BibitemShut {NoStop}%
\bibitem [{\citenamefont {Perivolaropoulos}\ and\ \citenamefont
  {Skara}(2022)}]{Perivolaropoulos:2021jda}%
  \BibitemOpen
  \bibfield  {author} {\bibinfo {author} {\bibfnamefont {L.}~\bibnamefont
  {Perivolaropoulos}}\ and\ \bibinfo {author} {\bibfnamefont {F.}~\bibnamefont
  {Skara}},\ }\bibfield  {title} {\bibinfo {title} {{Challenges for
  \ensuremath{\Lambda}CDM: An update}},\ }\href
  {https://doi.org/10.1016/j.newar.2022.101659} {\bibfield  {journal} {\bibinfo
   {journal} {New Astron. Rev.}\ }\textbf {\bibinfo {volume} {95}},\ \bibinfo
  {pages} {101659} (\bibinfo {year} {2022})},\ \Eprint
  {https://arxiv.org/abs/2105.05208} {arXiv:2105.05208 [astro-ph.CO]}
  \BibitemShut {NoStop}%
\bibitem [{\citenamefont {et~al.)}(2020)}]{Planck2018}%
  \BibitemOpen
  \bibfield  {author} {\bibinfo {author} {\bibfnamefont {P.~N.~A.}\
  \bibnamefont {et~al.)}},\ }\bibfield  {title} {\bibinfo {title} {{Planck 2018
  results. VI. Cosmological parameters}},\ }\href
  {https://doi.org/10.1051/0004-6361/201833910} {\bibfield  {journal} {\bibinfo
   {journal} {Astron. Astrophys.}\ }\textbf {\bibinfo {volume} {641}},\
  \bibinfo {pages} {A6} (\bibinfo {year} {2020})}\BibitemShut {NoStop}%
\bibitem [{\citenamefont {Riess}(2022)}]{Riess2022}%
  \BibitemOpen
  \bibfield  {author} {\bibinfo {author} {\bibfnamefont {A.~G. e.~a.}\
  \bibnamefont {Riess}},\ }\bibfield  {title} {\bibinfo {title} {{A
  Comprehensive Measurement of the Local Value of the Hubble Constant with 1 km
  s$^{-1}$ Mpc$^{-1}$ Uncertainty from HST and the SH0ES Team}},\ }\href
  {https://doi.org/10.3847/2041-8213/ac5c5b} {\bibfield  {journal} {\bibinfo
  {journal} {Astrophys. J. Lett.}\ }\textbf {\bibinfo {volume} {934}},\
  \bibinfo {pages} {L7} (\bibinfo {year} {2022})}\BibitemShut {NoStop}%
\bibitem [{\citenamefont {Asgari}(2021)}]{Asgari2021}%
  \BibitemOpen
  \bibfield  {author} {\bibinfo {author} {\bibfnamefont {M.~e.~a.}\
  \bibnamefont {Asgari}},\ }\bibfield  {title} {\bibinfo {title} {{KiDS-1000
  Cosmology: Cosmic shear constraints and comparison between two-point
  statistics}},\ }\href {https://doi.org/10.1051/0004-6361/202039070}
  {\bibfield  {journal} {\bibinfo  {journal} {Astron. Astrophys.}\ }\textbf
  {\bibinfo {volume} {645}},\ \bibinfo {pages} {A104} (\bibinfo {year}
  {2021})}\BibitemShut {NoStop}%
\bibitem [{\citenamefont {Abbott}(2022)}]{DESY3_Abbott2022}%
  \BibitemOpen
  \bibfield  {author} {\bibinfo {author} {\bibfnamefont {T.~M. C. e. a. D.~C.}\
  \bibnamefont {Abbott}},\ }\bibfield  {title} {\bibinfo {title} {{Dark Energy
  Survey Year 3 results: Cosmological constraints from galaxy clustering and
  weak lensing}},\ }\href {https://doi.org/10.1103/PhysRevD.105.023520}
  {\bibfield  {journal} {\bibinfo  {journal} {Phys. Rev. D}\ }\textbf {\bibinfo
  {volume} {105}},\ \bibinfo {pages} {023520} (\bibinfo {year}
  {2022})}\BibitemShut {NoStop}%
\bibitem [{\citenamefont {Handley}\ and\ \citenamefont
  {Lemos}(2021)}]{Handley2021}%
  \BibitemOpen
  \bibfield  {author} {\bibinfo {author} {\bibfnamefont {W.}~\bibnamefont
  {Handley}}\ and\ \bibinfo {author} {\bibfnamefont {P.}~\bibnamefont
  {Lemos}},\ }\bibfield  {title} {\bibinfo {title} {{Quantifying the global
  parameter tensions between ACT, SPT and Planck}},\ }\href
  {https://doi.org/10.1103/PhysRevD.103.063529} {\bibfield  {journal} {\bibinfo
   {journal} {Phys. Rev. D}\ }\textbf {\bibinfo {volume} {103}},\ \bibinfo
  {pages} {063529} (\bibinfo {year} {2021})}\BibitemShut {NoStop}%
\bibitem [{\citenamefont {Lodha}(2025)}]{Lodha2025}%
  \BibitemOpen
  \bibfield  {author} {\bibinfo {author} {\bibfnamefont {K.~e. a. D.~C.}\
  \bibnamefont {Lodha}},\ }\bibfield  {title} {\bibinfo {title} {{Extended Dark
  Energy analysis using DESI DR2 BAO measurements}},\ }\bibfield  {journal}
  {\bibinfo  {journal} {arXiv e-prints}\ }\href
  {https://doi.org/10.48550/arXiv.2503.14743} {10.48550/arXiv.2503.14743}
  (\bibinfo {year} {2025}),\ \Eprint {https://arxiv.org/abs/2503.14743}
  {arXiv:2503.14743} \BibitemShut {NoStop}%
\bibitem [{\citenamefont {Bahamonde}\ \emph {et~al.}(2023)\citenamefont
  {Bahamonde}, \citenamefont {Dialektopoulos}, \citenamefont
  {Escamilla-Rivera}, \citenamefont {Farrugia}, \citenamefont {Gakis},
  \citenamefont {Hendry}, \citenamefont {Hohmann}, \citenamefont {Levi~Said},
  \citenamefont {Mifsud},\ and\ \citenamefont
  {Di~Valentino}}]{Bahamonde:2021gfp}%
  \BibitemOpen
  \bibfield  {author} {\bibinfo {author} {\bibfnamefont {S.}~\bibnamefont
  {Bahamonde}}, \bibinfo {author} {\bibfnamefont {K.~F.}\ \bibnamefont
  {Dialektopoulos}}, \bibinfo {author} {\bibfnamefont {C.}~\bibnamefont
  {Escamilla-Rivera}}, \bibinfo {author} {\bibfnamefont {G.}~\bibnamefont
  {Farrugia}}, \bibinfo {author} {\bibfnamefont {V.}~\bibnamefont {Gakis}},
  \bibinfo {author} {\bibfnamefont {M.}~\bibnamefont {Hendry}}, \bibinfo
  {author} {\bibfnamefont {M.}~\bibnamefont {Hohmann}}, \bibinfo {author}
  {\bibfnamefont {J.}~\bibnamefont {Levi~Said}}, \bibinfo {author}
  {\bibfnamefont {J.}~\bibnamefont {Mifsud}},\ and\ \bibinfo {author}
  {\bibfnamefont {E.}~\bibnamefont {Di~Valentino}},\ }\bibfield  {title}
  {\bibinfo {title} {{Teleparallel gravity: from theory to cosmology}},\ }\href
  {https://doi.org/10.1088/1361-6633/ac9cef} {\bibfield  {journal} {\bibinfo
  {journal} {Rept. Prog. Phys.}\ }\textbf {\bibinfo {volume} {86}},\ \bibinfo
  {pages} {026901} (\bibinfo {year} {2023})},\ \Eprint
  {https://arxiv.org/abs/2106.13793} {arXiv:2106.13793 [gr-qc]} \BibitemShut
  {NoStop}%
\bibitem [{\citenamefont {Beltr\'an~Jim\'enez}\ \emph
  {et~al.}(2020)\citenamefont {Beltr\'an~Jim\'enez}, \citenamefont
  {Heisenberg}, \citenamefont {Iosifidis}, \citenamefont {Jim\'enez-Cano},\
  and\ \citenamefont {Koivisto}}]{BeltranJimenez:2019odq}%
  \BibitemOpen
  \bibfield  {author} {\bibinfo {author} {\bibfnamefont {J.}~\bibnamefont
  {Beltr\'an~Jim\'enez}}, \bibinfo {author} {\bibfnamefont {L.}~\bibnamefont
  {Heisenberg}}, \bibinfo {author} {\bibfnamefont {D.}~\bibnamefont
  {Iosifidis}}, \bibinfo {author} {\bibfnamefont {A.}~\bibnamefont
  {Jim\'enez-Cano}},\ and\ \bibinfo {author} {\bibfnamefont {T.~S.}\
  \bibnamefont {Koivisto}},\ }\bibfield  {title} {\bibinfo {title} {{General
  teleparallel quadratic gravity}},\ }\href
  {https://doi.org/10.1016/j.physletb.2020.135422} {\bibfield  {journal}
  {\bibinfo  {journal} {Phys. Lett. B}\ }\textbf {\bibinfo {volume} {805}},\
  \bibinfo {pages} {135422} (\bibinfo {year} {2020})},\ \Eprint
  {https://arxiv.org/abs/1909.09045} {arXiv:1909.09045 [gr-qc]} \BibitemShut
  {NoStop}%
\bibitem [{\citenamefont {Jimenez}\ \emph {et~al.}(2019)\citenamefont
  {Jimenez}, \citenamefont {Heisenberg},\ and\ \citenamefont
  {Koivisto}}]{jimenez2019geometrical}%
  \BibitemOpen
  \bibfield  {author} {\bibinfo {author} {\bibfnamefont {J.~B.}\ \bibnamefont
  {Jimenez}}, \bibinfo {author} {\bibfnamefont {L.}~\bibnamefont
  {Heisenberg}},\ and\ \bibinfo {author} {\bibfnamefont {T.~S.}\ \bibnamefont
  {Koivisto}},\ }\href@noop {} {\bibinfo {title} {The geometrical trinity of
  gravity}} (\bibinfo {year} {2019}),\ \Eprint
  {https://arxiv.org/abs/1903.06830} {arXiv:1903.06830 [hep-th]} \BibitemShut
  {NoStop}%
\bibitem [{\citenamefont {Weyl}(1918)}]{Weyl:1918}%
  \BibitemOpen
  \bibfield  {author} {\bibinfo {author} {\bibfnamefont {H.}~\bibnamefont
  {Weyl}},\ }\bibfield  {title} {\bibinfo {title} {{Gravitation and
  electricity}},\ }\href@noop {} {\bibfield  {journal} {\bibinfo  {journal}
  {Sitzungsber. Preuss. Akad. Wiss.}\ }\textbf {\bibinfo {volume} {26}},\
  \bibinfo {pages} {465} (\bibinfo {year} {1918})}\BibitemShut {NoStop}%
\bibitem [{\citenamefont {Weyl}(1922)}]{Weyl:1922}%
  \BibitemOpen
  \bibfield  {author} {\bibinfo {author} {\bibfnamefont {H.}~\bibnamefont
  {Weyl}},\ }\href@noop {} {\emph {\bibinfo {title} {{Space-Time-Matter}}}}\
  (\bibinfo  {publisher} {Dover Publications},\ \bibinfo {year}
  {1922})\BibitemShut {NoStop}%
\bibitem [{\citenamefont {Cartan}(1923)}]{Cartan:1923}%
  \BibitemOpen
  \bibfield  {author} {\bibinfo {author} {\bibfnamefont {{\'E}.}~\bibnamefont
  {Cartan}},\ }\bibfield  {title} {\bibinfo {title} {{Sur les vari{\'e}t{\'e}s
  {\`a} connexion affine et la th{\'e}orie de la relativit{\'e}
  g{\'e}n{\'e}ralis{\'e}e (premi{\`e}re partie)}},\ }in\ \href
  {https://doi.org/10.24033/asens.751} {\emph {\bibinfo {booktitle} {Annales
  scientifiques de l'{\'E}cole normale sup{\'e}rieure}}},\ Vol.~\bibinfo
  {volume} {40}\ (\bibinfo  {publisher} {Gauthier-Villars},\ \bibinfo {year}
  {1923})\ pp.\ \bibinfo {pages} {325--412}\BibitemShut {NoStop}%
\bibitem [{\citenamefont {Eddington}(1923)}]{Eddington:1923}%
  \BibitemOpen
  \bibfield  {author} {\bibinfo {author} {\bibfnamefont {A.~S.}\ \bibnamefont
  {Eddington}},\ }\href@noop {} {\emph {\bibinfo {title} {{The Mathematical
  Theory of Relativity}}}}\ (\bibinfo  {publisher} {Cambridge University
  Press},\ \bibinfo {year} {1923})\BibitemShut {NoStop}%
\bibitem [{\citenamefont {Cartan}(1924)}]{Cartan:1924}%
  \BibitemOpen
  \bibfield  {author} {\bibinfo {author} {\bibfnamefont {{\'E}.}~\bibnamefont
  {Cartan}},\ }\bibfield  {title} {\bibinfo {title} {{Sur les vari{\'e}t{\'e}s
  {\`a} connexion affine, et la th{\'e}orie de la relativit{\'e}
  g{\'e}n{\'e}ralis{\'e}e (premi{\`e}re partie)(suite)}},\ }in\ \href
  {https://doi.org/10.24033/asens.753} {\emph {\bibinfo {booktitle} {Annales
  scientifiques de l'{\'E}cole Normale Sup{\'e}rieure}}},\ Vol.~\bibinfo
  {volume} {41}\ (\bibinfo  {publisher} {Gauthier-Villars},\ \bibinfo {year}
  {1924})\ pp.\ \bibinfo {pages} {1--25}\BibitemShut {NoStop}%
\bibitem [{\citenamefont {Cartan}(1925)}]{Cartan:1925}%
  \BibitemOpen
  \bibfield  {author} {\bibinfo {author} {\bibfnamefont {{\'E}.}~\bibnamefont
  {Cartan}},\ }\bibfield  {title} {\bibinfo {title} {{Sur les vari{\'e}t{\'e}s
  {\`a} connexion affine, et la th{\'e}orie de la relativit{\'e}
  g{\'e}n{\'e}ralis{\'e}e (deuxi{\`e}me partie)}},\ }in\ \href
  {https://doi.org/10.24033/asens.761} {\emph {\bibinfo {booktitle} {Annales
  scientifiques de l'{\'E}cole normale sup{\'e}rieure}}},\ Vol.~\bibinfo
  {volume} {42}\ (\bibinfo  {publisher} {Gauthier-Villars},\ \bibinfo {year}
  {1925})\ pp.\ \bibinfo {pages} {17--88}\BibitemShut {NoStop}%
\bibitem [{\citenamefont {Einstein}(1925)}]{Einstein:1925}%
  \BibitemOpen
  \bibfield  {author} {\bibinfo {author} {\bibfnamefont {A.}~\bibnamefont
  {Einstein}},\ }\bibfield  {title} {\bibinfo {title} {{Einheitliche
  Feldtheorie von Gravitation und Elektrizit\"at}},\ }\href
  {https://doi.org/10.1002/3527608958.ch30} {\bibfield  {journal} {\bibinfo
  {journal} {Sitzungsber. Preuss. Akad. Wiss}\ }\textbf {\bibinfo {volume}
  {22}},\ \bibinfo {pages} {414} (\bibinfo {year} {1925})}\BibitemShut
  {NoStop}%
\bibitem [{\citenamefont {Einstein}(1928{\natexlab{a}})}]{Einstein:1928}%
  \BibitemOpen
  \bibfield  {author} {\bibinfo {author} {\bibfnamefont {A.}~\bibnamefont
  {Einstein}},\ }\bibfield  {title} {\bibinfo {title} {{Riemanngeometrie mit
  Aufrechterhaltung des Begriffes des Fern-Parallelismus}},\ }\href
  {https://doi.org/10.1002/3527608958.ch36} {\bibfield  {journal} {\bibinfo
  {journal} {Sitzungsber. Preuss. Akad. Wiss}\ }\textbf {\bibinfo {volume}
  {17}},\ \bibinfo {pages} {217} (\bibinfo {year}
  {1928}{\natexlab{a}})}\BibitemShut {NoStop}%
\bibitem [{\citenamefont {Einstein}(1928{\natexlab{b}})}]{Einstein:19282}%
  \BibitemOpen
  \bibfield  {author} {\bibinfo {author} {\bibfnamefont {A.}~\bibnamefont
  {Einstein}},\ }\bibfield  {title} {\bibinfo {title} {{Neue M{\"o}glichkeit
  f{\"u}r eine einheitliche Feldtheorie von Gravitation und
  Elektrizit{\"a}t}},\ }\href {https://doi.org/10.1002/3527608958.ch37}
  {\bibfield  {journal} {\bibinfo  {journal} {Sitzungsber. Preuss. Akad. Wiss}\
  }\textbf {\bibinfo {volume} {18}},\ \bibinfo {pages} {224} (\bibinfo {year}
  {1928}{\natexlab{b}})}\BibitemShut {NoStop}%
\bibitem [{\citenamefont {Utiyama}(1956)}]{Utiyama:1956sy}%
  \BibitemOpen
  \bibfield  {author} {\bibinfo {author} {\bibfnamefont {R.}~\bibnamefont
  {Utiyama}},\ }\bibfield  {title} {\bibinfo {title} {{Invariant theoretical
  interpretation of interaction}},\ }\href
  {https://doi.org/10.1103/PhysRev.101.1597} {\bibfield  {journal} {\bibinfo
  {journal} {Phys. Rev.}\ }\textbf {\bibinfo {volume} {101}},\ \bibinfo {pages}
  {1597} (\bibinfo {year} {1956})}\BibitemShut {NoStop}%
\bibitem [{\citenamefont {Sciama}(1962)}]{Sciama:1962}%
  \BibitemOpen
  \bibfield  {author} {\bibinfo {author} {\bibfnamefont {D.~W.}\ \bibnamefont
  {Sciama}},\ }\bibfield  {title} {\bibinfo {title} {On the analogy between
  charge and spin in general relativity},\ }in\ \href
  {https://ui.adsabs.harvard.edu/abs/1962rdgr.book..415S} {\emph {\bibinfo
  {booktitle} {Recent developments in general relativity}}}\ (\bibinfo
  {publisher} {Pergamon Press},\ \bibinfo {address} {Oxford},\ \bibinfo {year}
  {1962})\ p.\ \bibinfo {pages} {415}\BibitemShut {NoStop}%
\bibitem [{\citenamefont {Kibble}(1961)}]{Kibble:1961ba}%
  \BibitemOpen
  \bibfield  {author} {\bibinfo {author} {\bibfnamefont {T.~W.~B.}\
  \bibnamefont {Kibble}},\ }\bibfield  {title} {\bibinfo {title} {{Lorentz
  invariance and the gravitational field}},\ }\href
  {https://doi.org/10.1063/1.1703702} {\bibfield  {journal} {\bibinfo
  {journal} {J. Math. Phys.}\ }\textbf {\bibinfo {volume} {2}},\ \bibinfo
  {pages} {212} (\bibinfo {year} {1961})}\BibitemShut {NoStop}%
\bibitem [{\citenamefont {Blagojevic}(2002{\natexlab{a}})}]{Blagojevic:2002du}%
  \BibitemOpen
  \bibfield  {author} {\bibinfo {author} {\bibfnamefont {M.}~\bibnamefont
  {Blagojevic}},\ }\href {https://doi.org/10.1201/9781420034264} {\emph
  {\bibinfo {title} {{Gravitation and gauge symmetries}}}}\ (\bibinfo {year}
  {2002})\BibitemShut {NoStop}%
\bibitem [{\citenamefont {Salvio}(2018)}]{Salvio:2018crh}%
  \BibitemOpen
  \bibfield  {author} {\bibinfo {author} {\bibfnamefont {A.}~\bibnamefont
  {Salvio}},\ }\bibfield  {title} {\bibinfo {title} {{Quadratic Gravity}},\
  }\href {https://doi.org/10.3389/fphy.2018.00077} {\bibfield  {journal}
  {\bibinfo  {journal} {Front. in Phys.}\ }\textbf {\bibinfo {volume} {6}},\
  \bibinfo {pages} {77} (\bibinfo {year} {2018})},\ \Eprint
  {https://arxiv.org/abs/1804.09944} {arXiv:1804.09944 [hep-th]} \BibitemShut
  {NoStop}%
\bibitem [{\citenamefont {Salam}\ and\ \citenamefont
  {Strathdee}(1978)}]{Salam:1978}%
  \BibitemOpen
  \bibfield  {author} {\bibinfo {author} {\bibfnamefont {A.}~\bibnamefont
  {Salam}}\ and\ \bibinfo {author} {\bibfnamefont {J.}~\bibnamefont
  {Strathdee}},\ }\bibfield  {title} {\bibinfo {title} {Remarks on high-energy
  stability and renormalizability of gravity theory},\ }\href
  {https://doi.org/10.1103/PhysRevD.18.4480} {\bibfield  {journal} {\bibinfo
  {journal} {Phys. Rev. D}\ }\textbf {\bibinfo {volume} {18}},\ \bibinfo
  {pages} {4480} (\bibinfo {year} {1978})}\BibitemShut {NoStop}%
\bibitem [{\citenamefont {Stelle}(1977)}]{Stelle:1977}%
  \BibitemOpen
  \bibfield  {author} {\bibinfo {author} {\bibfnamefont {K.~S.}\ \bibnamefont
  {Stelle}},\ }\bibfield  {title} {\bibinfo {title} {Renormalization of
  higher-derivative quantum gravity},\ }\href
  {https://doi.org/10.1103/PhysRevD.16.953} {\bibfield  {journal} {\bibinfo
  {journal} {Phys. Rev. D}\ }\textbf {\bibinfo {volume} {16}},\ \bibinfo
  {pages} {953} (\bibinfo {year} {1977})}\BibitemShut {NoStop}%
\bibitem [{\citenamefont {Einhorn}\ and\ \citenamefont
  {Jones}(2017)}]{Einhorn:2017icw}%
  \BibitemOpen
  \bibfield  {author} {\bibinfo {author} {\bibfnamefont {M.~B.}\ \bibnamefont
  {Einhorn}}\ and\ \bibinfo {author} {\bibfnamefont {D.~R.~T.}\ \bibnamefont
  {Jones}},\ }\bibfield  {title} {\bibinfo {title} {{Renormalizable,
  asymptotically free gravity without ghosts or tachyons}},\ }\href
  {https://doi.org/10.1103/PhysRevD.96.124025} {\bibfield  {journal} {\bibinfo
  {journal} {Phys. Rev. D}\ }\textbf {\bibinfo {volume} {96}},\ \bibinfo
  {pages} {124025} (\bibinfo {year} {2017})},\ \Eprint
  {https://arxiv.org/abs/1710.03795} {arXiv:1710.03795 [hep-th]} \BibitemShut
  {NoStop}%
\bibitem [{\citenamefont {Donoghue}(1994)}]{Donoghue:1994dn}%
  \BibitemOpen
  \bibfield  {author} {\bibinfo {author} {\bibfnamefont {J.~F.}\ \bibnamefont
  {Donoghue}},\ }\bibfield  {title} {\bibinfo {title} {{General relativity as
  an effective field theory: The leading quantum corrections}},\ }\href
  {https://doi.org/10.1103/PhysRevD.50.3874} {\bibfield  {journal} {\bibinfo
  {journal} {Phys. Rev. D}\ }\textbf {\bibinfo {volume} {50}},\ \bibinfo
  {pages} {3874} (\bibinfo {year} {1994})},\ \Eprint
  {https://arxiv.org/abs/gr-qc/9405057} {arXiv:gr-qc/9405057} \BibitemShut
  {NoStop}%
\bibitem [{\citenamefont {Pagani}\ and\ \citenamefont
  {Percacci}(2015)}]{Pagani:2015ema}%
  \BibitemOpen
  \bibfield  {author} {\bibinfo {author} {\bibfnamefont {C.}~\bibnamefont
  {Pagani}}\ and\ \bibinfo {author} {\bibfnamefont {R.}~\bibnamefont
  {Percacci}},\ }\bibfield  {title} {\bibinfo {title} {{Quantum gravity with
  torsion and non-metricity}},\ }\href
  {https://doi.org/10.1088/0264-9381/32/19/195019} {\bibfield  {journal}
  {\bibinfo  {journal} {Class. Quant. Grav.}\ }\textbf {\bibinfo {volume}
  {32}},\ \bibinfo {pages} {195019} (\bibinfo {year} {2015})},\ \Eprint
  {https://arxiv.org/abs/1506.02882} {arXiv:1506.02882 [gr-qc]} \BibitemShut
  {NoStop}%
\bibitem [{\citenamefont {Percacci}(2020)}]{Percacci:2020bzf}%
  \BibitemOpen
  \bibfield  {author} {\bibinfo {author} {\bibfnamefont {R.}~\bibnamefont
  {Percacci}},\ }\bibfield  {title} {\bibinfo {title} {{Towards Metric-Affine
  Quantum Gravity}},\ }\href {https://doi.org/10.1142/S0219887820400034}
  {\bibfield  {journal} {\bibinfo  {journal} {Int. J. Geom. Meth. Mod. Phys.}\
  }\textbf {\bibinfo {volume} {17}},\ \bibinfo {pages} {2040003} (\bibinfo
  {year} {2020})},\ \Eprint {https://arxiv.org/abs/2003.09486}
  {arXiv:2003.09486 [gr-qc]} \BibitemShut {NoStop}%
\bibitem [{\citenamefont {Melichev}\ and\ \citenamefont
  {Percacci}(2024)}]{Melichev:2023lwj}%
  \BibitemOpen
  \bibfield  {author} {\bibinfo {author} {\bibfnamefont {O.}~\bibnamefont
  {Melichev}}\ and\ \bibinfo {author} {\bibfnamefont {R.}~\bibnamefont
  {Percacci}},\ }\bibfield  {title} {\bibinfo {title} {{On the renormalization
  of Poincar\'e gauge theories}},\ }\href
  {https://doi.org/10.1007/JHEP03(2024)133} {\bibfield  {journal} {\bibinfo
  {journal} {JHEP}\ }\textbf {\bibinfo {volume} {03}},\ \bibinfo {pages}
  {133}},\ \Eprint {https://arxiv.org/abs/2307.02336} {arXiv:2307.02336
  [hep-th]} \BibitemShut {NoStop}%
\bibitem [{\citenamefont {Melichev}(2024)}]{Melichev:2024hih}%
  \BibitemOpen
  \bibfield  {author} {\bibinfo {author} {\bibfnamefont {O.}~\bibnamefont
  {Melichev}},\ }\bibfield  {title} {\bibinfo {title} {{On the renormalization
  of Metric-Affine Gravity theories}},\ }\href@noop {} {\  (\bibinfo {year}
  {2024})},\ \Eprint {https://arxiv.org/abs/2406.14146} {arXiv:2406.14146
  [hep-th]} \BibitemShut {NoStop}%
\bibitem [{\citenamefont {Sezgin}(1981)}]{Sezgin:1981xs}%
  \BibitemOpen
  \bibfield  {author} {\bibinfo {author} {\bibfnamefont {E.}~\bibnamefont
  {Sezgin}},\ }\bibfield  {title} {\bibinfo {title} {{Class of Ghost Free
  Gravity Lagrangians With Massive or Massless Propagating Torsion}},\ }\href
  {https://doi.org/10.1103/PhysRevD.24.1677} {\bibfield  {journal} {\bibinfo
  {journal} {Phys. Rev. D}\ }\textbf {\bibinfo {volume} {24}},\ \bibinfo
  {pages} {1677} (\bibinfo {year} {1981})}\BibitemShut {NoStop}%
\bibitem [{\citenamefont {Blagojevic}\ and\ \citenamefont
  {Nikolic}(1983)}]{Blagojevic:1983zz}%
  \BibitemOpen
  \bibfield  {author} {\bibinfo {author} {\bibfnamefont {M.}~\bibnamefont
  {Blagojevic}}\ and\ \bibinfo {author} {\bibfnamefont {I.~A.}\ \bibnamefont
  {Nikolic}},\ }\bibfield  {title} {\bibinfo {title} {{Hamiltonian dynamics of
  Poincare gauge theory: General structure in the time gauge}},\ }\href
  {https://doi.org/10.1103/PhysRevD.28.2455} {\bibfield  {journal} {\bibinfo
  {journal} {Phys. Rev. D}\ }\textbf {\bibinfo {volume} {28}},\ \bibinfo
  {pages} {2455} (\bibinfo {year} {1983})}\BibitemShut {NoStop}%
\bibitem [{\citenamefont {Blagojevic}\ and\ \citenamefont
  {Vasilic}(1987)}]{Blagojevic:1986dm}%
  \BibitemOpen
  \bibfield  {author} {\bibinfo {author} {\bibfnamefont {M.}~\bibnamefont
  {Blagojevic}}\ and\ \bibinfo {author} {\bibfnamefont {M.}~\bibnamefont
  {Vasilic}},\ }\bibfield  {title} {\bibinfo {title} {Extra gauge symmetries in
  a weak field approximation of an {R + T**2 + R**2} theory of gravity},\
  }\href {https://doi.org/10.1103/PhysRevD.35.3748} {\bibfield  {journal}
  {\bibinfo  {journal} {Phys. Rev. D}\ }\textbf {\bibinfo {volume} {35}},\
  \bibinfo {pages} {3748} (\bibinfo {year} {1987})}\BibitemShut {NoStop}%
\bibitem [{\citenamefont {Kuhfuss}\ and\ \citenamefont
  {Nitsch}(1986)}]{Kuhfuss:1986rb}%
  \BibitemOpen
  \bibfield  {author} {\bibinfo {author} {\bibfnamefont {R.}~\bibnamefont
  {Kuhfuss}}\ and\ \bibinfo {author} {\bibfnamefont {J.}~\bibnamefont
  {Nitsch}},\ }\bibfield  {title} {\bibinfo {title} {{Propagating Modes in
  Gauge Field Theories of Gravity}},\ }\href
  {https://doi.org/10.1007/BF00763447} {\bibfield  {journal} {\bibinfo
  {journal} {Gen. Rel. Grav.}\ }\textbf {\bibinfo {volume} {18}},\ \bibinfo
  {pages} {1207} (\bibinfo {year} {1986})}\BibitemShut {NoStop}%
\bibitem [{\citenamefont {Yo}\ and\ \citenamefont {Nester}(1999)}]{Yo:1999ex}%
  \BibitemOpen
  \bibfield  {author} {\bibinfo {author} {\bibfnamefont {H.-j.}\ \bibnamefont
  {Yo}}\ and\ \bibinfo {author} {\bibfnamefont {J.~M.}\ \bibnamefont
  {Nester}},\ }\bibfield  {title} {\bibinfo {title} {{Hamiltonian analysis of
  Poincare gauge theory scalar modes}},\ }\href
  {https://doi.org/10.1142/S021827189900033X} {\bibfield  {journal} {\bibinfo
  {journal} {Int. J. Mod. Phys. D}\ }\textbf {\bibinfo {volume} {8}},\ \bibinfo
  {pages} {459} (\bibinfo {year} {1999})},\ \Eprint
  {https://arxiv.org/abs/gr-qc/9902032} {arXiv:gr-qc/9902032} \BibitemShut
  {NoStop}%
\bibitem [{\citenamefont {Yo}\ and\ \citenamefont {Nester}(2002)}]{Yo:2001sy}%
  \BibitemOpen
  \bibfield  {author} {\bibinfo {author} {\bibfnamefont {H.-J.}\ \bibnamefont
  {Yo}}\ and\ \bibinfo {author} {\bibfnamefont {J.~M.}\ \bibnamefont
  {Nester}},\ }\bibfield  {title} {\bibinfo {title} {{Hamiltonian analysis of
  Poincare gauge theory: Higher spin modes}},\ }\href
  {https://doi.org/10.1142/S0218271802001998} {\bibfield  {journal} {\bibinfo
  {journal} {Int. J. Mod. Phys. D}\ }\textbf {\bibinfo {volume} {11}},\
  \bibinfo {pages} {747} (\bibinfo {year} {2002})},\ \Eprint
  {https://arxiv.org/abs/gr-qc/0112030} {arXiv:gr-qc/0112030} \BibitemShut
  {NoStop}%
\bibitem [{\citenamefont {Blagojevic}(2002{\natexlab{b}})}]{Blagojevic:2002}%
  \BibitemOpen
  \bibfield  {author} {\bibinfo {author} {\bibfnamefont {M.}~\bibnamefont
  {Blagojevic}},\ }\href@noop {} {\emph {\bibinfo {title} {{Gravitation and
  gauge symmetries}}}}\ (\bibinfo  {publisher} {CRC Press},\ \bibinfo {address}
  {Bristol and Philadelphia},\ \bibinfo {year} {2002})\BibitemShut {NoStop}%
\bibitem [{\citenamefont {Puetzfeld}(2005)}]{Puetzfeld:2004yg}%
  \BibitemOpen
  \bibfield  {author} {\bibinfo {author} {\bibfnamefont {D.}~\bibnamefont
  {Puetzfeld}},\ }\bibfield  {title} {\bibinfo {title} {{Status of
  non-Riemannian cosmology}},\ }\href
  {https://doi.org/10.1016/j.newar.2005.01.022} {\bibfield  {journal} {\bibinfo
   {journal} {New Astron. Rev.}\ }\textbf {\bibinfo {volume} {49}},\ \bibinfo
  {pages} {59} (\bibinfo {year} {2005})},\ \Eprint
  {https://arxiv.org/abs/gr-qc/0404119} {arXiv:gr-qc/0404119} \BibitemShut
  {NoStop}%
\bibitem [{\citenamefont {Yo}\ and\ \citenamefont {Nester}(2007)}]{Yo:2006qs}%
  \BibitemOpen
  \bibfield  {author} {\bibinfo {author} {\bibfnamefont {H.-J.}\ \bibnamefont
  {Yo}}\ and\ \bibinfo {author} {\bibfnamefont {J.~M.}\ \bibnamefont
  {Nester}},\ }\bibfield  {title} {\bibinfo {title} {{Dynamic Scalar Torsion
  and an Oscillating Universe}},\ }\href
  {https://doi.org/10.1142/S0217732307025303} {\bibfield  {journal} {\bibinfo
  {journal} {Mod. Phys. Lett. A}\ }\textbf {\bibinfo {volume} {22}},\ \bibinfo
  {pages} {2057} (\bibinfo {year} {2007})},\ \Eprint
  {https://arxiv.org/abs/astro-ph/0612738} {arXiv:astro-ph/0612738}
  \BibitemShut {NoStop}%
\bibitem [{\citenamefont {Shie}\ \emph {et~al.}(2008)\citenamefont {Shie},
  \citenamefont {Nester},\ and\ \citenamefont {Yo}}]{Shie:2008ms}%
  \BibitemOpen
  \bibfield  {author} {\bibinfo {author} {\bibfnamefont {K.-F.}\ \bibnamefont
  {Shie}}, \bibinfo {author} {\bibfnamefont {J.~M.}\ \bibnamefont {Nester}},\
  and\ \bibinfo {author} {\bibfnamefont {H.-J.}\ \bibnamefont {Yo}},\
  }\bibfield  {title} {\bibinfo {title} {{Torsion Cosmology and the
  Accelerating Universe}},\ }\href {https://doi.org/10.1103/PhysRevD.78.023522}
  {\bibfield  {journal} {\bibinfo  {journal} {Phys. Rev. D}\ }\textbf {\bibinfo
  {volume} {78}},\ \bibinfo {pages} {023522} (\bibinfo {year} {2008})},\
  \Eprint {https://arxiv.org/abs/0805.3834} {arXiv:0805.3834 [gr-qc]}
  \BibitemShut {NoStop}%
\bibitem [{\citenamefont {Nair}\ \emph {et~al.}(2009)\citenamefont {Nair},
  \citenamefont {Randjbar-Daemi},\ and\ \citenamefont {Rubakov}}]{Nair:2008yh}%
  \BibitemOpen
  \bibfield  {author} {\bibinfo {author} {\bibfnamefont {V.~P.}\ \bibnamefont
  {Nair}}, \bibinfo {author} {\bibfnamefont {S.}~\bibnamefont
  {Randjbar-Daemi}},\ and\ \bibinfo {author} {\bibfnamefont {V.}~\bibnamefont
  {Rubakov}},\ }\bibfield  {title} {\bibinfo {title} {{Massive Spin-2 fields of
  Geometric Origin in Curved Spacetimes}},\ }\href
  {https://doi.org/10.1103/PhysRevD.80.104031} {\bibfield  {journal} {\bibinfo
  {journal} {Phys. Rev. D}\ }\textbf {\bibinfo {volume} {80}},\ \bibinfo
  {pages} {104031} (\bibinfo {year} {2009})},\ \Eprint
  {https://arxiv.org/abs/0811.3781} {arXiv:0811.3781 [hep-th]} \BibitemShut
  {NoStop}%
\bibitem [{\citenamefont {Nikiforova}\ \emph {et~al.}(2009)\citenamefont
  {Nikiforova}, \citenamefont {Randjbar-Daemi},\ and\ \citenamefont
  {Rubakov}}]{Nikiforova:2009qr}%
  \BibitemOpen
  \bibfield  {author} {\bibinfo {author} {\bibfnamefont {V.}~\bibnamefont
  {Nikiforova}}, \bibinfo {author} {\bibfnamefont {S.}~\bibnamefont
  {Randjbar-Daemi}},\ and\ \bibinfo {author} {\bibfnamefont {V.}~\bibnamefont
  {Rubakov}},\ }\bibfield  {title} {\bibinfo {title} {{Infrared Modified
  Gravity with Dynamical Torsion}},\ }\href
  {https://doi.org/10.1103/PhysRevD.80.124050} {\bibfield  {journal} {\bibinfo
  {journal} {Phys. Rev. D}\ }\textbf {\bibinfo {volume} {80}},\ \bibinfo
  {pages} {124050} (\bibinfo {year} {2009})},\ \Eprint
  {https://arxiv.org/abs/0905.3732} {arXiv:0905.3732 [hep-th]} \BibitemShut
  {NoStop}%
\bibitem [{\citenamefont {Chen}\ \emph {et~al.}(2009)\citenamefont {Chen},
  \citenamefont {Ho}, \citenamefont {Nester}, \citenamefont {Wang},\ and\
  \citenamefont {Yo}}]{Chen:2009at}%
  \BibitemOpen
  \bibfield  {author} {\bibinfo {author} {\bibfnamefont {H.}~\bibnamefont
  {Chen}}, \bibinfo {author} {\bibfnamefont {F.-H.}\ \bibnamefont {Ho}},
  \bibinfo {author} {\bibfnamefont {J.~M.}\ \bibnamefont {Nester}}, \bibinfo
  {author} {\bibfnamefont {C.-H.}\ \bibnamefont {Wang}},\ and\ \bibinfo
  {author} {\bibfnamefont {H.-J.}\ \bibnamefont {Yo}},\ }\bibfield  {title}
  {\bibinfo {title} {{Cosmological dynamics with propagating Lorentz connection
  modes of spin zero}},\ }\href {https://doi.org/10.1088/1475-7516/2009/10/027}
  {\bibfield  {journal} {\bibinfo  {journal} {JCAP}\ }\textbf {\bibinfo
  {volume} {10}},\ \bibinfo {pages} {027}},\ \Eprint
  {https://arxiv.org/abs/0908.3323} {arXiv:0908.3323 [gr-qc]} \BibitemShut
  {NoStop}%
\bibitem [{\citenamefont {Ni}(2010)}]{Ni:2009fg}%
  \BibitemOpen
  \bibfield  {author} {\bibinfo {author} {\bibfnamefont {W.-T.}\ \bibnamefont
  {Ni}},\ }\bibfield  {title} {\bibinfo {title} {{Searches for the role of spin
  and polarization in gravity}},\ }\href
  {https://doi.org/10.1088/0034-4885/73/5/056901} {\bibfield  {journal}
  {\bibinfo  {journal} {Rept. Prog. Phys.}\ }\textbf {\bibinfo {volume} {73}},\
  \bibinfo {pages} {056901} (\bibinfo {year} {2010})},\ \Eprint
  {https://arxiv.org/abs/0912.5057} {arXiv:0912.5057 [gr-qc]} \BibitemShut
  {NoStop}%
\bibitem [{\citenamefont {Baekler}\ \emph {et~al.}(2011)\citenamefont
  {Baekler}, \citenamefont {Hehl},\ and\ \citenamefont
  {Nester}}]{Baekler:2010fr}%
  \BibitemOpen
  \bibfield  {author} {\bibinfo {author} {\bibfnamefont {P.}~\bibnamefont
  {Baekler}}, \bibinfo {author} {\bibfnamefont {F.~W.}\ \bibnamefont {Hehl}},\
  and\ \bibinfo {author} {\bibfnamefont {J.~M.}\ \bibnamefont {Nester}},\
  }\bibfield  {title} {\bibinfo {title} {{Poincare gauge theory of gravity:
  Friedman cosmology with even and odd parity modes. Analytic part}},\ }\href
  {https://doi.org/10.1103/PhysRevD.83.024001} {\bibfield  {journal} {\bibinfo
  {journal} {Phys. Rev. D}\ }\textbf {\bibinfo {volume} {83}},\ \bibinfo
  {pages} {024001} (\bibinfo {year} {2011})},\ \Eprint
  {https://arxiv.org/abs/1009.5112} {arXiv:1009.5112 [gr-qc]} \BibitemShut
  {NoStop}%
\bibitem [{\citenamefont {Ho}\ and\ \citenamefont {Nester}(2011)}]{Ho:2011qn}%
  \BibitemOpen
  \bibfield  {author} {\bibinfo {author} {\bibfnamefont {F.-H.}\ \bibnamefont
  {Ho}}\ and\ \bibinfo {author} {\bibfnamefont {J.~M.}\ \bibnamefont
  {Nester}},\ }\bibfield  {title} {\bibinfo {title} {{Poincar\'e gauge theory
  with even and odd parity dynamic connection modes: isotropic Bianchi
  cosmological models}},\ }\href
  {https://doi.org/10.1088/1742-6596/330/1/012005} {\bibfield  {journal}
  {\bibinfo  {journal} {J. Phys. Conf. Ser.}\ }\textbf {\bibinfo {volume}
  {330}},\ \bibinfo {pages} {012005} (\bibinfo {year} {2011})},\ \Eprint
  {https://arxiv.org/abs/1105.5001} {arXiv:1105.5001 [gr-qc]} \BibitemShut
  {NoStop}%
\bibitem [{\citenamefont {Ho}\ and\ \citenamefont {Nester}(2012)}]{Ho:2011xf}%
  \BibitemOpen
  \bibfield  {author} {\bibinfo {author} {\bibfnamefont {F.-H.}\ \bibnamefont
  {Ho}}\ and\ \bibinfo {author} {\bibfnamefont {J.~M.}\ \bibnamefont
  {Nester}},\ }\bibfield  {title} {\bibinfo {title} {{Poincar\'e Gauge Theory
  With Coupled Even And Odd Parity Dynamic Spin-0 Modes: Dynamic Equations For
  Isotropic Bianchi Cosmologies}},\ }\href
  {https://doi.org/10.1002/andp.201100101} {\bibfield  {journal} {\bibinfo
  {journal} {Annalen Phys.}\ }\textbf {\bibinfo {volume} {524}},\ \bibinfo
  {pages} {97} (\bibinfo {year} {2012})},\ \Eprint
  {https://arxiv.org/abs/1106.0711} {arXiv:1106.0711 [gr-qc]} \BibitemShut
  {NoStop}%
\bibitem [{\citenamefont {Ong}\ \emph {et~al.}(2013)\citenamefont {Ong},
  \citenamefont {Izumi}, \citenamefont {Nester},\ and\ \citenamefont
  {Chen}}]{Ong:2013qja}%
  \BibitemOpen
  \bibfield  {author} {\bibinfo {author} {\bibfnamefont {Y.~C.}\ \bibnamefont
  {Ong}}, \bibinfo {author} {\bibfnamefont {K.}~\bibnamefont {Izumi}}, \bibinfo
  {author} {\bibfnamefont {J.~M.}\ \bibnamefont {Nester}},\ and\ \bibinfo
  {author} {\bibfnamefont {P.}~\bibnamefont {Chen}},\ }\bibfield  {title}
  {\bibinfo {title} {{Problems with Propagation and Time Evolution in f(T)
  Gravity}},\ }\href {https://doi.org/10.1103/PhysRevD.88.024019} {\bibfield
  {journal} {\bibinfo  {journal} {Phys. Rev. D}\ }\textbf {\bibinfo {volume}
  {88}},\ \bibinfo {pages} {024019} (\bibinfo {year} {2013})},\ \Eprint
  {https://arxiv.org/abs/1303.0993} {arXiv:1303.0993 [gr-qc]} \BibitemShut
  {NoStop}%
\bibitem [{\citenamefont {Puetzfeld}\ and\ \citenamefont
  {Obukhov}(2014)}]{Puetzfeld:2014sja}%
  \BibitemOpen
  \bibfield  {author} {\bibinfo {author} {\bibfnamefont {D.}~\bibnamefont
  {Puetzfeld}}\ and\ \bibinfo {author} {\bibfnamefont {Y.~N.}\ \bibnamefont
  {Obukhov}},\ }\bibfield  {title} {\bibinfo {title} {{Prospects of detecting
  spacetime torsion}},\ }\href {https://doi.org/10.1142/S0218271814420048}
  {\bibfield  {journal} {\bibinfo  {journal} {Int. J. Mod. Phys. D}\ }\textbf
  {\bibinfo {volume} {23}},\ \bibinfo {pages} {1442004} (\bibinfo {year}
  {2014})},\ \Eprint {https://arxiv.org/abs/1405.4137} {arXiv:1405.4137
  [gr-qc]} \BibitemShut {NoStop}%
\bibitem [{\citenamefont {Karananas}(2015)}]{Karananas:2014pxa}%
  \BibitemOpen
  \bibfield  {author} {\bibinfo {author} {\bibfnamefont {G.~K.}\ \bibnamefont
  {Karananas}},\ }\bibfield  {title} {\bibinfo {title} {{The particle spectrum
  of parity-violating Poincar\'e gravitational theory}},\ }\href
  {https://doi.org/10.1088/0264-9381/32/5/055012} {\bibfield  {journal}
  {\bibinfo  {journal} {Class. Quant. Grav.}\ }\textbf {\bibinfo {volume}
  {32}},\ \bibinfo {pages} {055012} (\bibinfo {year} {2015})},\ \Eprint
  {https://arxiv.org/abs/1411.5613} {arXiv:1411.5613 [gr-qc]} \BibitemShut
  {NoStop}%
\bibitem [{\citenamefont {Ni}(2016)}]{Ni:2015poa}%
  \BibitemOpen
  \bibfield  {author} {\bibinfo {author} {\bibfnamefont {W.-T.}\ \bibnamefont
  {Ni}},\ }\bibfield  {title} {\bibinfo {title} {{Searches for the role of spin
  and polarization in gravity: a five-year update}},\ }\href
  {https://doi.org/10.1142/S2010194516600107} {\bibfield  {journal} {\bibinfo
  {journal} {Int. J. Mod. Phys. Conf. Ser.}\ }\textbf {\bibinfo {volume}
  {40}},\ \bibinfo {pages} {1660010} (\bibinfo {year} {2016})},\ \Eprint
  {https://arxiv.org/abs/1501.07696} {arXiv:1501.07696 [hep-ph]} \BibitemShut
  {NoStop}%
\bibitem [{\citenamefont {Ho}\ \emph {et~al.}(2015)\citenamefont {Ho},
  \citenamefont {Chen}, \citenamefont {Nester},\ and\ \citenamefont
  {Yo}}]{Ho:2015ulu}%
  \BibitemOpen
  \bibfield  {author} {\bibinfo {author} {\bibfnamefont {F.-H.}\ \bibnamefont
  {Ho}}, \bibinfo {author} {\bibfnamefont {H.}~\bibnamefont {Chen}}, \bibinfo
  {author} {\bibfnamefont {J.~M.}\ \bibnamefont {Nester}},\ and\ \bibinfo
  {author} {\bibfnamefont {H.-J.}\ \bibnamefont {Yo}},\ }\bibfield  {title}
  {\bibinfo {title} {{General Poincar\'e Gauge Theory Cosmology}},\ }\href
  {https://doi.org/10.6122/CJP.20151014} {\bibfield  {journal} {\bibinfo
  {journal} {Chin. J. Phys.}\ }\textbf {\bibinfo {volume} {53}},\ \bibinfo
  {pages} {110109} (\bibinfo {year} {2015})},\ \Eprint
  {https://arxiv.org/abs/1512.01202} {arXiv:1512.01202 [gr-qc]} \BibitemShut
  {NoStop}%
\bibitem [{\citenamefont {Karananas}(2016)}]{Karananas:2016ltn}%
  \BibitemOpen
  \bibfield  {author} {\bibinfo {author} {\bibfnamefont {G.~K.}\ \bibnamefont
  {Karananas}},\ }\emph {\bibinfo {title} {{Poincar\'e, Scale and Conformal
  Symmetries Gauge Perspective and Cosmological Ramifications}}},\ \href
  {https://doi.org/10.5075/epfl-thesis-7173} {Ph.D. thesis},\ \bibinfo
  {school} {Ecole Polytechnique, Lausanne} (\bibinfo {year} {2016}),\ \Eprint
  {https://arxiv.org/abs/1608.08451} {arXiv:1608.08451 [hep-th]} \BibitemShut
  {NoStop}%
\bibitem [{\citenamefont {Obukhov}(2017)}]{Obukhov:2017pxa}%
  \BibitemOpen
  \bibfield  {author} {\bibinfo {author} {\bibfnamefont {Y.~N.}\ \bibnamefont
  {Obukhov}},\ }\bibfield  {title} {\bibinfo {title} {{Gravitational waves in
  Poincar\'e gauge gravity theory}},\ }\href
  {https://doi.org/10.1103/PhysRevD.95.084028} {\bibfield  {journal} {\bibinfo
  {journal} {Phys. Rev. D}\ }\textbf {\bibinfo {volume} {95}},\ \bibinfo
  {pages} {084028} (\bibinfo {year} {2017})},\ \Eprint
  {https://arxiv.org/abs/1702.05185} {arXiv:1702.05185 [gr-qc]} \BibitemShut
  {NoStop}%
\bibitem [{\citenamefont {Blagojevi\'c}\ \emph {et~al.}(2017)\citenamefont
  {Blagojevi\'c}, \citenamefont {Cvetkovi\'c},\ and\ \citenamefont
  {Obukhov}}]{Blagojevic:2017ssv}%
  \BibitemOpen
  \bibfield  {author} {\bibinfo {author} {\bibfnamefont {M.}~\bibnamefont
  {Blagojevi\'c}}, \bibinfo {author} {\bibfnamefont {B.}~\bibnamefont
  {Cvetkovi\'c}},\ and\ \bibinfo {author} {\bibfnamefont {Y.~N.}\ \bibnamefont
  {Obukhov}},\ }\bibfield  {title} {\bibinfo {title} {{Generalized plane waves
  in Poincar\'e gauge theory of gravity}},\ }\href
  {https://doi.org/10.1103/PhysRevD.96.064031} {\bibfield  {journal} {\bibinfo
  {journal} {Phys. Rev. D}\ }\textbf {\bibinfo {volume} {96}},\ \bibinfo
  {pages} {064031} (\bibinfo {year} {2017})},\ \Eprint
  {https://arxiv.org/abs/1708.08766} {arXiv:1708.08766 [gr-qc]} \BibitemShut
  {NoStop}%
\bibitem [{\citenamefont {Blagojevi\'c}\ and\ \citenamefont
  {Cvetkovi\'c}(2018)}]{Blagojevic:2018dpz}%
  \BibitemOpen
  \bibfield  {author} {\bibinfo {author} {\bibfnamefont {M.}~\bibnamefont
  {Blagojevi\'c}}\ and\ \bibinfo {author} {\bibfnamefont {B.}~\bibnamefont
  {Cvetkovi\'c}},\ }\bibfield  {title} {\bibinfo {title} {{General Poincar\'e
  gauge theory: Hamiltonian structure and particle spectrum}},\ }\href
  {https://doi.org/10.1103/PhysRevD.98.024014} {\bibfield  {journal} {\bibinfo
  {journal} {Phys. Rev. D}\ }\textbf {\bibinfo {volume} {98}},\ \bibinfo
  {pages} {024014} (\bibinfo {year} {2018})},\ \Eprint
  {https://arxiv.org/abs/1804.05556} {arXiv:1804.05556 [gr-qc]} \BibitemShut
  {NoStop}%
\bibitem [{\citenamefont {Tseng}(2018)}]{Tseng:2018feo}%
  \BibitemOpen
  \bibfield  {author} {\bibinfo {author} {\bibfnamefont {H.-H.}\ \bibnamefont
  {Tseng}},\ }\emph {\bibinfo {title} {{Gravitational Theories with
  Torsion}}},\ \href@noop {} {Ph.D. thesis},\ \bibinfo  {school} {Taiwan, Natl.
  Tsing Hua U.} (\bibinfo {year} {2018}),\ \Eprint
  {https://arxiv.org/abs/1812.00314} {arXiv:1812.00314 [gr-qc]} \BibitemShut
  {NoStop}%
\bibitem [{\citenamefont {Lin}\ \emph {et~al.}(2019)\citenamefont {Lin},
  \citenamefont {Hobson},\ and\ \citenamefont {Lasenby}}]{Lin:2018awc}%
  \BibitemOpen
  \bibfield  {author} {\bibinfo {author} {\bibfnamefont {Y.-C.}\ \bibnamefont
  {Lin}}, \bibinfo {author} {\bibfnamefont {M.~P.}\ \bibnamefont {Hobson}},\
  and\ \bibinfo {author} {\bibfnamefont {A.~N.}\ \bibnamefont {Lasenby}},\
  }\bibfield  {title} {\bibinfo {title} {{Ghost and tachyon free Poincar\'e
  gauge theories: A systematic approach}},\ }\href
  {https://doi.org/10.1103/PhysRevD.99.064001} {\bibfield  {journal} {\bibinfo
  {journal} {Phys. Rev. D}\ }\textbf {\bibinfo {volume} {99}},\ \bibinfo
  {pages} {064001} (\bibinfo {year} {2019})},\ \Eprint
  {https://arxiv.org/abs/1812.02675} {arXiv:1812.02675 [gr-qc]} \BibitemShut
  {NoStop}%
\bibitem [{\citenamefont {Beltr\'an~Jim\'enez}\ and\ \citenamefont
  {Delhom}(2019)}]{BeltranJimenez:2019acz}%
  \BibitemOpen
  \bibfield  {author} {\bibinfo {author} {\bibfnamefont {J.}~\bibnamefont
  {Beltr\'an~Jim\'enez}}\ and\ \bibinfo {author} {\bibfnamefont
  {A.}~\bibnamefont {Delhom}},\ }\bibfield  {title} {\bibinfo {title} {{Ghosts
  in metric-affine higher order curvature gravity}},\ }\href
  {https://doi.org/10.1140/epjc/s10052-019-7149-x} {\bibfield  {journal}
  {\bibinfo  {journal} {Eur. Phys. J. C}\ }\textbf {\bibinfo {volume} {79}},\
  \bibinfo {pages} {656} (\bibinfo {year} {2019})},\ \Eprint
  {https://arxiv.org/abs/1901.08988} {arXiv:1901.08988 [gr-qc]} \BibitemShut
  {NoStop}%
\bibitem [{\citenamefont {Zhang}\ and\ \citenamefont
  {Xu}(2019)}]{Zhang:2019mhd}%
  \BibitemOpen
  \bibfield  {author} {\bibinfo {author} {\bibfnamefont {H.}~\bibnamefont
  {Zhang}}\ and\ \bibinfo {author} {\bibfnamefont {L.}~\bibnamefont {Xu}},\
  }\bibfield  {title} {\bibinfo {title} {{Late-time acceleration and inflation
  in a Poincar\'e gauge cosmological model}},\ }\href
  {https://doi.org/10.1088/1475-7516/2019/09/050} {\bibfield  {journal}
  {\bibinfo  {journal} {JCAP}\ }\textbf {\bibinfo {volume} {09}},\ \bibinfo
  {pages} {050}},\ \Eprint {https://arxiv.org/abs/1904.03545} {arXiv:1904.03545
  [gr-qc]} \BibitemShut {NoStop}%
\bibitem [{\citenamefont {Aoki}\ and\ \citenamefont
  {Shimada}(2019)}]{Aoki:2019rvi}%
  \BibitemOpen
  \bibfield  {author} {\bibinfo {author} {\bibfnamefont {K.}~\bibnamefont
  {Aoki}}\ and\ \bibinfo {author} {\bibfnamefont {K.}~\bibnamefont {Shimada}},\
  }\bibfield  {title} {\bibinfo {title} {{Scalar-metric-affine theories: Can we
  get ghost-free theories from symmetry?}},\ }\href
  {https://doi.org/10.1103/PhysRevD.100.044037} {\bibfield  {journal} {\bibinfo
   {journal} {Phys. Rev. D}\ }\textbf {\bibinfo {volume} {100}},\ \bibinfo
  {pages} {044037} (\bibinfo {year} {2019})},\ \Eprint
  {https://arxiv.org/abs/1904.10175} {arXiv:1904.10175 [hep-th]} \BibitemShut
  {NoStop}%
\bibitem [{\citenamefont {Zhang}\ and\ \citenamefont
  {Xu}(2020)}]{Zhang:2019xek}%
  \BibitemOpen
  \bibfield  {author} {\bibinfo {author} {\bibfnamefont {H.}~\bibnamefont
  {Zhang}}\ and\ \bibinfo {author} {\bibfnamefont {L.}~\bibnamefont {Xu}},\
  }\bibfield  {title} {\bibinfo {title} {{Inflation in the parity-conserving
  Poincar\'e gauge cosmology}},\ }\href
  {https://doi.org/10.1088/1475-7516/2020/10/003} {\bibfield  {journal}
  {\bibinfo  {journal} {JCAP}\ }\textbf {\bibinfo {volume} {10}},\ \bibinfo
  {pages} {003}},\ \Eprint {https://arxiv.org/abs/1906.04340} {arXiv:1906.04340
  [gr-qc]} \BibitemShut {NoStop}%
\bibitem [{\citenamefont {Beltr\'an~Jim\'enez}\ and\ \citenamefont
  {Maldonado~Torralba}(2020)}]{Jimenez:2019qjc}%
  \BibitemOpen
  \bibfield  {author} {\bibinfo {author} {\bibfnamefont {J.}~\bibnamefont
  {Beltr\'an~Jim\'enez}}\ and\ \bibinfo {author} {\bibfnamefont {F.~J.}\
  \bibnamefont {Maldonado~Torralba}},\ }\bibfield  {title} {\bibinfo {title}
  {{Revisiting the stability of quadratic Poincar\'e gauge gravity}},\ }\href
  {https://doi.org/10.1140/epjc/s10052-020-8163-8} {\bibfield  {journal}
  {\bibinfo  {journal} {Eur. Phys. J. C}\ }\textbf {\bibinfo {volume} {80}},\
  \bibinfo {pages} {611} (\bibinfo {year} {2020})},\ \Eprint
  {https://arxiv.org/abs/1910.07506} {arXiv:1910.07506 [gr-qc]} \BibitemShut
  {NoStop}%
\bibitem [{\citenamefont {Lin}\ \emph {et~al.}(2020)\citenamefont {Lin},
  \citenamefont {Hobson},\ and\ \citenamefont {Lasenby}}]{Lin:2019ugq}%
  \BibitemOpen
  \bibfield  {author} {\bibinfo {author} {\bibfnamefont {Y.-C.}\ \bibnamefont
  {Lin}}, \bibinfo {author} {\bibfnamefont {M.~P.}\ \bibnamefont {Hobson}},\
  and\ \bibinfo {author} {\bibfnamefont {A.~N.}\ \bibnamefont {Lasenby}},\
  }\bibfield  {title} {\bibinfo {title} {{Power-counting renormalizable,
  ghost-and-tachyon-free Poincar\'e gauge theories}},\ }\href
  {https://doi.org/10.1103/PhysRevD.101.064038} {\bibfield  {journal} {\bibinfo
   {journal} {Phys. Rev. D}\ }\textbf {\bibinfo {volume} {101}},\ \bibinfo
  {pages} {064038} (\bibinfo {year} {2020})},\ \Eprint
  {https://arxiv.org/abs/1910.14197} {arXiv:1910.14197 [gr-qc]} \BibitemShut
  {NoStop}%
\bibitem [{\citenamefont {Percacci}\ and\ \citenamefont
  {Sezgin}(2020{\natexlab{b}})}]{Percacci:2019hxn}%
  \BibitemOpen
  \bibfield  {author} {\bibinfo {author} {\bibfnamefont {R.}~\bibnamefont
  {Percacci}}\ and\ \bibinfo {author} {\bibfnamefont {E.}~\bibnamefont
  {Sezgin}},\ }\bibfield  {title} {\bibinfo {title} {{New class of ghost- and
  tachyon-free metric affine gravities}},\ }\href
  {https://doi.org/10.1103/PhysRevD.101.084040} {\bibfield  {journal} {\bibinfo
   {journal} {Phys. Rev. D}\ }\textbf {\bibinfo {volume} {101}},\ \bibinfo
  {pages} {084040} (\bibinfo {year} {2020}{\natexlab{b}})},\ \Eprint
  {https://arxiv.org/abs/1912.01023} {arXiv:1912.01023 [hep-th]} \BibitemShut
  {NoStop}%
\bibitem [{\citenamefont {Barker}\ \emph
  {et~al.}(2020{\natexlab{a}})\citenamefont {Barker}, \citenamefont {Lasenby},
  \citenamefont {Hobson},\ and\ \citenamefont {Handley}}]{Barker:2020gcp}%
  \BibitemOpen
  \bibfield  {author} {\bibinfo {author} {\bibfnamefont {W.~E.~V.}\
  \bibnamefont {Barker}}, \bibinfo {author} {\bibfnamefont {A.~N.}\
  \bibnamefont {Lasenby}}, \bibinfo {author} {\bibfnamefont {M.~P.}\
  \bibnamefont {Hobson}},\ and\ \bibinfo {author} {\bibfnamefont {W.~J.}\
  \bibnamefont {Handley}},\ }\bibfield  {title} {\bibinfo {title} {{Systematic
  study of background cosmology in unitary Poincar\'e gauge theories with
  application to emergent dark radiation and $H_0$ tension}},\ }\href
  {https://doi.org/10.1103/PhysRevD.102.024048} {\bibfield  {journal} {\bibinfo
   {journal} {Phys. Rev. D}\ }\textbf {\bibinfo {volume} {102}},\ \bibinfo
  {pages} {024048} (\bibinfo {year} {2020}{\natexlab{a}})},\ \Eprint
  {https://arxiv.org/abs/2003.02690} {arXiv:2003.02690 [gr-qc]} \BibitemShut
  {NoStop}%
\bibitem [{\citenamefont {Beltr\'an~Jim\'enez}\ and\ \citenamefont
  {Delhom}(2020)}]{BeltranJimenez:2020sqf}%
  \BibitemOpen
  \bibfield  {author} {\bibinfo {author} {\bibfnamefont {J.}~\bibnamefont
  {Beltr\'an~Jim\'enez}}\ and\ \bibinfo {author} {\bibfnamefont
  {A.}~\bibnamefont {Delhom}},\ }\bibfield  {title} {\bibinfo {title}
  {{Instabilities in metric-affine theories of gravity with higher order
  curvature terms}},\ }\href {https://doi.org/10.1140/epjc/s10052-020-8143-z}
  {\bibfield  {journal} {\bibinfo  {journal} {Eur. Phys. J. C}\ }\textbf
  {\bibinfo {volume} {80}},\ \bibinfo {pages} {585} (\bibinfo {year} {2020})},\
  \Eprint {https://arxiv.org/abs/2004.11357} {arXiv:2004.11357 [gr-qc]}
  \BibitemShut {NoStop}%
\bibitem [{\citenamefont
  {Maldonado~Torralba}(2020)}]{MaldonadoTorralba:2020mbh}%
  \BibitemOpen
  \bibfield  {author} {\bibinfo {author} {\bibfnamefont {F.~J.}\ \bibnamefont
  {Maldonado~Torralba}},\ }\emph {\bibinfo {title} {{New effective theories of
  gravitation and their phenomenological consequences}}},\ \href
  {https://doi.org/10.33612/diss.143961423} {Ph.D. thesis},\ \bibinfo  {school}
  {Cape Town U., Dept. Math.} (\bibinfo {year} {2020}),\ \Eprint
  {https://arxiv.org/abs/2101.11523} {arXiv:2101.11523 [gr-qc]} \BibitemShut
  {NoStop}%
\bibitem [{\citenamefont {Barker}\ \emph {et~al.}(2021)\citenamefont {Barker},
  \citenamefont {Lasenby}, \citenamefont {Hobson},\ and\ \citenamefont
  {Handley}}]{Barker:2021oez}%
  \BibitemOpen
  \bibfield  {author} {\bibinfo {author} {\bibfnamefont {W.~E.~V.}\
  \bibnamefont {Barker}}, \bibinfo {author} {\bibfnamefont {A.~N.}\
  \bibnamefont {Lasenby}}, \bibinfo {author} {\bibfnamefont {M.~P.}\
  \bibnamefont {Hobson}},\ and\ \bibinfo {author} {\bibfnamefont {W.~J.}\
  \bibnamefont {Handley}},\ }\bibfield  {title} {\bibinfo {title} {{Nonlinear
  Hamiltonian analysis of new quadratic torsion theories: Cases with
  curvature-free constraints}},\ }\href
  {https://doi.org/10.1103/PhysRevD.104.084036} {\bibfield  {journal} {\bibinfo
   {journal} {Phys. Rev. D}\ }\textbf {\bibinfo {volume} {104}},\ \bibinfo
  {pages} {084036} (\bibinfo {year} {2021})},\ \Eprint
  {https://arxiv.org/abs/2101.02645} {arXiv:2101.02645 [gr-qc]} \BibitemShut
  {NoStop}%
\bibitem [{\citenamefont {Marzo}(2022{\natexlab{a}})}]{Marzo:2021esg}%
  \BibitemOpen
  \bibfield  {author} {\bibinfo {author} {\bibfnamefont {C.}~\bibnamefont
  {Marzo}},\ }\bibfield  {title} {\bibinfo {title} {{Ghost and tachyon free
  propagation up to spin 3 in Lorentz invariant field theories}},\ }\href
  {https://doi.org/10.1103/PhysRevD.105.065017} {\bibfield  {journal} {\bibinfo
   {journal} {Phys. Rev. D}\ }\textbf {\bibinfo {volume} {105}},\ \bibinfo
  {pages} {065017} (\bibinfo {year} {2022}{\natexlab{a}})},\ \Eprint
  {https://arxiv.org/abs/2108.11982} {arXiv:2108.11982 [hep-ph]} \BibitemShut
  {NoStop}%
\bibitem [{\citenamefont {Marzo}(2022{\natexlab{b}})}]{Marzo:2021iok}%
  \BibitemOpen
  \bibfield  {author} {\bibinfo {author} {\bibfnamefont {C.}~\bibnamefont
  {Marzo}},\ }\bibfield  {title} {\bibinfo {title} {{Radiatively stable ghost
  and tachyon freedom in metric affine gravity}},\ }\href
  {https://doi.org/10.1103/PhysRevD.106.024045} {\bibfield  {journal} {\bibinfo
   {journal} {Phys. Rev. D}\ }\textbf {\bibinfo {volume} {106}},\ \bibinfo
  {pages} {024045} (\bibinfo {year} {2022}{\natexlab{b}})},\ \Eprint
  {https://arxiv.org/abs/2110.14788} {arXiv:2110.14788 [hep-th]} \BibitemShut
  {NoStop}%
\bibitem [{\citenamefont {de~la Cruz~Dombriz}\ \emph
  {et~al.}(2022)\citenamefont {de~la Cruz~Dombriz}, \citenamefont
  {Maldonado~Torralba},\ and\ \citenamefont {Mota}}]{delaCruzDombriz:2021nrg}%
  \BibitemOpen
  \bibfield  {author} {\bibinfo {author} {\bibfnamefont {A.}~\bibnamefont
  {de~la Cruz~Dombriz}}, \bibinfo {author} {\bibfnamefont {F.~J.}\ \bibnamefont
  {Maldonado~Torralba}},\ and\ \bibinfo {author} {\bibfnamefont {D.~F.}\
  \bibnamefont {Mota}},\ }\bibfield  {title} {\bibinfo {title} {{Dark matter
  candidate from torsion}},\ }\href
  {https://doi.org/10.1016/j.physletb.2022.137488} {\bibfield  {journal}
  {\bibinfo  {journal} {Phys. Lett. B}\ }\textbf {\bibinfo {volume} {834}},\
  \bibinfo {pages} {137488} (\bibinfo {year} {2022})},\ \Eprint
  {https://arxiv.org/abs/2112.03957} {arXiv:2112.03957 [gr-qc]} \BibitemShut
  {NoStop}%
\bibitem [{\citenamefont {Annala}\ and\ \citenamefont
  {Rasanen}(2023)}]{Annala:2022gtl}%
  \BibitemOpen
  \bibfield  {author} {\bibinfo {author} {\bibfnamefont {J.}~\bibnamefont
  {Annala}}\ and\ \bibinfo {author} {\bibfnamefont {S.}~\bibnamefont
  {Rasanen}},\ }\bibfield  {title} {\bibinfo {title} {{Stability of
  non-degenerate Ricci-type Palatini theories}},\ }\href
  {https://doi.org/10.1088/1475-7516/2023/04/014} {\bibfield  {journal}
  {\bibinfo  {journal} {JCAP}\ }\textbf {\bibinfo {volume} {04}},\ \bibinfo
  {pages} {014}},\ \bibinfo {note} {[Erratum: JCAP 08, E02 (2023)]},\ \Eprint
  {https://arxiv.org/abs/2212.09820} {arXiv:2212.09820 [gr-qc]} \BibitemShut
  {NoStop}%
\bibitem [{\citenamefont {Mikura}\ \emph {et~al.}(2024)\citenamefont {Mikura},
  \citenamefont {Naso},\ and\ \citenamefont {Percacci}}]{Mikura:2023ruz}%
  \BibitemOpen
  \bibfield  {author} {\bibinfo {author} {\bibfnamefont {Y.}~\bibnamefont
  {Mikura}}, \bibinfo {author} {\bibfnamefont {V.}~\bibnamefont {Naso}},\ and\
  \bibinfo {author} {\bibfnamefont {R.}~\bibnamefont {Percacci}},\ }\bibfield
  {title} {\bibinfo {title} {{Some simple theories of gravity with propagating
  torsion}},\ }\href {https://doi.org/10.1103/PhysRevD.109.104071} {\bibfield
  {journal} {\bibinfo  {journal} {Phys. Rev. D}\ }\textbf {\bibinfo {volume}
  {109}},\ \bibinfo {pages} {104071} (\bibinfo {year} {2024})},\ \Eprint
  {https://arxiv.org/abs/2312.10249} {arXiv:2312.10249 [gr-qc]} \BibitemShut
  {NoStop}%
\bibitem [{\citenamefont {Mikura}\ and\ \citenamefont
  {Percacci}(2024)}]{Mikura:2024mji}%
  \BibitemOpen
  \bibfield  {author} {\bibinfo {author} {\bibfnamefont {Y.}~\bibnamefont
  {Mikura}}\ and\ \bibinfo {author} {\bibfnamefont {R.}~\bibnamefont
  {Percacci}},\ }\bibfield  {title} {\bibinfo {title} {{Some simple theories of
  gravity with propagating nonmetricity}},\ }\href@noop {} {\  (\bibinfo {year}
  {2024})},\ \Eprint {https://arxiv.org/abs/2401.10097} {arXiv:2401.10097
  [gr-qc]} \BibitemShut {NoStop}%
\bibitem [{\citenamefont {Barker}\ and\ \citenamefont
  {Marzo}(2024{\natexlab{b}})}]{Barker:2024ydb}%
  \BibitemOpen
  \bibfield  {author} {\bibinfo {author} {\bibfnamefont {W.}~\bibnamefont
  {Barker}}\ and\ \bibinfo {author} {\bibfnamefont {C.}~\bibnamefont {Marzo}},\
  }\bibfield  {title} {\bibinfo {title} {{Particle spectra of general
  Ricci-type Palatini or metric-affine theories}},\ }\href
  {https://doi.org/10.1103/PhysRevD.109.104017} {\bibfield  {journal} {\bibinfo
   {journal} {Phys. Rev. D}\ }\textbf {\bibinfo {volume} {109}},\ \bibinfo
  {pages} {104017} (\bibinfo {year} {2024}{\natexlab{b}})},\ \Eprint
  {https://arxiv.org/abs/2402.07641} {arXiv:2402.07641 [hep-th]} \BibitemShut
  {NoStop}%
\bibitem [{\citenamefont {M{\o}ller}(1961)}]{Moller:1961}%
  \BibitemOpen
  \bibfield  {author} {\bibinfo {author} {\bibfnamefont {C.}~\bibnamefont
  {M{\o}ller}},\ }\bibfield  {title} {\bibinfo {title} {{Conservation Law and
  Absolute Parallelism in General Relativity}},\ }\href@noop {} {\bibfield
  {journal} {\bibinfo  {journal} {K. Dan. Vidensk. Selsk. Mat. Fys. Skr.}\
  }\textbf {\bibinfo {volume} {1}},\ \bibinfo {pages} {1} (\bibinfo {year}
  {1961})}\BibitemShut {NoStop}%
\bibitem [{\citenamefont {Pellegrini}\ and\ \citenamefont
  {Plebanski}(1963)}]{Pellegrini:1963}%
  \BibitemOpen
  \bibfield  {author} {\bibinfo {author} {\bibfnamefont {C.}~\bibnamefont
  {Pellegrini}}\ and\ \bibinfo {author} {\bibfnamefont {J.}~\bibnamefont
  {Plebanski}},\ }\bibfield  {title} {\bibinfo {title} {{Tetrad Fields and
  Gravitational Fields}},\ }\href {https://www.osti.gov/biblio/4688687}
  {\bibfield  {journal} {\bibinfo  {journal} {K. Dan. Vidensk. Selsk. Mat. Fys.
  Skr.}\ }\textbf {\bibinfo {volume} {2}},\ \bibinfo {pages} {1} (\bibinfo
  {year} {1963})}\BibitemShut {NoStop}%
\bibitem [{\citenamefont {Hayashi}\ and\ \citenamefont
  {Nakano}(1967)}]{Hayashi:1967se}%
  \BibitemOpen
  \bibfield  {author} {\bibinfo {author} {\bibfnamefont {K.}~\bibnamefont
  {Hayashi}}\ and\ \bibinfo {author} {\bibfnamefont {T.}~\bibnamefont
  {Nakano}},\ }\bibfield  {title} {\bibinfo {title} {{Extended translation
  invariance and associated gauge fields}},\ }\href
  {https://doi.org/10.1143/PTP.38.491} {\bibfield  {journal} {\bibinfo
  {journal} {Prog. Theor. Phys.}\ }\textbf {\bibinfo {volume} {38}},\ \bibinfo
  {pages} {491} (\bibinfo {year} {1967})}\BibitemShut {NoStop}%
\bibitem [{\citenamefont {Cho}(1976)}]{Cho:1975dh}%
  \BibitemOpen
  \bibfield  {author} {\bibinfo {author} {\bibfnamefont {Y.~M.}\ \bibnamefont
  {Cho}},\ }\bibfield  {title} {\bibinfo {title} {{Einstein Lagrangian as the
  Translational Yang-Mills Lagrangian}},\ }\href
  {https://doi.org/10.1103/PhysRevD.14.2521} {\bibfield  {journal} {\bibinfo
  {journal} {Phys. Rev. D}\ }\textbf {\bibinfo {volume} {14}},\ \bibinfo
  {pages} {2521} (\bibinfo {year} {1976})}\BibitemShut {NoStop}%
\bibitem [{\citenamefont {Hayashi}\ and\ \citenamefont
  {Shirafuji}(1979)}]{Hayashi:1979qx}%
  \BibitemOpen
  \bibfield  {author} {\bibinfo {author} {\bibfnamefont {K.}~\bibnamefont
  {Hayashi}}\ and\ \bibinfo {author} {\bibfnamefont {T.}~\bibnamefont
  {Shirafuji}},\ }\bibfield  {title} {\bibinfo {title} {{New general
  relativity.}},\ }\href {https://doi.org/10.1103/PhysRevD.19.3524} {\bibfield
  {journal} {\bibinfo  {journal} {Phys. Rev. D}\ }\textbf {\bibinfo {volume}
  {19}},\ \bibinfo {pages} {3524} (\bibinfo {year} {1979})},\ \bibinfo {note}
  {[Addendum: Phys.Rev.D 24, 3312--3314 (1982)]}\BibitemShut {NoStop}%
\bibitem [{\citenamefont {Dimakis}(1989{\natexlab{a}})}]{Dimakis:1989az}%
  \BibitemOpen
  \bibfield  {author} {\bibinfo {author} {\bibfnamefont {A.}~\bibnamefont
  {Dimakis}},\ }\bibfield  {title} {\bibinfo {title} {{The Initial Value
  Problem of the Poincare Gauge Theory in Vacuum. 1: Second Order Formalism}},\
  }\href {http://eudml.org/doc/76473} {\bibfield  {journal} {\bibinfo
  {journal} {Ann. Inst. H. Poincare Phys. Theor.}\ }\textbf {\bibinfo {volume}
  {51}},\ \bibinfo {pages} {371} (\bibinfo {year}
  {1989}{\natexlab{a}})}\BibitemShut {NoStop}%
\bibitem [{\citenamefont {Dimakis}(1989{\natexlab{b}})}]{Dimakis:1989ba}%
  \BibitemOpen
  \bibfield  {author} {\bibinfo {author} {\bibfnamefont {A.}~\bibnamefont
  {Dimakis}},\ }\bibfield  {title} {\bibinfo {title} {{The Initial Value
  Problem of the Poincare Gauge Theory in Vacuum. 2: First order formalism}},\
  }\href {http://eudml.org/doc/76474} {\bibfield  {journal} {\bibinfo
  {journal} {Ann. Inst. H. Poincare Phys. Theor.}\ }\textbf {\bibinfo {volume}
  {51}},\ \bibinfo {pages} {389} (\bibinfo {year}
  {1989}{\natexlab{b}})}\BibitemShut {NoStop}%
\bibitem [{\citenamefont {Lemke}(1990)}]{Lemke:1990su}%
  \BibitemOpen
  \bibfield  {author} {\bibinfo {author} {\bibfnamefont {J.}~\bibnamefont
  {Lemke}},\ }\bibfield  {title} {\bibinfo {title} {{Shock waves in the
  Poincare gauge theory of gravitation}},\ }\href
  {https://doi.org/10.1016/0375-9601(90)90789-Q} {\bibfield  {journal}
  {\bibinfo  {journal} {Phys. Lett. A}\ }\textbf {\bibinfo {volume} {143}},\
  \bibinfo {pages} {13} (\bibinfo {year} {1990})}\BibitemShut {NoStop}%
\bibitem [{\citenamefont {Hecht}\ \emph {et~al.}(1990)\citenamefont {Hecht},
  \citenamefont {Lemke},\ and\ \citenamefont {Wallner}}]{Hecht:1990wn}%
  \BibitemOpen
  \bibfield  {author} {\bibinfo {author} {\bibfnamefont {R.~D.}\ \bibnamefont
  {Hecht}}, \bibinfo {author} {\bibfnamefont {J.}~\bibnamefont {Lemke}},\ and\
  \bibinfo {author} {\bibfnamefont {R.~P.}\ \bibnamefont {Wallner}},\
  }\bibfield  {title} {\bibinfo {title} {{Tachyonic torsion shock waves in
  Poincare gauge theory}},\ }\href
  {https://doi.org/10.1016/0375-9601(90)90837-E} {\bibfield  {journal}
  {\bibinfo  {journal} {Phys. Lett. A}\ }\textbf {\bibinfo {volume} {151}},\
  \bibinfo {pages} {12} (\bibinfo {year} {1990})}\BibitemShut {NoStop}%
\bibitem [{\citenamefont {Hecht}\ \emph {et~al.}(1991)\citenamefont {Hecht},
  \citenamefont {Lemke},\ and\ \citenamefont {Wallner}}]{Hecht:1991jh}%
  \BibitemOpen
  \bibfield  {author} {\bibinfo {author} {\bibfnamefont {R.~D.}\ \bibnamefont
  {Hecht}}, \bibinfo {author} {\bibfnamefont {J.}~\bibnamefont {Lemke}},\ and\
  \bibinfo {author} {\bibfnamefont {R.~P.}\ \bibnamefont {Wallner}},\
  }\bibfield  {title} {\bibinfo {title} {{Can Poincare gauge theory be
  saved?}},\ }\href {https://doi.org/10.1103/PhysRevD.44.2442} {\bibfield
  {journal} {\bibinfo  {journal} {Phys. Rev. D}\ }\textbf {\bibinfo {volume}
  {44}},\ \bibinfo {pages} {2442} (\bibinfo {year} {1991})}\BibitemShut
  {NoStop}%
\bibitem [{\citenamefont {Afshordi}\ \emph {et~al.}(2007)\citenamefont
  {Afshordi}, \citenamefont {Chung},\ and\ \citenamefont
  {Geshnizjani}}]{Afshordi:2006ad}%
  \BibitemOpen
  \bibfield  {author} {\bibinfo {author} {\bibfnamefont {N.}~\bibnamefont
  {Afshordi}}, \bibinfo {author} {\bibfnamefont {D.~J.~H.}\ \bibnamefont
  {Chung}},\ and\ \bibinfo {author} {\bibfnamefont {G.}~\bibnamefont
  {Geshnizjani}},\ }\bibfield  {title} {\bibinfo {title} {{Cuscuton: A Causal
  Field Theory with an Infinite Speed of Sound}},\ }\href
  {https://doi.org/10.1103/PhysRevD.75.083513} {\bibfield  {journal} {\bibinfo
  {journal} {Phys. Rev. D}\ }\textbf {\bibinfo {volume} {75}},\ \bibinfo
  {pages} {083513} (\bibinfo {year} {2007})},\ \Eprint
  {https://arxiv.org/abs/hep-th/0609150} {arXiv:hep-th/0609150} \BibitemShut
  {NoStop}%
\bibitem [{\citenamefont {Magueijo}(2009)}]{Magueijo:2008sx}%
  \BibitemOpen
  \bibfield  {author} {\bibinfo {author} {\bibfnamefont {J.}~\bibnamefont
  {Magueijo}},\ }\bibfield  {title} {\bibinfo {title} {{Bimetric varying speed
  of light theories and primordial fluctuations}},\ }\href
  {https://doi.org/10.1103/PhysRevD.79.043525} {\bibfield  {journal} {\bibinfo
  {journal} {Phys. Rev. D}\ }\textbf {\bibinfo {volume} {79}},\ \bibinfo
  {pages} {043525} (\bibinfo {year} {2009})},\ \Eprint
  {https://arxiv.org/abs/0807.1689} {arXiv:0807.1689 [gr-qc]} \BibitemShut
  {NoStop}%
\bibitem [{\citenamefont {Charmousis}\ and\ \citenamefont
  {Padilla}(2008)}]{Charmousis:2008ce}%
  \BibitemOpen
  \bibfield  {author} {\bibinfo {author} {\bibfnamefont {C.}~\bibnamefont
  {Charmousis}}\ and\ \bibinfo {author} {\bibfnamefont {A.}~\bibnamefont
  {Padilla}},\ }\bibfield  {title} {\bibinfo {title} {{The Instability of Vacua
  in Gauss-Bonnet Gravity}},\ }\href
  {https://doi.org/10.1088/1126-6708/2008/12/038} {\bibfield  {journal}
  {\bibinfo  {journal} {JHEP}\ }\textbf {\bibinfo {volume} {12}},\ \bibinfo
  {pages} {038}},\ \Eprint {https://arxiv.org/abs/0807.2864} {arXiv:0807.2864
  [hep-th]} \BibitemShut {NoStop}%
\bibitem [{\citenamefont {Charmousis}\ \emph {et~al.}(2009)\citenamefont
  {Charmousis}, \citenamefont {Niz}, \citenamefont {Padilla},\ and\
  \citenamefont {Saffin}}]{Charmousis:2009tc}%
  \BibitemOpen
  \bibfield  {author} {\bibinfo {author} {\bibfnamefont {C.}~\bibnamefont
  {Charmousis}}, \bibinfo {author} {\bibfnamefont {G.}~\bibnamefont {Niz}},
  \bibinfo {author} {\bibfnamefont {A.}~\bibnamefont {Padilla}},\ and\ \bibinfo
  {author} {\bibfnamefont {P.~M.}\ \bibnamefont {Saffin}},\ }\bibfield  {title}
  {\bibinfo {title} {{Strong coupling in Horava gravity}},\ }\href
  {https://doi.org/10.1088/1126-6708/2009/08/070} {\bibfield  {journal}
  {\bibinfo  {journal} {JHEP}\ }\textbf {\bibinfo {volume} {08}},\ \bibinfo
  {pages} {070}},\ \Eprint {https://arxiv.org/abs/0905.2579} {arXiv:0905.2579
  [hep-th]} \BibitemShut {NoStop}%
\bibitem [{\citenamefont {Papazoglou}\ and\ \citenamefont
  {Sotiriou}(2010)}]{Papazoglou:2009fj}%
  \BibitemOpen
  \bibfield  {author} {\bibinfo {author} {\bibfnamefont {A.}~\bibnamefont
  {Papazoglou}}\ and\ \bibinfo {author} {\bibfnamefont {T.~P.}\ \bibnamefont
  {Sotiriou}},\ }\bibfield  {title} {\bibinfo {title} {{Strong coupling in
  extended Horava-Lifshitz gravity}},\ }\href
  {https://doi.org/10.1016/j.physletb.2010.01.054} {\bibfield  {journal}
  {\bibinfo  {journal} {Phys. Lett. B}\ }\textbf {\bibinfo {volume} {685}},\
  \bibinfo {pages} {197} (\bibinfo {year} {2010})},\ \Eprint
  {https://arxiv.org/abs/0911.1299} {arXiv:0911.1299 [hep-th]} \BibitemShut
  {NoStop}%
\bibitem [{\citenamefont {Baumann}\ \emph {et~al.}(2011)\citenamefont
  {Baumann}, \citenamefont {Senatore},\ and\ \citenamefont
  {Zaldarriaga}}]{Baumann:2011dt}%
  \BibitemOpen
  \bibfield  {author} {\bibinfo {author} {\bibfnamefont {D.}~\bibnamefont
  {Baumann}}, \bibinfo {author} {\bibfnamefont {L.}~\bibnamefont {Senatore}},\
  and\ \bibinfo {author} {\bibfnamefont {M.}~\bibnamefont {Zaldarriaga}},\
  }\bibfield  {title} {\bibinfo {title} {{Scale-Invariance and the Strong
  Coupling Problem}},\ }\href {https://doi.org/10.1088/1475-7516/2011/05/004}
  {\bibfield  {journal} {\bibinfo  {journal} {JCAP}\ }\textbf {\bibinfo
  {volume} {05}},\ \bibinfo {pages} {004}},\ \Eprint
  {https://arxiv.org/abs/1101.3320} {arXiv:1101.3320 [hep-th]} \BibitemShut
  {NoStop}%
\bibitem [{\citenamefont {D'Amico}\ \emph {et~al.}(2011)\citenamefont
  {D'Amico}, \citenamefont {de~Rham}, \citenamefont {Dubovsky}, \citenamefont
  {Gabadadze}, \citenamefont {Pirtskhalava},\ and\ \citenamefont
  {Tolley}}]{DAmico:2011eto}%
  \BibitemOpen
  \bibfield  {author} {\bibinfo {author} {\bibfnamefont {G.}~\bibnamefont
  {D'Amico}}, \bibinfo {author} {\bibfnamefont {C.}~\bibnamefont {de~Rham}},
  \bibinfo {author} {\bibfnamefont {S.}~\bibnamefont {Dubovsky}}, \bibinfo
  {author} {\bibfnamefont {G.}~\bibnamefont {Gabadadze}}, \bibinfo {author}
  {\bibfnamefont {D.}~\bibnamefont {Pirtskhalava}},\ and\ \bibinfo {author}
  {\bibfnamefont {A.~J.}\ \bibnamefont {Tolley}},\ }\bibfield  {title}
  {\bibinfo {title} {{Massive Cosmologies}},\ }\href
  {https://doi.org/10.1103/PhysRevD.84.124046} {\bibfield  {journal} {\bibinfo
  {journal} {Phys. Rev. D}\ }\textbf {\bibinfo {volume} {84}},\ \bibinfo
  {pages} {124046} (\bibinfo {year} {2011})},\ \Eprint
  {https://arxiv.org/abs/1108.5231} {arXiv:1108.5231 [hep-th]} \BibitemShut
  {NoStop}%
\bibitem [{\citenamefont {Gumrukcuoglu}\ \emph {et~al.}(2012)\citenamefont
  {Gumrukcuoglu}, \citenamefont {Lin},\ and\ \citenamefont
  {Mukohyama}}]{Gumrukcuoglu:2012aa}%
  \BibitemOpen
  \bibfield  {author} {\bibinfo {author} {\bibfnamefont {A.~E.}\ \bibnamefont
  {Gumrukcuoglu}}, \bibinfo {author} {\bibfnamefont {C.}~\bibnamefont {Lin}},\
  and\ \bibinfo {author} {\bibfnamefont {S.}~\bibnamefont {Mukohyama}},\
  }\bibfield  {title} {\bibinfo {title} {{Anisotropic
  Friedmann-Robertson-Walker universe from nonlinear massive gravity}},\ }\href
  {https://doi.org/10.1016/j.physletb.2012.09.049} {\bibfield  {journal}
  {\bibinfo  {journal} {Phys. Lett. B}\ }\textbf {\bibinfo {volume} {717}},\
  \bibinfo {pages} {295} (\bibinfo {year} {2012})},\ \Eprint
  {https://arxiv.org/abs/1206.2723} {arXiv:1206.2723 [hep-th]} \BibitemShut
  {NoStop}%
\bibitem [{\citenamefont {Wang}(2017)}]{Wang:2017brl}%
  \BibitemOpen
  \bibfield  {author} {\bibinfo {author} {\bibfnamefont {A.}~\bibnamefont
  {Wang}},\ }\bibfield  {title} {\bibinfo {title} {{Ho\v{r}ava gravity at a
  Lifshitz point: A progress report}},\ }\href
  {https://doi.org/10.1142/S0218271817300142} {\bibfield  {journal} {\bibinfo
  {journal} {Int. J. Mod. Phys. D}\ }\textbf {\bibinfo {volume} {26}},\
  \bibinfo {pages} {1730014} (\bibinfo {year} {2017})},\ \Eprint
  {https://arxiv.org/abs/1701.06087} {arXiv:1701.06087 [gr-qc]} \BibitemShut
  {NoStop}%
\bibitem [{\citenamefont {Mazuet}\ \emph {et~al.}(2017)\citenamefont {Mazuet},
  \citenamefont {Mukohyama},\ and\ \citenamefont {Volkov}}]{Mazuet:2017rgq}%
  \BibitemOpen
  \bibfield  {author} {\bibinfo {author} {\bibfnamefont {C.}~\bibnamefont
  {Mazuet}}, \bibinfo {author} {\bibfnamefont {S.}~\bibnamefont {Mukohyama}},\
  and\ \bibinfo {author} {\bibfnamefont {M.~S.}\ \bibnamefont {Volkov}},\
  }\bibfield  {title} {\bibinfo {title} {{Anisotropic deformations of spatially
  open cosmology in massive gravity theory}},\ }\href
  {https://doi.org/10.1088/1475-7516/2017/04/039} {\bibfield  {journal}
  {\bibinfo  {journal} {JCAP}\ }\textbf {\bibinfo {volume} {04}},\ \bibinfo
  {pages} {039}},\ \Eprint {https://arxiv.org/abs/1702.04205} {arXiv:1702.04205
  [hep-th]} \BibitemShut {NoStop}%
\bibitem [{\citenamefont {Beltr\'an~Jim\'enez}\ and\ \citenamefont
  {Jim\'enez-Cano}(2021)}]{BeltranJimenez:2020lee}%
  \BibitemOpen
  \bibfield  {author} {\bibinfo {author} {\bibfnamefont {J.}~\bibnamefont
  {Beltr\'an~Jim\'enez}}\ and\ \bibinfo {author} {\bibfnamefont
  {A.}~\bibnamefont {Jim\'enez-Cano}},\ }\bibfield  {title} {\bibinfo {title}
  {{On the strong coupling of Einsteinian Cubic Gravity and its
  generalisations}},\ }\href {https://doi.org/10.1088/1475-7516/2021/01/069}
  {\bibfield  {journal} {\bibinfo  {journal} {JCAP}\ }\textbf {\bibinfo
  {volume} {01}},\ \bibinfo {pages} {069}},\ \Eprint
  {https://arxiv.org/abs/2009.08197} {arXiv:2009.08197 [gr-qc]} \BibitemShut
  {NoStop}%
\bibitem [{\citenamefont {Barker}(2023{\natexlab{a}})}]{Barker:2022kdk}%
  \BibitemOpen
  \bibfield  {author} {\bibinfo {author} {\bibfnamefont {W.~E.~V.}\
  \bibnamefont {Barker}},\ }\bibfield  {title} {\bibinfo {title}
  {{Supercomputers against strong coupling in gravity with curvature and
  torsion}},\ }\href {https://doi.org/10.1140/epjc/s10052-023-11179-6}
  {\bibfield  {journal} {\bibinfo  {journal} {Eur. Phys. J. C}\ }\textbf
  {\bibinfo {volume} {83}},\ \bibinfo {pages} {228} (\bibinfo {year}
  {2023}{\natexlab{a}})},\ \Eprint {https://arxiv.org/abs/2206.00658}
  {arXiv:2206.00658 [gr-qc]} \BibitemShut {NoStop}%
\bibitem [{\citenamefont {Delhom}\ \emph {et~al.}(2022)\citenamefont {Delhom},
  \citenamefont {Jim\'enez-Cano},\ and\ \citenamefont
  {Maldonado~Torralba}}]{Delhom:2022vae}%
  \BibitemOpen
  \bibfield  {author} {\bibinfo {author} {\bibfnamefont {A.}~\bibnamefont
  {Delhom}}, \bibinfo {author} {\bibfnamefont {A.}~\bibnamefont
  {Jim\'enez-Cano}},\ and\ \bibinfo {author} {\bibfnamefont {F.~J.}\
  \bibnamefont {Maldonado~Torralba}},\ }\bibfield  {title} {\bibinfo {title}
  {{Instabilities in Field Theory: A Primer with Applications in Modified
  Gravity}}\ }(\bibinfo  {publisher} {Springer},\ \bibinfo {year} {2022})\
  \Eprint {https://arxiv.org/abs/2207.13431} {arXiv:2207.13431 [gr-qc]}
  \BibitemShut {NoStop}%
\bibitem [{\citenamefont {Barker}\ and\ \citenamefont
  {Zell}(2024{\natexlab{b}})}]{Barker:2023fem}%
  \BibitemOpen
  \bibfield  {author} {\bibinfo {author} {\bibfnamefont {W.}~\bibnamefont
  {Barker}}\ and\ \bibinfo {author} {\bibfnamefont {S.}~\bibnamefont {Zell}},\
  }\bibfield  {title} {\bibinfo {title} {{Einstein-Proca theory from the
  Einstein-Cartan formulation}},\ }\href
  {https://doi.org/10.1103/PhysRevD.109.024007} {\bibfield  {journal} {\bibinfo
   {journal} {Phys. Rev. D}\ }\textbf {\bibinfo {volume} {109}},\ \bibinfo
  {pages} {024007} (\bibinfo {year} {2024}{\natexlab{b}})},\ \Eprint
  {https://arxiv.org/abs/2306.14953} {arXiv:2306.14953 [hep-th]} \BibitemShut
  {NoStop}%
\bibitem [{\citenamefont {Iosifidis}\ and\ \citenamefont
  {Pallikaris}(2024)}]{iosifidis2024magverse}%
  \BibitemOpen
  \bibfield  {author} {\bibinfo {author} {\bibfnamefont {D.}~\bibnamefont
  {Iosifidis}}\ and\ \bibinfo {author} {\bibfnamefont {K.}~\bibnamefont
  {Pallikaris}},\ }\href@noop {} {\bibinfo {title} {Into the mag-verse or:
  Cosmology of the complete quadratic metric-affine gravity}} (\bibinfo {year}
  {2024}),\ \Eprint {https://arxiv.org/abs/2404.19498} {arXiv:2404.19498
  [gr-qc]} \BibitemShut {NoStop}%
\bibitem [{\citenamefont {Shimada}\ \emph {et~al.}(2019)\citenamefont
  {Shimada}, \citenamefont {Aoki},\ and\ \citenamefont
  {Maeda}}]{Shimada:2018lnm}%
  \BibitemOpen
  \bibfield  {author} {\bibinfo {author} {\bibfnamefont {K.}~\bibnamefont
  {Shimada}}, \bibinfo {author} {\bibfnamefont {K.}~\bibnamefont {Aoki}},\ and\
  \bibinfo {author} {\bibfnamefont {K.-i.}\ \bibnamefont {Maeda}},\ }\bibfield
  {title} {\bibinfo {title} {{Metric-affine Gravity and Inflation}},\ }\href
  {https://doi.org/10.1103/PhysRevD.99.104020} {\bibfield  {journal} {\bibinfo
  {journal} {Phys. Rev. D}\ }\textbf {\bibinfo {volume} {99}},\ \bibinfo
  {pages} {104020} (\bibinfo {year} {2019})},\ \Eprint
  {https://arxiv.org/abs/1812.03420} {arXiv:1812.03420 [gr-qc]} \BibitemShut
  {NoStop}%
\bibitem [{\citenamefont {Gialamas}\ and\ \citenamefont
  {Tamvakis}(2023)}]{Gialamas:2022xtt}%
  \BibitemOpen
  \bibfield  {author} {\bibinfo {author} {\bibfnamefont {I.~D.}\ \bibnamefont
  {Gialamas}}\ and\ \bibinfo {author} {\bibfnamefont {K.}~\bibnamefont
  {Tamvakis}},\ }\bibfield  {title} {\bibinfo {title} {{Inflation in
  metric-affine quadratic gravity}},\ }\href
  {https://doi.org/10.1088/1475-7516/2023/03/042} {\bibfield  {journal}
  {\bibinfo  {journal} {JCAP}\ }\textbf {\bibinfo {volume} {03}},\ \bibinfo
  {pages} {042}},\ \Eprint {https://arxiv.org/abs/2212.09896} {arXiv:2212.09896
  [gr-qc]} \BibitemShut {NoStop}%
\bibitem [{\citenamefont {Minkevich}\ and\ \citenamefont
  {Garkun}(1998)}]{minkevich1998isotropic}%
  \BibitemOpen
  \bibfield  {author} {\bibinfo {author} {\bibfnamefont {A.~V.}\ \bibnamefont
  {Minkevich}}\ and\ \bibinfo {author} {\bibfnamefont {A.~S.}\ \bibnamefont
  {Garkun}},\ }\href@noop {} {\bibinfo {title} {Isotropic cosmology in
  metric-affine gauge theory of gravity}} (\bibinfo {year} {1998}),\ \Eprint
  {https://arxiv.org/abs/gr-qc/9805007} {arXiv:gr-qc/9805007 [gr-qc]}
  \BibitemShut {NoStop}%
\bibitem [{\citenamefont {Obukhov}\ \emph {et~al.}(1997)\citenamefont
  {Obukhov}, \citenamefont {Vlachynsky}, \citenamefont {Esser},\ and\
  \citenamefont {Hehl}}]{obukhovQuad}%
  \BibitemOpen
  \bibfield  {author} {\bibinfo {author} {\bibfnamefont {Y.~N.}\ \bibnamefont
  {Obukhov}}, \bibinfo {author} {\bibfnamefont {E.~J.}\ \bibnamefont
  {Vlachynsky}}, \bibinfo {author} {\bibfnamefont {W.}~\bibnamefont {Esser}},\
  and\ \bibinfo {author} {\bibfnamefont {F.~W.}\ \bibnamefont {Hehl}},\
  }\bibfield  {title} {\bibinfo {title} {Irreducible decompositions in
  metric-affine gravity models},\ }\href@noop {} {\  (\bibinfo {year}
  {1997})},\ \Eprint {https://arxiv.org/abs/gr-qc/9705039} {arXiv:gr-qc/9705039
  [gr-qc]} \BibitemShut {NoStop}%
\bibitem [{\citenamefont {Iosifidis}\ and\ \citenamefont
  {Ravera}(2022)}]{Iosifidis_2022}%
  \BibitemOpen
  \bibfield  {author} {\bibinfo {author} {\bibfnamefont {D.}~\bibnamefont
  {Iosifidis}}\ and\ \bibinfo {author} {\bibfnamefont {L.}~\bibnamefont
  {Ravera}},\ }\bibfield  {title} {\bibinfo {title} {Cosmology of quadratic
  metric-affine gravity},\ }\bibfield  {journal} {\bibinfo  {journal} {Physical
  Review D}\ }\textbf {\bibinfo {volume} {105}},\ \href
  {https://doi.org/10.1103/physrevd.105.024007} {10.1103/physrevd.105.024007}
  (\bibinfo {year} {2022})\BibitemShut {NoStop}%
\bibitem [{\citenamefont {Gonner}\ and\ \citenamefont
  {Mueller-Hoissen}(1984)}]{Gonner:1984rw}%
  \BibitemOpen
  \bibfield  {author} {\bibinfo {author} {\bibfnamefont {H.}~\bibnamefont
  {Gonner}}\ and\ \bibinfo {author} {\bibfnamefont {F.}~\bibnamefont
  {Mueller-Hoissen}},\ }\bibfield  {title} {\bibinfo {title} {{Spatially
  Homogeneous and Isotropic Spaces in Theories of Gravitation With Torsion}},\
  }\href {https://doi.org/10.1088/0264-9381/1/6/010} {\bibfield  {journal}
  {\bibinfo  {journal} {Class. Quant. Grav.}\ }\textbf {\bibinfo {volume}
  {1}},\ \bibinfo {pages} {651} (\bibinfo {year} {1984})}\BibitemShut {NoStop}%
\bibitem [{\citenamefont {Iosifidis}(2022)}]{IosifidisQuad}%
  \BibitemOpen
  \bibfield  {author} {\bibinfo {author} {\bibfnamefont {D.}~\bibnamefont
  {Iosifidis}},\ }\bibfield  {title} {\bibinfo {title} {The full quadratic
  metric-affine gravity (including parity odd terms): exact solutions for the
  affine-connection},\ }\href {https://doi.org/10.1088/1361-6382/ac6058}
  {\bibfield  {journal} {\bibinfo  {journal} {Class. Quant. Grav.}\ }\textbf
  {\bibinfo {volume} {39}},\ \bibinfo {pages} {095002} (\bibinfo {year}
  {2022})},\ \Eprint {https://arxiv.org/abs/2112.09154} {arXiv:2112.09154
  [gr-qc]} \BibitemShut {NoStop}%
\bibitem [{\citenamefont {Heisenberg}\ and\ \citenamefont
  {Hohmann}(2023)}]{HeisenbergHohmann2023}%
  \BibitemOpen
  \bibfield  {author} {\bibinfo {author} {\bibfnamefont {L.}~\bibnamefont
  {Heisenberg}}\ and\ \bibinfo {author} {\bibfnamefont {M.}~\bibnamefont
  {Hohmann}},\ }\bibfield  {title} {\bibinfo {title} {Gauge-invariant
  cosmological perturbations in general teleparallel gravity},\ }\href
  {https://doi.org/10.1088/1361-6382/acad48} {\bibfield  {journal} {\bibinfo
  {journal} {Class. Quantum Grav.}\ }\textbf {\bibinfo {volume} {41}},\
  \bibinfo {pages} {015025} (\bibinfo {year} {2023})},\ \Eprint
  {https://arxiv.org/abs/2311.05597} {arXiv:2311.05597 [gr-qc]} \BibitemShut
  {NoStop}%
\bibitem [{\citenamefont {Castillo-Felisola}\ \emph {et~al.}(2024)\citenamefont
  {Castillo-Felisola}, \citenamefont {Gannouji}, \citenamefont
  {Morocho-L\'opez},\ and\ \citenamefont {Rozas-Rojas}}]{CastilloFelisola2024}%
  \BibitemOpen
  \bibfield  {author} {\bibinfo {author} {\bibfnamefont {O.}~\bibnamefont
  {Castillo-Felisola}}, \bibinfo {author} {\bibfnamefont {R.}~\bibnamefont
  {Gannouji}}, \bibinfo {author} {\bibfnamefont {M.}~\bibnamefont
  {Morocho-L\'opez}},\ and\ \bibinfo {author} {\bibfnamefont {M.}~\bibnamefont
  {Rozas-Rojas}},\ }\bibfield  {title} {\bibinfo {title} {A model for
  cosmological perturbations in affine gravity},\ }\href@noop {} {\bibfield
  {journal} {\bibinfo  {journal} {arXiv e-prints}\ } (\bibinfo {year}
  {2024})},\ \Eprint {https://arxiv.org/abs/2401.08198} {arXiv:2401.08198
  [gr-qc]} \BibitemShut {NoStop}%
\bibitem [{\citenamefont {Aoki}\ \emph {et~al.}(2024)\citenamefont {Aoki},
  \citenamefont {Bahamonde}, \citenamefont {Gigante~Valcarcel},\ and\
  \citenamefont {Gorji}}]{Aoki2024}%
  \BibitemOpen
  \bibfield  {author} {\bibinfo {author} {\bibfnamefont {K.}~\bibnamefont
  {Aoki}}, \bibinfo {author} {\bibfnamefont {S.}~\bibnamefont {Bahamonde}},
  \bibinfo {author} {\bibfnamefont {J.}~\bibnamefont {Gigante~Valcarcel}},\
  and\ \bibinfo {author} {\bibfnamefont {M.~A.}\ \bibnamefont {Gorji}},\
  }\bibfield  {title} {\bibinfo {title} {Cosmological perturbation theory in
  metric-affine gravity},\ }\href {https://doi.org/10.1103/PhysRevD.110.024017}
  {\bibfield  {journal} {\bibinfo  {journal} {Phys. Rev. D}\ }\textbf {\bibinfo
  {volume} {110}},\ \bibinfo {pages} {024017} (\bibinfo {year} {2024})},\
  \Eprint {https://arxiv.org/abs/2310.16007} {arXiv:2310.16007 [gr-qc]}
  \BibitemShut {NoStop}%
\bibitem [{\citenamefont {Tsamparlis}(1979)}]{Tsamparlis_1979}%
  \BibitemOpen
  \bibfield  {author} {\bibinfo {author} {\bibfnamefont {M.}~\bibnamefont
  {Tsamparlis}},\ }\bibfield  {title} {\bibinfo {title} {Cosmological principle
  and torsion},\ }\href
  {https://doi.org/https://doi.org/10.1016/0375-9601(79)90265-2} {\bibfield
  {journal} {\bibinfo  {journal} {Physics Letters A}\ }\textbf {\bibinfo
  {volume} {75}},\ \bibinfo {pages} {27} (\bibinfo {year} {1979})}\BibitemShut
  {NoStop}%
\bibitem [{\citenamefont {Iosifidis}(2020{\natexlab{b}})}]{Iosifidis:2019fsh}%
  \BibitemOpen
  \bibfield  {author} {\bibinfo {author} {\bibfnamefont {D.}~\bibnamefont
  {Iosifidis}},\ }\bibfield  {title} {\bibinfo {title} {{Linear Transformations
  on Affine-Connections}},\ }\href {https://doi.org/10.1088/1361-6382/ab778d}
  {\bibfield  {journal} {\bibinfo  {journal} {Class. Quant. Grav.}\ }\textbf
  {\bibinfo {volume} {37}},\ \bibinfo {pages} {085010} (\bibinfo {year}
  {2020}{\natexlab{b}})},\ \Eprint {https://arxiv.org/abs/1911.04535}
  {arXiv:1911.04535 [gr-qc]} \BibitemShut {NoStop}%
\bibitem [{\citenamefont {Remmen}\ and\ \citenamefont
  {Carroll}(2013)}]{Remmen:2013eja}%
  \BibitemOpen
  \bibfield  {author} {\bibinfo {author} {\bibfnamefont {G.~N.}\ \bibnamefont
  {Remmen}}\ and\ \bibinfo {author} {\bibfnamefont {S.~M.}\ \bibnamefont
  {Carroll}},\ }\bibfield  {title} {\bibinfo {title} {{Attractor Solutions in
  Scalar-Field Cosmology}},\ }\href
  {https://doi.org/10.1103/PhysRevD.88.083518} {\bibfield  {journal} {\bibinfo
  {journal} {Phys. Rev. D}\ }\textbf {\bibinfo {volume} {88}},\ \bibinfo
  {pages} {083518} (\bibinfo {year} {2013})},\ \Eprint
  {https://arxiv.org/abs/1309.2611} {arXiv:1309.2611 [gr-qc]} \BibitemShut
  {NoStop}%
\bibitem [{\citenamefont {Achour}(2021)}]{Achour:2021lqq}%
  \BibitemOpen
  \bibfield  {author} {\bibinfo {author} {\bibfnamefont {J.~B.}\ \bibnamefont
  {Achour}},\ }\bibfield  {title} {\bibinfo {title} {{Proper time
  reparametrization in cosmology: M\"obius symmetry and Kodama charges}},\
  }\href {https://doi.org/10.1088/1475-7516/2021/12/005} {\bibfield  {journal}
  {\bibinfo  {journal} {JCAP}\ }\textbf {\bibinfo {volume} {12}}\bibfield
  {number} {\bibinfo  {number} { (12)},\ \bibinfo {pages} {005}},\ }\Eprint
  {https://arxiv.org/abs/2103.10700} {arXiv:2103.10700 [gr-qc]} \BibitemShut
  {NoStop}%
\bibitem [{\citenamefont {Deser}\ and\ \citenamefont
  {Tekin}(2003)}]{Deser:2003up}%
  \BibitemOpen
  \bibfield  {author} {\bibinfo {author} {\bibfnamefont {S.}~\bibnamefont
  {Deser}}\ and\ \bibinfo {author} {\bibfnamefont {B.}~\bibnamefont {Tekin}},\
  }\bibfield  {title} {\bibinfo {title} {{Shortcuts to high symmetry solutions
  in gravitational theories}},\ }\href
  {https://doi.org/10.1088/0264-9381/20/22/011} {\bibfield  {journal} {\bibinfo
   {journal} {Class. Quant. Grav.}\ }\textbf {\bibinfo {volume} {20}},\
  \bibinfo {pages} {4877} (\bibinfo {year} {2003})},\ \Eprint
  {https://arxiv.org/abs/gr-qc/0306114} {arXiv:gr-qc/0306114} \BibitemShut
  {NoStop}%
\bibitem [{\citenamefont {Maccallum}\ and\ \citenamefont
  {Taub}(1972)}]{Maccallum:1972er}%
  \BibitemOpen
  \bibfield  {author} {\bibinfo {author} {\bibfnamefont {M.~A.~H.}\
  \bibnamefont {Maccallum}}\ and\ \bibinfo {author} {\bibfnamefont {A.~H.}\
  \bibnamefont {Taub}},\ }\bibfield  {title} {\bibinfo {title} {{Variational
  principles and spatially-homogeneous universes, including rotation}},\ }\href
  {https://doi.org/10.1007/BF01877589} {\bibfield  {journal} {\bibinfo
  {journal} {Commun. Math. Phys.}\ }\textbf {\bibinfo {volume} {25}},\ \bibinfo
  {pages} {173} (\bibinfo {year} {1972})}\BibitemShut {NoStop}%
\bibitem [{\citenamefont {Fels}\ and\ \citenamefont {Torre}(2001)}]{Fels:2001}%
  \BibitemOpen
  \bibfield  {author} {\bibinfo {author} {\bibfnamefont {M.}~\bibnamefont
  {Fels}}\ and\ \bibinfo {author} {\bibfnamefont {C.}~\bibnamefont {Torre}},\
  }\bibfield  {title} {\bibinfo {title} {The principle of symmetric criticality
  in general relativity},\ }\href {https://doi.org/10.1088/0264-9381/19/4/303}
  {\bibfield  {journal} {\bibinfo  {journal} {Classical and Quantum Gravity}\
  }\textbf {\bibinfo {volume} {19}} (\bibinfo {year} {2001})}\BibitemShut
  {NoStop}%
\bibitem [{\citenamefont {Hawking}(1969)}]{Hawking:1969}%
  \BibitemOpen
  \bibfield  {author} {\bibinfo {author} {\bibfnamefont {S.}~\bibnamefont
  {Hawking}},\ }\bibfield  {title} {\bibinfo {title} {{On the Rotation of the
  Universe}},\ }\href {https://doi.org/10.1093/mnras/142.2.129} {\bibfield
  {journal} {\bibinfo  {journal} {Monthly Notices of the Royal Astronomical
  Society}\ }\textbf {\bibinfo {volume} {142}},\ \bibinfo {pages} {129}
  (\bibinfo {year} {1969})}\BibitemShut {NoStop}%
\bibitem [{\citenamefont {Cox}\ \emph {et~al.}(1992)\citenamefont {Cox},
  \citenamefont {Little},\ and\ \citenamefont {O'Shea}}]{CommAlg}%
  \BibitemOpen
  \bibfield  {author} {\bibinfo {author} {\bibfnamefont {D.~A.}\ \bibnamefont
  {Cox}}, \bibinfo {author} {\bibfnamefont {J.~B.}\ \bibnamefont {Little}},\
  and\ \bibinfo {author} {\bibfnamefont {D.}~\bibnamefont {O'Shea}},\
  }\href@noop {} {\emph {\bibinfo {title} {Ideals, varieties, and algorithms :
  an introduction to computational algebraic geometry and commutative
  algebra}}},\ \bibinfo {edition} {1st}\ ed.,\ Undergraduate texts in
  mathematics\ (\bibinfo  {publisher} {Springer New York},\ \bibinfo {year}
  {1992})\BibitemShut {NoStop}%
\bibitem [{\citenamefont {Grayson}\ and\ \citenamefont {Stillman}()}]{M2}%
  \BibitemOpen
  \bibfield  {author} {\bibinfo {author} {\bibfnamefont {D.~R.}\ \bibnamefont
  {Grayson}}\ and\ \bibinfo {author} {\bibfnamefont {M.~E.}\ \bibnamefont
  {Stillman}},\ }\href {http://www2.macaulay2.com} {\bibinfo {title}
  {Macaulay2, a software system for research in algebraic geometry}},\ \bibinfo
  {howpublished} {Available at \url{http://www2.macaulay2.com}}\BibitemShut
  {NoStop}%
\bibitem [{\citenamefont {Kolekar}(2022)}]{Kolekar:2022vkp}%
  \BibitemOpen
  \bibfield  {author} {\bibinfo {author} {\bibfnamefont {S.}~\bibnamefont
  {Kolekar}},\ }\bibfield  {title} {\bibinfo {title} {{On the Bianchi identity
  in generalized theories of gravity}},\ }\href
  {https://doi.org/10.1007/s10714-022-02978-5} {\bibfield  {journal} {\bibinfo
  {journal} {Gen. Rel. Grav.}\ }\textbf {\bibinfo {volume} {54}},\ \bibinfo
  {pages} {92} (\bibinfo {year} {2022})}\BibitemShut {NoStop}%
\bibitem [{\citenamefont {Iosifidis}\ and\ \citenamefont
  {Koivisto}(2019)}]{Iosifidis:2018zwo}%
  \BibitemOpen
  \bibfield  {author} {\bibinfo {author} {\bibfnamefont {D.}~\bibnamefont
  {Iosifidis}}\ and\ \bibinfo {author} {\bibfnamefont {T.}~\bibnamefont
  {Koivisto}},\ }\bibfield  {title} {\bibinfo {title} {{Scale transformations
  in metric-affine geometry}},\ }\href
  {https://doi.org/10.3390/universe5030082} {\bibfield  {journal} {\bibinfo
  {journal} {Universe}\ }\textbf {\bibinfo {volume} {5}},\ \bibinfo {pages}
  {82} (\bibinfo {year} {2019})},\ \Eprint {https://arxiv.org/abs/1810.12276}
  {arXiv:1810.12276 [gr-qc]} \BibitemShut {NoStop}%
\bibitem [{\citenamefont {Koivisto}\ \emph {et~al.}(2019)\citenamefont
  {Koivisto}, \citenamefont {Hohmann},\ and\ \citenamefont
  {Z\l{}o\'snik}}]{Koivisto:2019ejt}%
  \BibitemOpen
  \bibfield  {author} {\bibinfo {author} {\bibfnamefont {T.}~\bibnamefont
  {Koivisto}}, \bibinfo {author} {\bibfnamefont {M.}~\bibnamefont {Hohmann}},\
  and\ \bibinfo {author} {\bibfnamefont {T.}~\bibnamefont {Z\l{}o\'snik}},\
  }\bibfield  {title} {\bibinfo {title} {{The General Linear Cartan Khronon}},\
  }\href {https://doi.org/10.3390/universe5070168} {\bibfield  {journal}
  {\bibinfo  {journal} {Universe}\ }\textbf {\bibinfo {volume} {5}},\ \bibinfo
  {pages} {168} (\bibinfo {year} {2019})},\ \Eprint
  {https://arxiv.org/abs/1905.02967} {arXiv:1905.02967 [gr-qc]} \BibitemShut
  {NoStop}%
\bibitem [{\citenamefont {{Barker}}()}]{PSALTer}%
  \BibitemOpen
  \bibfield  {author} {\bibinfo {author} {\bibfnamefont {W.}~\bibnamefont
  {{Barker}}},\ }\href@noop {} {\emph {\bibinfo {title} {{Particle Spectrum for
  Any Tensor Lagrangian (PSALTer)}}}},\ \bibinfo {note}
  {download:~\href{https://github.com/wevbarker/PSALTer}{github.com/wevbarker/PSALTer}}\BibitemShut
  {NoStop}%
\bibitem [{\citenamefont {{Dyer}}\ \emph {et~al.}()\citenamefont {{Dyer}},
  \citenamefont {{Iosifidis}},\ and\ \citenamefont {{Barker}}}]{Supplement}%
  \BibitemOpen
  \bibfield  {author} {\bibinfo {author} {\bibfnamefont {T.}~\bibnamefont
  {{Dyer}}}, \bibinfo {author} {\bibfnamefont {D.}~\bibnamefont
  {{Iosifidis}}},\ and\ \bibinfo {author} {\bibfnamefont {W.}~\bibnamefont
  {{Barker}}},\ }\href@noop {} {\emph {\bibinfo {title} {{Supplemental
  materials at
  \href{https://github.com/wevbarker/SupplementalMaterials-2412}{www.github.com/wevbarker/SupplementalMaterials-2412}.}}}}\BibitemShut
  {Stop}%
\bibitem [{\citenamefont {Barker}\ \emph
  {et~al.}(2020{\natexlab{b}})\citenamefont {Barker}, \citenamefont {Lasenby},
  \citenamefont {Hobson},\ and\ \citenamefont {Handley}}]{Barker:2020elg}%
  \BibitemOpen
  \bibfield  {author} {\bibinfo {author} {\bibfnamefont {W.~E.~V.}\
  \bibnamefont {Barker}}, \bibinfo {author} {\bibfnamefont {A.~N.}\
  \bibnamefont {Lasenby}}, \bibinfo {author} {\bibfnamefont {M.~P.}\
  \bibnamefont {Hobson}},\ and\ \bibinfo {author} {\bibfnamefont {W.~J.}\
  \bibnamefont {Handley}},\ }\bibfield  {title} {\bibinfo {title} {{Mapping
  Poincar\'e gauge cosmology to Horndeski theory for emergent dark energy}},\
  }\href {https://doi.org/10.1103/PhysRevD.102.084002} {\bibfield  {journal}
  {\bibinfo  {journal} {Phys. Rev. D}\ }\textbf {\bibinfo {volume} {102}},\
  \bibinfo {pages} {084002} (\bibinfo {year} {2020}{\natexlab{b}})},\ \Eprint
  {https://arxiv.org/abs/2006.03581} {arXiv:2006.03581 [gr-qc]} \BibitemShut
  {NoStop}%
\bibitem [{\citenamefont {Rew}\ and\ \citenamefont
  {Barker}(2023)}]{Rew:2023zxy}%
  \BibitemOpen
  \bibfield  {author} {\bibinfo {author} {\bibfnamefont {C.}~\bibnamefont
  {Rew}}\ and\ \bibinfo {author} {\bibfnamefont {W.~E.~V.}\ \bibnamefont
  {Barker}},\ }\bibfield  {title} {\bibinfo {title} {{The effective
  inflationary potential of constant-torsion emergent gravity}},\ }\href@noop
  {} {\  (\bibinfo {year} {2023})},\ \Eprint {https://arxiv.org/abs/2302.07250}
  {arXiv:2302.07250 [gr-qc]} \BibitemShut {NoStop}%
\bibitem [{\citenamefont {Barker}(2023{\natexlab{b}})}]{Barker:2023bmr}%
  \BibitemOpen
  \bibfield  {author} {\bibinfo {author} {\bibfnamefont {W.}~\bibnamefont
  {Barker}},\ }\bibfield  {title} {\bibinfo {title} {{Particle spectra of
  gravity based on internal symmetry of quantum fields}},\ }\href@noop {} {\
  (\bibinfo {year} {2023}{\natexlab{b}})},\ \Eprint
  {https://arxiv.org/abs/2311.11790} {arXiv:2311.11790 [hep-th]} \BibitemShut
  {NoStop}%
\bibitem [{\citenamefont {Barker}\ \emph
  {et~al.}(2024{\natexlab{a}})\citenamefont {Barker}, \citenamefont {Marzo},\
  and\ \citenamefont {Rigouzzo}}]{Barker:2024juc}%
  \BibitemOpen
  \bibfield  {author} {\bibinfo {author} {\bibfnamefont {W.}~\bibnamefont
  {Barker}}, \bibinfo {author} {\bibfnamefont {C.}~\bibnamefont {Marzo}},\ and\
  \bibinfo {author} {\bibfnamefont {C.}~\bibnamefont {Rigouzzo}},\ }\bibfield
  {title} {\bibinfo {title} {{PSALTer: Particle Spectrum for Any Tensor
  Lagrangian}},\ }\href@noop {} {\  (\bibinfo {year} {2024}{\natexlab{a}})},\
  \Eprint {https://arxiv.org/abs/2406.09500} {arXiv:2406.09500 [hep-th]}
  \BibitemShut {NoStop}%
\bibitem [{\citenamefont {Barker}\ \emph
  {et~al.}(2024{\natexlab{b}})\citenamefont {Barker}, \citenamefont {Hobson},
  \citenamefont {Lasenby}, \citenamefont {Lin},\ and\ \citenamefont
  {Wei}}]{Barker:2024goa}%
  \BibitemOpen
  \bibfield  {author} {\bibinfo {author} {\bibfnamefont {W.}~\bibnamefont
  {Barker}}, \bibinfo {author} {\bibfnamefont {M.}~\bibnamefont {Hobson}},
  \bibinfo {author} {\bibfnamefont {A.}~\bibnamefont {Lasenby}}, \bibinfo
  {author} {\bibfnamefont {Y.-C.}\ \bibnamefont {Lin}},\ and\ \bibinfo {author}
  {\bibfnamefont {Z.}~\bibnamefont {Wei}},\ }\bibfield  {title} {\bibinfo
  {title} {{Every Poincar\'e gauge theory is conformal: a compelling case for
  dynamical vector torsion}},\ }\href@noop {} {\  (\bibinfo {year}
  {2024}{\natexlab{b}})},\ \Eprint {https://arxiv.org/abs/2406.12826}
  {arXiv:2406.12826 [hep-th]} \BibitemShut {NoStop}%
\end{thebibliography}%

\appendix
\begin{widetext}

\section{Quadratic and cubic relations}\label{AppendixInvar}
\paragraph*{Distortion invariants} In this appendix we provide additional algebraic identities obeyed by the quadratic invariants in~\cref{fullAction} in a cosmological setting. These take the form of identically vanishing sums of squares or cubes of the quadratic invariants formed from the curvature, torsion and non-metricity tensors. The means for obtaining these relations is discussed in~\cref{Invariants}. There is precisely one relation of fourth order in the non-metricity, i.e. quadratic in the quadratic non-metricity invariants
\begin{equation}\label{NonRiemRel3}
    \AInvar{6}\AInvar{7}-\AInvar{8}^2 = 0.
\end{equation}
\paragraph*{Curvature invariants} Meanwhile, there are two relations of fourth order in the curvature, and one sixth order relation
        \begin{align}
            &\CInvar{1}^2-2 \CInvar{1} \CInvar{3}-\CInvar{1} \CInvar{4}-2 \CInvar{1} \CInvar{5}-\CInvar{1} \CInvar{6}-\CInvar{2}^2-2 \CInvar{2} \CInvar{3}+\CInvar{2} \CInvar{4}+2 \CInvar{2} \CInvar{5}+\CInvar{2} \CInvar{6}+2 \CInvar{3} \CInvar{4}+4 \CInvar{3} \CInvar{5}+2 \CInvar{3} \CInvar{6}\nonumber\\
&-2 \CInvar{4}^2+4 \CInvar{4} \CInvar{6}-2 \CInvar{6}^2=0,\label{CurvRel3}
	    \\
            &\CInvar{1} \CInvar{3}-4 \CInvar{1} \CInvar{10}-6 \CInvar{1} \CInvar{12}+3 \CInvar{1} \CInvar{2}+3 \CInvar{1} \CInvar{4}+8 \CInvar{1} \CInvar{5}+3 \CInvar{1} \CInvar{6}+4 \CInvar{10}^2+8 \CInvar{10} \CInvar{12}-4 \CInvar{10} \CInvar{2}-4 \CInvar{10} \CInvar{4}-8 \CInvar{10} \CInvar{5}\nonumber\\
&-4 \CInvar{10} \CInvar{6}+4 \CInvar{12}^2-6 \CInvar{12} \CInvar{2}-6 \CInvar{12} \CInvar{4}-12 \CInvar{12} \CInvar{5}-6 \CInvar{12} \CInvar{6}+3 \CInvar{2}^2+\CInvar{2} \CInvar{3}+2 \CInvar{2} \CInvar{4}+6 \CInvar{2} \CInvar{5}+2 \CInvar{2} \CInvar{6}-3 \CInvar{3} \CInvar{4}\nonumber\\
&-6 \CInvar{3} \CInvar{5}-3 \CInvar{3} \CInvar{6}+3 \CInvar{4}^2+6 \CInvar{4} \CInvar{5}-2 \CInvar{4} \CInvar{6}+8 \CInvar{5}^2+6 \CInvar{5} \CInvar{6}+3 \CInvar{6}^2=0,\label{CurvRel1}
	    \\
            &8\CInvar{1} \CInvar{10} \CInvar{5}-3 \CInvar{1} \CInvar{10}^2-6 \CInvar{1} \CInvar{10} \CInvar{12}+4 \CInvar{1} \CInvar{10} \CInvar{4}+4 \CInvar{1} \CInvar{10} \CInvar{6}-3 \CInvar{1} \CInvar{12}^2+6 \CInvar{1} \CInvar{12} \CInvar{4}+12 \CInvar{1} \CInvar{12} \CInvar{5}+6 \CInvar{1} \CInvar{12} \CInvar{6}\nonumber\\
&+2 \CInvar{1} \CInvar{3} \CInvar{4}+4 \CInvar{1} \CInvar{3} \CInvar{5}+2 \CInvar{1} \CInvar{3} \CInvar{6}-3 \CInvar{1} \CInvar{4}^2-8 \CInvar{1} \CInvar{4} \CInvar{5}-10 \CInvar{1} \CInvar{5}^2-8 \CInvar{1} \CInvar{5} \CInvar{6}-3 \CInvar{1} \CInvar{6}^2+3 \CInvar{10}^2 \CInvar{2}+6 \CInvar{10}^2 \CInvar{3}\nonumber\\
&+2 \CInvar{10}^2 \CInvar{4}+4 \CInvar{10}^2 \CInvar{5}+2 \CInvar{10}^2 \CInvar{6}+6 \CInvar{10} \CInvar{12} \CInvar{2}+12 \CInvar{10} \CInvar{12} \CInvar{3}+4 \CInvar{10} \CInvar{12} \CInvar{4}+8 \CInvar{10} \CInvar{12} \CInvar{5}+4 \CInvar{10} \CInvar{12} \CInvar{6}-8 \CInvar{10} \CInvar{2} \CInvar{4}\nonumber\\
&-16 \CInvar{10} \CInvar{2} \CInvar{5}-8 \CInvar{10} \CInvar{2} \CInvar{6}-12 \CInvar{10} \CInvar{3} \CInvar{4}-24 \CInvar{10} \CInvar{3} \CInvar{5}-12 \CInvar{10} \CInvar{3} \CInvar{6}+4 \CInvar{10} \CInvar{4}^2-8 \CInvar{10} \CInvar{4} \CInvar{5}-16 \CInvar{10} \CInvar{4} \CInvar{6}\nonumber\\
&-8 \CInvar{10} \CInvar{5}^2-8 \CInvar{10} \CInvar{5} \CInvar{6}+4 \CInvar{10} \CInvar{6}^2+3 \CInvar{12}^2 \CInvar{2}+6 \CInvar{12}^2 \CInvar{3}+2 \CInvar{12}^2 \CInvar{4}+4 \CInvar{12}^2 \CInvar{5}+2 \CInvar{12}^2 \CInvar{6}-12 \CInvar{12} \CInvar{2} \CInvar{4}\nonumber\\
&-24 \CInvar{12} \CInvar{2} \CInvar{5}-12 \CInvar{12} \CInvar{2} \CInvar{6}-18 \CInvar{12} \CInvar{3} \CInvar{4}-36 \CInvar{12} \CInvar{3} \CInvar{5}-18 \CInvar{12} \CInvar{3} \CInvar{6}+6 \CInvar{12} \CInvar{4}^2-12 \CInvar{12} \CInvar{4} \CInvar{5}-24 \CInvar{12} \CInvar{4} \CInvar{6}\nonumber\\
&-12 \CInvar{12} \CInvar{5} \CInvar{6}-12 \CInvar{12} \CInvar{5}^2+6 \CInvar{12} \CInvar{6}^2+3 \CInvar{2}^2 \CInvar{4}+6 \CInvar{2}^2 \CInvar{5}+3 \CInvar{2}^2 \CInvar{6}+2 \CInvar{2} \CInvar{3} \CInvar{4}+4 \CInvar{2} \CInvar{3} \CInvar{5}+2 \CInvar{2} \CInvar{3} \CInvar{6}+\CInvar{2} \CInvar{4}^2\nonumber\\
&+20 \CInvar{2} \CInvar{4} \CInvar{5}+14 \CInvar{2} \CInvar{4} \CInvar{6}+24 \CInvar{2} \CInvar{5}^2+20 \CInvar{2} \CInvar{5} \CInvar{6}+\CInvar{2} \CInvar{6}^2-3 \CInvar{3}^2 \CInvar{4}-6 \CInvar{3}^2 \CInvar{5}-3 \CInvar{3}^2 \CInvar{6}+6 \CInvar{3} \CInvar{4}^2+24 \CInvar{3} \CInvar{4} \CInvar{5}\nonumber\\
&+6 \CInvar{3} \CInvar{4} \CInvar{6}+30 \CInvar{3} \CInvar{5}^2+24 \CInvar{3} \CInvar{5} \CInvar{6}+6 \CInvar{3} \CInvar{6}^2-3 \CInvar{4}^3-6 \CInvar{4}^2 \CInvar{5}+5 \CInvar{4}^2 \CInvar{6}+10 \CInvar{4} \CInvar{5}^2+28 \CInvar{4} \CInvar{5} \CInvar{6}+5 \CInvar{4} \CInvar{6}^2\nonumber\\
&+8 \CInvar{5}^3+10 \CInvar{5}^2 \CInvar{6}-6 \CInvar{5} \CInvar{6}^2-3 \CInvar{6}^3=0.\label{CurvRel2}
        \end{align}

\section{The field equations in full}\label{AppendixFried}
\paragraph*{Friedmann equations} In this appendix we display the background field equations of~\cref{fullAction} subject only to the condition~$K=0$ of vanishing spatial curvature, corresponding to~\cref{FLRWBasic} --- the full equations, with general~$K\neq 0$ corresponding to~\cref{FLRWCurved}, are made available in the supplemental materials~\cite{Supplement}. The means by which these equations are obtained is described in~\cref{MSS}. The first and second modified Friedmann equations are respectively:
\begin{align}
    6a_{0}H^2&+d_{1}\left(2H\ddot{H}+6H^2\dot{H}-\dot{H}^2\right)+\left(\mathcal{A}_{ij}N_{i}\dot{N_{j}}+\ddot{\Xi}\right)H+\left(\tfrac{1}{2}\mathcal{B}_{ij}N_{i}N_{j}+f_1(\dot{N})\right)H^2 + 3f_2(N)H^3 \nonumber\\& -\dot{\Xi}\dot{H}+\tfrac{1}{2}\mathcal{C}_{ij}\dot{N_{i}}\dot{N_{j}}+\tfrac{1}{2}\mathcal{G}_{ij}N_{i}N_{j}+F_1(N^4)=0,\label{Friedmann1}
    \\
    12a_{0}\dot{H}&+2d_{1}\left(2\dot{H}^2-H\ddot{H}\right)+\left(\mathcal{D}_{ij}N_{i}\dot{N_{j}}-\ddot{\Xi}\right)H+\left(\tfrac{1}{2}\mathcal{M}_{ij}N_{i}N_{j}+4\dot{\Xi}\right)\dot{H}-f_2(\dot{N})H^2+f_2(N)H\dot{H}\nonumber\\& -\Xi\ddot{H}+\mathcal{K}_{ij}N_{i}\ddot{N_{j}}+\mathcal{J}_{ij}\dot{N_{i}}\dot{N_{j}}+f_3(\dot{N})+F_2(N^2,\dot{N})=0.\label{Friedmann2}
\end{align}
In~\cref{Friedmann1,Friedmann2} we define the following quantities:
$$
N_{i} \in \{X,Y,V,Z,W\}, \quad \Xi \equiv \alpha X + \beta Y, \quad U \equiv V+Z,
$$
$$
\alpha \equiv 3d_1-d_3-\tfrac{d_2}{2}, \quad \beta \equiv d_1-d_3-\tfrac{d_2}{2}, \quad \gamma \equiv d_3-2d_4, \quad \delta \equiv 2d_4,
$$
$$
f_1(\dot{N})=\left(3\alpha+\beta\right) \dot{X} +\left(\alpha+3\beta\right) \dot{Y}+\left(\alpha+\beta\right)\dot{U}, \quad
f_2(N) = \beta X+ \alpha Y + (\alpha+\beta)U,
$$
$$
f_2(\dot{N}) = \beta \dot{X}+ \alpha \dot{Y} +
(\alpha+\beta)\dot{U}, \quad f_3(\dot{N})=\left(6 a_0+2 d_9-d_{10}\right)\dot{X}+\left(-6 a_0+d_{10}-2 d_{11}\right)\dot{Y} +\left(d_{12}-d_{13}\right)\dot{U},
$$

\begin{equation*}
\mathcal{A}_{ij}=
\begin{pmatrix}
\gamma & \alpha+\gamma & \alpha & \alpha & 0 \\
\gamma-\beta & \gamma & -\beta & -\beta & 0 \\
2\left(\delta + \beta + 2 \gamma\right) & 2\left(\delta + \beta + 2 \gamma\right) & 0 & 0 & 0 \\
2\left(\delta + \beta + 2 \gamma\right) & 2\left(\delta + \beta + 2 \gamma\right) & 0 & 0 & 0 \\
0 & 0 & 0 & 0 & d_2-2d_5+d_6-d_7
\end{pmatrix},
\end{equation*}

\begin{equation*}
\mathcal{B}_{ij}=
\left(
\begin{array}{ccccc}
 -\delta & 3 \alpha -4 \beta -\delta & 3 \alpha + 2 \beta + 3 \gamma + \delta & 3 \alpha +2 \beta +3 \gamma +\delta & 0 \\
 3 \alpha -4 \beta -\delta & -\alpha -\beta -\delta & 3 \gamma -\alpha -2 \beta +\delta & 3 \gamma -\alpha -2 \beta +\delta \\
 3 \alpha +2 \beta +3 \gamma +\delta & 3 \gamma -\alpha -2 \beta +\delta & 5 \beta-\alpha +8 \gamma +4 \delta & 5 \beta-\alpha +8 \gamma +4 \delta & 0 \\
 3 \alpha +2 \beta +3 \gamma +\delta & 3 \gamma -\alpha -2 \beta +\delta & 5 \beta-\alpha +8 \gamma +4 \delta & 5 \beta-\alpha +8 \gamma +4 \delta \\
 0 & 0 & 0 & 0 & 2 d_2-d_5 \\
\end{array}
\right),
\end{equation*}

\begin{equation*}
\mathcal{C}_{ij}=
\begin{pmatrix}
\beta-\alpha -2 \gamma -\delta & \beta +2 \gamma +\delta & 0 & 0 & 0 \\
\beta +2 \gamma +\delta & 2\beta + 2\gamma  +\delta & 0 & 0 & 0 \\
0 & 0 & 0 & 0 & 0 \\
0 & 0 & 0 & 0 & 0 \\
0 & 0 & 0 & 0 & \tfrac{1}{3} (d_{2}-2 d_{1}+d_{5}-d_{6}+d_{7})
\end{pmatrix},
\end{equation*}

\begingroup
    \fontsize{8pt}{12pt}\selectfont

\begin{equation*}
\mathcal{G}_{ij}=
\left(
\begin{array}{ccccc}
 2d_{9} & d_{10}-6 a_0 & 3 a_0+d_{12} & 3 a_0+d_{15} & 0 \\d_{10}-6 a_0 & 2d_{11} & 3 a_0+d_{13} & 3 a_0-2 d_{11}+3 d_{12}-3 d_{13}+d_{15}+2 d_{9} & 0 \\
3 a_0+d_{12} & 3 a_0+d_{13}& d_{14} &d_{10}+2 d_{12}-d_{13}-6 d_{14}+d_{15}+d_{9} & 0 \\
3 a_0+d_{15} &  3 a_0-2 d_{11}+3 d_{12}-3 d_{13}+d_{15}+2 d_{9} & d_{10}+2 d_{12}-d_{13}-6 d_{14}+d_{15}+d_{9} & -d_{10}+d_{11}-3 d_{12}+3 d_{13}+9 d_{14}-2 d_{15}-3 d_{9} & 0 \\
 0 & 0 & 0 & 0 & 2\left(d_{16}-6 a_0\right) \\
\end{array}
\right),
\end{equation*}

\endgroup

\begin{equation*}
\mathcal{D}_{ij}=
\left(
\begin{array}{ccccc}
 -\gamma  & -2 \alpha -\beta -\gamma  & 2 (-\alpha +\beta +2 \gamma +\delta ) & 2 (-\alpha +\beta +2 \gamma +\delta ) & 0 \\
 \alpha +2 \beta -\gamma  & -\gamma  & 4 (\beta +\gamma )+2 \delta  & 4 (\beta +\gamma )+2 \delta  & 0 \\
 \alpha -4 (\beta +2 \gamma +\delta ) & -5 \beta -4 (2 \gamma +\delta ) & 0 & 0 & 0 \\
 \alpha -4 (\beta +2 \gamma +\delta ) & -5 \beta -4 (2 \gamma +\delta ) & 0 & 0 & 0 \\
 0 & 0 & 0 & 0 & 2d_5-d_2-d_6+d_7 \\
\end{array}
\right),
\end{equation*}

\begin{equation*}
\mathcal{M}_{ij}=
\left(
\begin{array}{ccccc}
 2 \gamma  & \alpha -\beta +2 \gamma  & \alpha +2 (\beta +2 \gamma +\delta ) & \alpha +2 (\beta +2 \gamma +\delta ) & 0 \\
 \alpha -\beta +2 \gamma  & 2 \gamma  & \beta +4 \gamma +2 \delta  & \beta +4 \gamma +2 \delta  & 0 \\
 \alpha +2 (\beta +2 \gamma +\delta ) & \beta +4 \gamma +2 \delta  & 0 & 0 & 0 \\
 \alpha +2 (\beta +2 \gamma +\delta ) & \beta +4 \gamma +2 \delta  & 0 & 0 & 0 \\
 0 & 0 & 0 & 0 & d_2-2 d_5+d_6-d_7 \\
\end{array}
\right),
\end{equation*}

\begin{equation*}
\mathcal{K}_{ij}=
\left(
\begin{array}{ccccc}
 \alpha -\beta -2 \gamma -\delta  & \beta +2 \gamma +\delta  & 0 & 0 & 0 \\
 \beta +2 \gamma +\delta  & 2 \beta +2 \gamma +\delta  & 0 & 0 & 0 \\
 0 & 0 & 0 & 0 & 0 \\
 0 & 0 & 0 & 0 & 0 \\
 0 & 0 & 0 & 0 & -\frac{2}{3}\left(2 d_1-d_2-d_5+d_6-d_7\right) \\
\end{array}
\right),
\end{equation*}

\begin{equation*}
\mathcal{J}_{ij}=
\left(
\begin{array}{ccccc}
 2 (\alpha -\beta -2 \gamma -\delta ) & -2 (\beta +2 \gamma +\delta ) & 0 & 0 & 0 \\
 -2 (\beta +2 \gamma +\delta ) & -2 (2 \beta +2 \gamma +\delta ) & 0 & 0 & 0 \\
 0 & 0 & 0 & 0 & 0 \\
 0 & 0 & 0 & 0 & 0 \\
 0 & 0 & 0 & 0 & \frac{4}{3} \left(2 d_1-d_2-d_5+d_6-d_7\right)\\
\end{array}
\right),
\end{equation*}

\begin{equation*}
    \begin{aligned}
        F_1(N^4)&=\tfrac{1}{2} \left(\alpha -\beta \right) W^4-\left(2 \beta +\gamma +\delta \right) U Y W^2+ \left(\alpha -\beta -\gamma -\delta \right)U X W^2-\tfrac{1}{2}\left( 2\beta +2\gamma+ \delta\right) U^2 Y^2-\left(2 \beta +\gamma +\delta \right) U X Y^2\\&+\tfrac{1}{2}\left(\alpha -\beta \right)X^2 Y^2 +\left(\alpha -\beta -\gamma -\delta \right)U X^2 Y +\tfrac{1}{2} \left(\alpha -\beta -2 \gamma -\delta \right)U^2 X^2 + \left(\beta +2 \gamma +\delta \right)U^2 X Y\\&+\tfrac{1}{3}  \left(4 d_5-7 d_6+d_7+6 d_8+6 \alpha -8 \left(\beta +\gamma +\delta \right)\right) X Y W^2+\left(\alpha +\beta -d_6-d_7+d_8\right) Y^2 W^2+ d_8 X^2W^2,
    \end{aligned}
\end{equation*}

\begin{equation*}
    \begin{aligned}
        F_2(N^2,\dot{N})&=(-\alpha +\beta +2 \gamma +\delta )X^2\dot{U}-(2\beta + 2\gamma +\delta ) Y^2 \dot{U}-(2 \beta +\gamma +\delta ) Y^2 \dot{X}+(\alpha -\beta -\gamma -\delta )XY\dot{X} + (\alpha -\beta -2 \gamma -\delta )UX\dot{X}\\&+3(\beta +2 \gamma +\delta )UY\dot{X} - 3(\beta +2 \gamma +\delta )UX\dot{Y} + (-\alpha +\beta +\gamma +\delta )X^2\dot{Y}+(2 \beta +\gamma +\delta )XY\dot{Y} + (2\beta + 2\gamma+\delta )UY\dot{Y}\\& + \tfrac{1}{3} \left(-3 \alpha +\beta +\gamma +\delta +4 d_5-4 d_6+d_7\right)W\left(X\dot{W}-W\dot{X}\right)+\tfrac{1}{3}\left(4 \beta +\gamma +\delta +4 d_5-d_6+4 d_7\right)W\left(W\dot{Y}-Y\dot{W}\right).
    \end{aligned}
\end{equation*}

\paragraph*{Distortion equations} The remaining five equations of motion correspond to variations of the distortion fields:
   \begin{align}
        &\alpha\ddot{X}+\beta\ddot{Y} + \left(\alpha  X+\beta  Y\right)\dot{H} + 2d_1\left(4 Η\dot{H}+\ddot{H}\right)-\left(6 a_0+2 d_9-d_{10}\right)X+ \left(6 a_0-d_{10}+2 d_{11}\right)Y+\left(d_{13}-d_{12}\right) U+2 \left(\right.d_9-d_{11}\nonumber\\
&+2 d_{12}-2 d_{13}\left.\right) Z+\Big[\left(3 \alpha -9 \beta -6 (\gamma +\delta )-d_6+d_7\right)W^2+(5 \beta -\alpha +8 \gamma +4 \delta )U^2 + (5 \alpha +2 (\beta +\gamma ))UX- (2 \alpha +5 \beta -2 \gamma )UY \nonumber\\
& -(\gamma +\delta)X^2+(5 \alpha -7 \beta -2 (\gamma +\delta ))XY+(-\alpha -\beta -\gamma -\delta )Y^2\Big]H+\Big[(4 \alpha +\beta ) \dot{X}+(\alpha +4 \beta )\dot{Y}+(\alpha +\beta ) \dot{U}\Big]H \nonumber\\
&+4\Big[(\alpha +\beta )U+\beta  X+\alpha  Y\Big]H^2 +(\beta -\alpha )X Y^2+ (2 (\beta +2 \gamma +\delta )-\alpha )U^2 X- (3 \beta +4 \gamma +2 \delta )U^2 Y+ (\alpha -\beta )X^2 Y- (\alpha +\beta )U W^2\nonumber\\
&+\tfrac{1}{3} \left(6 \alpha -8 (\beta +\gamma +\delta )+4 d_5-7 d_6+d_7\right)W^2 X+\tfrac{1}{3} \left(14 \beta +8 (\gamma +\delta )-4 d_5+d_6-7 d_7\right)W^2Y+ (\alpha -\beta -\gamma -\delta )U X^2\nonumber\\
&-2  (\alpha +\beta )U X Y+ (2 \beta +\gamma +\delta )U Y^2+ 2(\beta +2 \gamma +\delta ) U \dot{Y}-\beta  Y \dot{U}+2  (\beta +2 \gamma +\delta )U \dot{X}+\alpha  X \dot{U}+ (\alpha -\beta -\gamma -\delta )X \dot{X}- \left(\right.2 \beta \nonumber\\
&+\gamma +\delta \left.\right)Y \dot{X} + (\alpha -\beta -\gamma -\delta )X \dot{Y}- (2 \beta +\gamma +\delta )Y \dot{Y}-\tfrac{1}{3}\left(-3 \alpha +5 \beta +2 (\gamma +\delta )+8 d_5-5 d_6+5 d_7\right)W\dot{W} = 0\label{EqY-EqX}
\\
        &(\beta +2 \gamma +\delta )\ddot{X}+(2 (\beta +\gamma )+\delta )\ddot{Y}+ \Big[(\alpha +\gamma )X+\gamma Y +2(\beta +2 \gamma +\delta )U\Big]\dot{H}+(\alpha -3 \beta ) H\dot{H}-\beta  \ddot{H}+\left(d_{10}-6 a_0\right)X+2 d_{11} Y\nonumber\\
&+ \left(3 a_0+d_{13}\right)U+\left(2 d_9-2 d_{11}+3 d_{12}-4 d_{13}+d_{15}\right) Z+ \Big[ \left(d_7-3 (2 \beta +\gamma +\delta )\right)W^2+ (3 \beta +4 \gamma +2 \delta )U^2+ (2 \alpha +5 \beta +6 \gamma \nonumber\\
&+3 \delta )U X+(-6 \beta -4 \gamma -3 \delta )U Y+ (-\gamma -\delta )X^2+ (2 \alpha -4 \beta -\gamma -\delta )X Y\Big]H+\Big[(\alpha +4 \beta +6 \gamma +3 \delta )\dot{X} + 3(2 (\beta +\gamma )+\delta )\dot{Y}\nonumber\\
&+ (3 \beta +4 \gamma +2 \delta )\dot{U}\Big]H+\Big[(\alpha +8 \beta +9 \gamma +5 \delta )U+ (4 \beta +3 \gamma +\delta )X+ (\alpha +\beta +3 \gamma +\delta )Y\Big]H^2+\tfrac{1}{3} \big(6 \alpha -8 (\beta +\gamma +\delta )+4 d_5\nonumber\\
&-7 d_6+d_7+6 d_8\big)W^2 X+2\left(\alpha +\beta -d_6-d_7+d_8\right)W^2 Y+ (\beta +2 \gamma +\delta )U^2 X- (2 (\beta +\gamma )+\delta )U^2 Y- (2 \beta +\gamma +\delta )U W^2\nonumber\\
&+(\alpha -\beta -\gamma -\delta )U X^2+2(\beta +2 \gamma +\delta )U \dot{X}-2(2 \beta +\gamma +\delta ) U X Y + (\alpha -\beta )X^2 Y+(\beta +2 \gamma +\delta )X \dot{U}-(2 (\beta +\gamma )+\delta )Y \dot{U} \nonumber\\
& + (\alpha -\beta -\gamma -\delta )X \dot{X}- (2 \beta +\gamma +\delta )Y \dot{X}-\tfrac{1}{3}\left(4 \beta +\gamma +\delta +4 d_5-d_6+4 d_7\right)W \dot{W} = 0,\label{EqY}
\\
        &(\alpha +\beta )H\dot{H}+\left(3 a_0+d_{12}\right)X+ \left(3 a_0+d_{13}\right)Y+2 d_{14} U+\left(2 d_9+d_{10}+2 d_{12}-d_{13}-8 d_{14}+d_{15}\right)Z+ \Big[(\alpha -5 \beta -4 (2 \gamma +\delta ))U\nonumber\\
&+(-2 \beta -3 \gamma -\delta )X+(\alpha -\beta -3 \gamma -\delta )Y\Big]H^2+\Big[2(\alpha -2 (\beta +2 \gamma +\delta ))UX + (6 \beta +8 \gamma +4 \delta )UY + (\alpha +\beta )W^2- \gamma X^2\nonumber\\
&+2(\alpha +\beta )XY+\gamma Y^2+ (\alpha -2 (\beta +2 \gamma +\delta ))\dot{X} + (-3 \beta -4 \gamma -2 \delta )\dot{Y}\Big]H+(\alpha X- \beta  Y)\dot{H}+ (\alpha -\beta -2 \gamma -\delta )U X^2\nonumber\\
&+2(\beta +2 \gamma +\delta )U X Y- (2 (\beta +\gamma )+\delta )U Y^2+ (\alpha -\beta -\gamma -\delta )W^2 X- (2 \beta +\gamma +\delta )W^2 Y+ (\alpha -\beta -\gamma -\delta )X^2 Y\nonumber\\
&- (2 \beta +\gamma +\delta )X Y^2+ (\alpha -\beta -2 \gamma -\delta )X \dot{X}+ (\beta +2 \gamma +\delta )Y \dot{X}- (\beta +2 \gamma +\delta )X \dot{Y}+ (2 (\beta +\gamma )+\delta )Y \dot{Y} = 0,\label{EqV}
	\\
    &\left(d_{12}-d_{15}\right) X+\left(-2 d_9+2 d_{11}-3 d_{12}+4 d_{13}-d_{15}\right) Y-\left(2 d_9+d_{10}+2 d_{12}-d_{13}-8 d_{14}+d_{15}\right)V\nonumber\\
&+\left(8 d_9+3 d_{10}-2 d_{11}+8 d_{12}-7 d_{13}-24 d_{14}+5 d_{15}\right) Z = 0,\label{EqV-EqZ}
\\
        &2 \left(-6 a_0+d_{16}\right) W+\left(-4 d_5+3 d_6-3 d_7-\alpha+3 \beta+2 \left(\gamma+\delta\right)\right) H^2 W+2 \left(\alpha-\beta\right) W^3+2 \left(\alpha-\beta-\gamma-\delta\right) U W X+2 d_8 W X^2\nonumber\\
&-2 \left(2 \beta+\gamma+\delta\right) U W Y+\tfrac{2}{3}\left(4 d_5-7 d_6+d_7+6 d_8+6 \alpha-8 \left(\beta+\gamma+\delta\right)\right) W X Y+2 \left(-d_6-d_7+d_8+\alpha+\beta\right) W Y^2\nonumber\\
&+\left(d_6-2d_5-d_7+\alpha-3 \beta-2 \left(\gamma+\delta\right)\right) W \dot{H}+\Big[2 \left(\alpha+\beta\right) U W-\left(4 d_5-6 d_6+d_7+3 \alpha-5 \left(\beta+\gamma+\delta\right)\right) W X\nonumber\\
&+\left(4 d_5-d_6+6 d_7-8 \beta-5 \left(\gamma+\delta\right)\right) W Y+2 \left(d_5-d_6+d_7-2 \left(\beta+\gamma+\delta\right)\right) \dot{W}\Big]H-\tfrac{1}{3}\left(4 d_5-4 d_6+d_7-3 \alpha+\beta+\gamma+\delta\right) W \dot{X}\nonumber\\
&+\tfrac{1}{3}\left(4 d_5-d_6+4 d_7+4 \beta+\gamma+\delta\right) W \dot{Y}+\tfrac{1}{3}\left(d_5-d_6+d_7-2 \left(\beta+\gamma+\delta\right)\right) \ddot{W}= 0.\label{EqW}
\end{align}
Specifically~\cref{EqY-EqX} is~$\delta S/\delta Y - \delta S/\delta X= 0$,~\cref{EqY} is~$\delta S/\delta Y= 0$,~\cref{EqV} is~$\delta S/\delta V= 0$,~\cref{EqV-EqZ} is~$\delta S/\delta V - \delta S/\delta Z= 0$, and~\cref{EqW} is~$\delta S/\delta W= 0$.

\section{Particle spectrographs}\label{ParticleSpectra}

\paragraph*{Conventions} For computing particle spectra, we employ the \emph{Particle Spectrum for Any Tensor Lagrangian} (\PSALTer{}) software~\cite{Barker:2023bmr,Barker:2024ydb,Barker:2024dhb,Barker:2024juc,Barker:2024goa}. In the vicinity of Minkowski spacetime, we define the metric perturbation as~$\FieldH{_{\mu\nu}}\equiv\FieldG{_{\mu\nu}}-\FieldEta{_{\mu\nu}}$. Within the MAG framework, the connection~$\MAGA{_{\mu\nu}^\rho}$ is treated as fundamentally perturbative (that is, the background connection vanishes). The field~$\FieldH{_{\mu\nu}}$ couples to the linearised stress-energy tensor~$\tensor{T}{^{\mu\nu}}$, while~$\MAGA{_{\mu\nu}^\rho}$ couples to the linearised hypermomentum~$\FieldHyp{^{\mu\nu}_\rho}$. We collectively represent these sources as the vector~$\mathsf{J}$, while the fields~$\FieldH{_{\mu\nu}}$ and~$\MAGA{_{\mu\nu}^\rho}$ are grouped into another vector~$\zeta$, allowing the entire matter Lagrangian to be written as~$\sim\zeta^{\text{T}}\cdot\mathsf{J}$ at first order in perturbations. Working in position space, the action from~\cref{fullAction} may be written schematically using these source and field vectors as
\begin{equation}\label{GenLag}
	\mathcal{S}=\int\mathrm{d}^4x\ \zeta^{\text{T}}(x)\cdot\left[\mathsf{O}(\partial)\cdot\zeta(x)-\mathsf{J}(x)\right]+\mathcal{O}(\zeta^3),
\end{equation}
where we call the operator matrix~$\mathsf{O}(\partial)$ the (linearized) \emph{wave operator}. This matrix represents a tensor structure that is polynomial in the partial derivative~$\tensor{\partial}{_\mu}$ up to second order. The matrix depends on the Lagrangian coupling coefficients~$\left\{a_{0},\ldots,a_{11}\right\}$ and~$\left\{c_{1},\ldots,c_{16}\right\}$. Starting from this action representation, we obtain the saturated propagator from~\cref{GenLag} by transitioning to momentum space through~$\tensor{\partial}{_\mu}\mapsto -i\tensor{k}{_\mu}$, then `inverting' the wave operator and placing it between the conserved sources via the formula
\begin{equation}\label{Propagator}
	\Pi(k)\equiv\mathsf{J}^\dagger(k)\cdot\mathsf{O}^{-1}(k)\cdot\mathsf{J}(k).
\end{equation}
Characteristic of gravitational theories, QMAG and its special cases exhibit multiple gauge invariances: consequently, the putative `inverse' in~\cref{Propagator} lacks proper definition due to degeneracies within the wave operator's matrix representation. As a fundamental principle, these degeneracies correspond one-to-one with gauge symmetries that impose constraints on the allowable physical values of the source currents. Therefore, the `sandwiching' operation in~\cref{Propagator} results in precise cancellation between the singular components of the matrix inverse and the source zeroes. Remaining within~$k$-space, established techniques from polology demonstrate how pole positions in~$\Pi(k)$ characterize the (squared) masses of persistent quantum states in the free theory. In particular, real masses are required for non-tachyonic (physical) particles. Additionally, the residue of~$\Pi(k)$ at each pole must be positive-definite to ensure the state is not a ghost. For massive quantum states, employing a spin-parity ($J^P$) basis proves optimal for casting the wave operator inversion problem in basic linear algebra terms. For massless states, conversely, helicity quantum numbers provide the appropriate description of spin. For massless computations, the spectrum is calculated via brute-force methods using individual tensor components. While this component-based approach successfully extracts polarization counts and associated unitarity constraints, the massless~$J$ and~$P$ quantum numbers are not always evident. Through projection using the unit-timelike vector~$\tensor{n}{_\mu}\tensor{n}{^\mu}\equiv 1$, with~$\tensor{n}{^\mu}\equiv\tensor{k}{^\mu}/k$ for~$k^2\equiv\tensor{k}{^\mu}\tensor{k}{_\mu}$ and~$\tensor{k}{^\mu}$ representing the massive four-momentum, we display the different~$\mathrm{SO}(3)$ irreducible components of the sources in~\cref{FieldKinematicsMetricPerturbation,FieldKinematicsConnection}. These carry spin-parity ($J^P$) labels for identification purposes. Multiple~$J^P$ states may occur, which we differentiate using additional labels~`$\#1$',~`$\#2$', etc. We conducted the calculations in~\crefrange{ParticleSpectrographKScreeningNoA2m1p1mA1mB}{ParticleSpectrographMaxwellNoHigherSpinNo1m} using 64 AMD\textsuperscript{\textregistered} \emph{Ryzen Threadripper} CPUs. Additional information can be found in the supplemental materials~\cite{Supplement}.

\begin{figure*}[h!]
\includegraphics[width=\textwidth,height=\textheight,keepaspectratio]{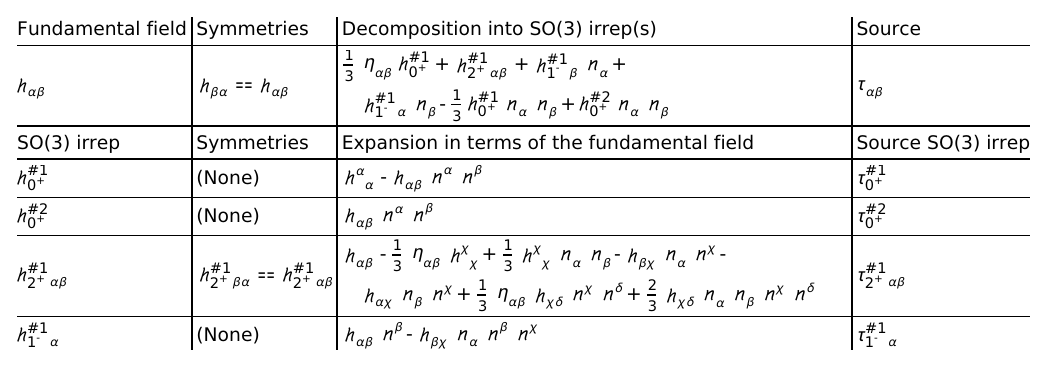}
	\caption{\label{FieldKinematicsMetricPerturbation} Kinematic structure of the metric perturbation~$\FieldH{_{\mu\nu}}$. These definitions are used in~\crefrange{ParticleSpectrographKScreeningNoA2m1p1mA1mB}{ParticleSpectrographMaxwellNoHigherSpinNo1m}. See~\cite{Barker:2024juc} for further notational details. This is a vector graphic: all details are visible under magnification.}
\end{figure*}
\begin{figure*}[h!]
\includegraphics[width=\textwidth,height=\textheight,keepaspectratio]{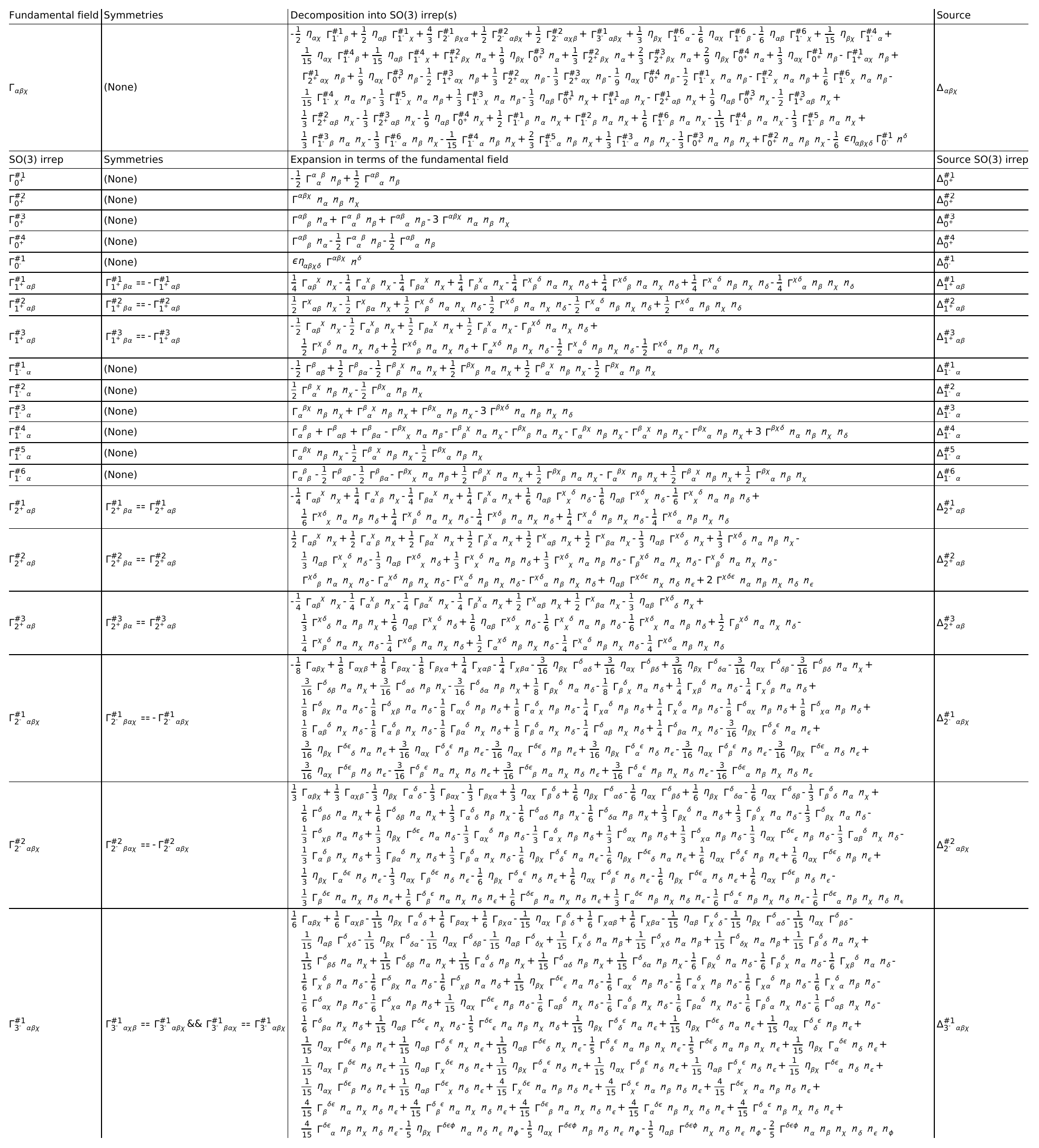}
	\caption{\label{FieldKinematicsConnection} Kinematic structure of the affine connection~$\MAGA{_{\mu\nu}^\rho}$. These definitions are used in~\crefrange{ParticleSpectrographKScreeningNoA2m1p1mA1mB}{ParticleSpectrographMaxwellNoHigherSpinNo1m}. See~\cite{Barker:2024juc} for further notational details. This is a vector graphic: all details are visible under magnification.}
\end{figure*}

\paragraph*{Gravity with~$K$-screening} We will first study the model considered in~\cref{spectra}. The conditions in~\cref{kscreened} may be imposed with the rules
\begin{equation}\label{KScreeningRules}
	a_{0}\mapsto 0, \quad c_{6}\mapsto -2c_{1}+6c_{16}+2c_{2}-2c_{3}-c_{4}+c_{5}, \quad c_{8}\mapsto \frac{c_{11}}{2}+\frac{c_{12}}{2}-3c_{16}-c_{7}, \quad c_{9}\mapsto -c_{10}+\frac{c_{11}}{2}+\frac{c_{12}}{2}-3c_{16}.
\end{equation}
To simplify the formulae for the masses of the various propagating particles, we consider the case where only~$a_{1}$ and~$a_{4}$ are non-zero. By comparing with~\cref{fullAction}, we see that these parameters modulate the `index-for-index squares' of the torsion and non-metricity, respectively. Thus, they are somewhat analagous to the Kretschmann scalar for curvature, and we take them to be representative of the mass-dimension-two parameters that may be included in the model. Accordingly, we impose the conditions
\begin{equation}\label{NoA}
	a_{2}\mapsto 0,\quad
	a_{3}\mapsto 0,\quad
	a_{5}\mapsto 0,\quad
	a_{6}\mapsto 0,\quad
	a_{7}\mapsto 0,\quad
	a_{8}\mapsto 0,\quad
	a_{9}\mapsto 0,\quad
	a_{10}\mapsto 0,\quad
	a_{11}\mapsto 0.
\end{equation}
When~\cref{KScreeningRules,NoA} are applied to~\cref{fullAction}, the abundance of unconstrained parameters makes the computation of the full spectrum difficult (see the supplemental materials~\cite{Supplement}), and the~$1^-$ sector is entirely neglected at this stage. Despite this, there is enough information about the remaining massive~$J^P$ sectors to begin constraining the model. Beginning with the~$2^-$ sector, we find that the denominator of the saturated propagator (the pseudo-determinant of the wave operator matrix) is a quadratic in~$k^2$ -- we thus expect \emph{two} massive poles. Generically, the expressions for the square masses in terms of the remaining Lagrangian couplings will contain radicals owing to the quadratic formula, and these are awkward points about which to compute pole residues when determining the unitarity of the model. To proceed, we impose further constraints so as to remvoe the coefficient of~$k^4$. This coefficient is a quadratic form in the mass-dimension-zero couplings, which is problematic because it is computationally preferable to impose \emph{linear} constraints among the couplings. An impartial (i.e. parameterisation-independent) approach is to diagonalise the quadratic form as a weighted sum of squares, and insist that each term vanish separately. This approach yields the constraints
\begin{equation}\label{Rule2m}
	c_{4}\mapsto 2c_{2}, \quad c_{5}\mapsto 4c_{1}, \quad c_{16}\mapsto -\frac{c_{1}}{3}+\frac{c_{2}}{3}+\frac{c_{3}}{3},
\end{equation}
and, in fact, these eliminate not only the~$k^4$ coefficient but also the coefficient of~$k^2$: there is accordingly no massive~$2^-$ particle from this point forward. After imposing~\cref{Rule2m}, the determinants for the remaining~$J^P$ sectors are naturally much simplified. Proceeding to the~$1^+$ sector, a similar quartic system is attempted to be reduced to a single massive pole using the same method, yielding the constraints
\begin{equation}\label{Rule1p}
	c_{2}\mapsto -c_{1}+c_{10}-c_{11}+c_{7}, \quad c_{3}\mapsto 2c_{1}-c_{10}+c_{7}, \quad c_{12}\mapsto 2c_{10}-3c_{11}+6c_{7},
\end{equation}
but once again this removes the~$1^+$ particle entirely. The conditions in~\cref{Rule2m,Rule1p} suggest that one massive particle will still arise from each of the~$0^+$,~$0^-$~$2^+$ and~$3^-$ sectors. However, the~$1^-$ sector was not yet considered at all. When~\cref{KScreeningRules,NoA,Rule2m,Rule1p} are applied to~\cref{fullAction}, the partial particle spectrum for the~$1^-$ sector is simple enough to be obtainable, but still too large to present (see the supplemental materials~\cite{Supplement}). Commensurate with the large size of its wave operator and saturated propagator matrices, the~$1^-$ sector contains a \emph{quartic} in~$k^2$, suggesting that there will be \emph{four} massive poles. It is easiest to begin `half-way through' by eliminating the~$k^4$ coefficient once again. This coefficient is a sum of quadratic forms in the mass-dimension-zero couplings, one multiplied by~$a_{1}a_{4}^3$ and the other by~$a_{1}^2a_{4}^2$. Eliminating them both independently yields, respectively
\begin{subequations}
\begin{gather}
	c_{11}\mapsto 2c_{7}, \quad c_{14}\mapsto 3c_{10}+2c_{13}, \quad c_{15}\mapsto 3c_{10}+2c_{13},\label{Rule1mA}
	\\
	c_{7}\mapsto 0, \quad c_{10}\mapsto 0, \quad c_{13}\mapsto 0.\label{Rule1mB}
\end{gather}
\end{subequations}
As was the experience with~\cref{Rule2m,Rule1p}, the whole of the~$1^-$ sector is actually rendered non-propagating by~\cref{Rule1mA,Rule1mB}, although this was not our intention. When~\cref{Rule1mA,Rule1mB} are applied sequentially on~\cref{ParticleSpectrographKScreeningNoA2m1p1mA1mB}, all the massive poles vanish except for that arising from the~$0^-$ sector: the system is now simple enough that the full spectral analysis may be performed and presented. When~\cref{KScreeningRules,NoA,Rule2m,Rule1p,Rule1mA,Rule1mB} are applied to~\cref{fullAction}, the resulting particle spectrograph is shown in~\cref{ParticleSpectrographKScreeningNoA2m1p1mA1mB}. The~$0^-$ sector contains a single massive pole, as expected. The outcome of the massless analysis is very much harder to anticipate from the computations above. In this case, the software tells us that there are \emph{no massless particles whatever}. The paucity of propagating d.o.f in the final model means that unitarity is easy to achieve, indeed the no-tachyon and no-ghost conditions are respectively
\begin{equation}\label{UnitarityConditionsKScreened}
	a_{1}<0<c_{1}.
\end{equation}
As explained in~\cref{spectra}, the model we have been forced to construct, through the mandatory requirements of a unitary spectrum near Minkowski spacetime, is simply the square of the Holst operator, with added mass terms.

\begin{figure*}[ht]
\includegraphics[width=\textwidth]{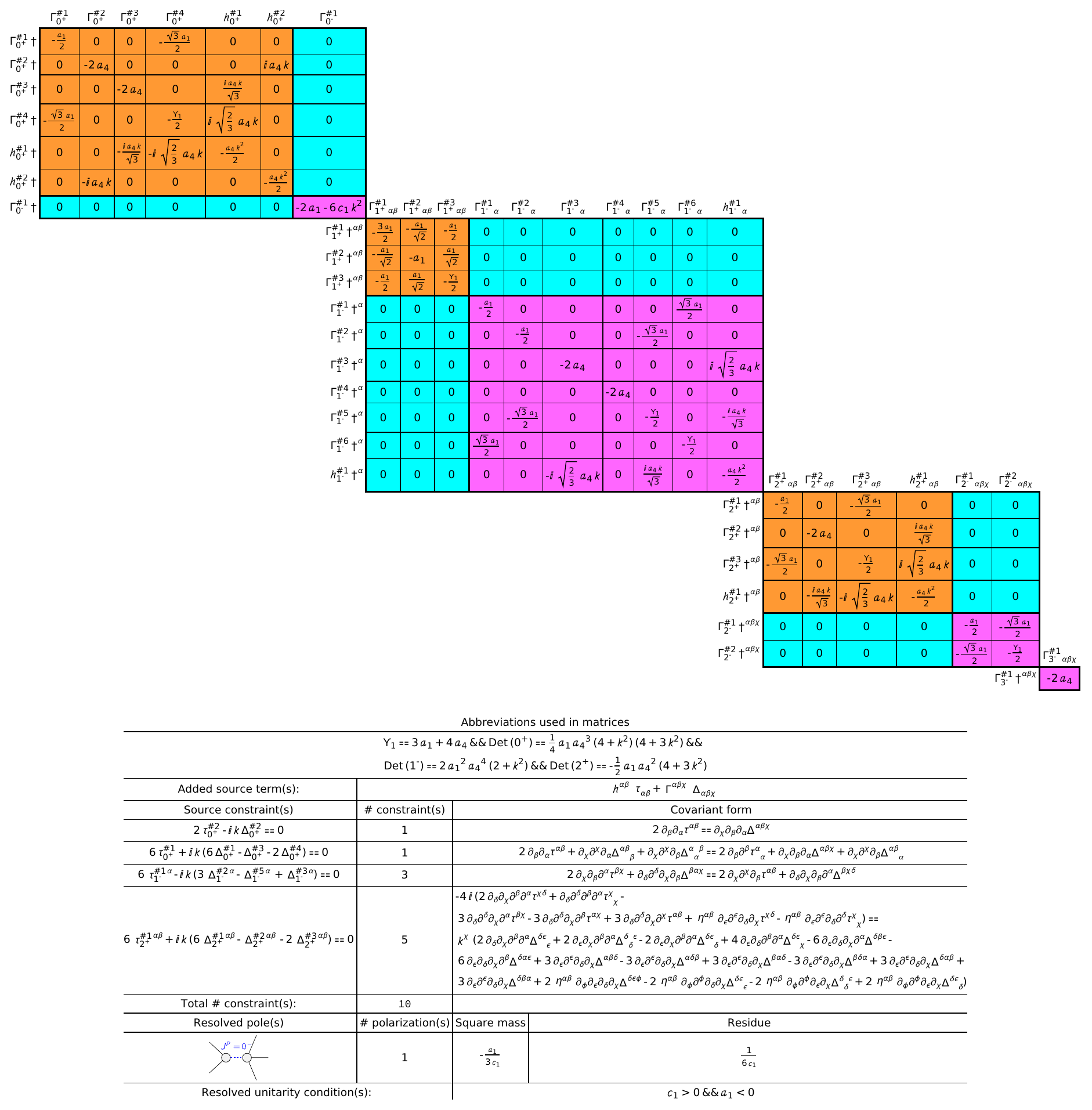}
\caption{\label{ParticleSpectrographKScreeningNoA2m1p1mA1mB} Complete particle spectrograph of~\cref{fullAction} after the sequential imposition of~\cref{KScreeningRules,NoA,Rule2m,Rule1p,Rule1mA,Rule1mB}. The collection of matrices at the top encode the wave operator $\mathsf{O}(\partial)$ of the theory, as defined in~\cref{GenLag}, presented in $k$-space. The matrices are divided into blocks of spin $J=0$, $J=1$, $J=2$ and $J=3$. Each block is further separated into parity-even $P=1$ and odd $P=-1$ sectors. Even-even and odd-odd mixings are shown in orange and purple, respectively. Mixing between even and odd sectors is shown in teal; since the version of QMAG we consider is parity-preserving, all these matrix elements vanish identically. The table below the matrices summarises the spectrum. The constraints on the sources (associated with gauge symmetries, i.e. degeneracies of the wave  operator) are shown in $J^P$-notation. There are a total of ten source constraints, i.e. ten gauge generators. Four of these are associated with the diffeomorphism invariance of the model. All other quantities are defined in~\cref{FieldKinematicsMetricPerturbation,FieldKinematicsConnection}. See~\cite{Barker:2024juc} for further notational details. This is a vector graphic: all details are visible under magnification.}
\end{figure*}

\paragraph*{Integrable gravity and its Maxwell limit} We now turn to the model considered in~\cref{EvenHubble}. The condition that all the~$\left\{d_{1},\ldots,d_{8}\right\}$ vanish may be imposed with the replacements
\begin{equation}\label{AntiCosmologicalRules}
	\begin{gathered}
	c_{5}\mapsto 4c_{6}+2c_{7}+2c_{8},\quad 
	c_{4}\mapsto c_{6},\quad
	c_{3}\mapsto -c_{6}-c_{7}-c_{8},\quad
	c_{2}\mapsto 2c_{6}+c_{7}+c_{8},\quad
	c_{16}\mapsto -c_{6}-c_{7}-c_{8},
	\\
	c_{11}\mapsto -c_{12}-6c_{6}-4c_{7}-4c_{8},\quad
	c_{10}\mapsto c_{7}+c_{8}-c_{9},\quad
	c_{1}\mapsto c_{6},
	\end{gathered}
\end{equation}
and the condition that the~$\left\{a_{1},\ldots,a_{11}\right\}$ vanish may be imposed with the replacements
\begin{equation}\label{AllAZero}
	a_{1}\mapsto 0,\quad
	a_{2}\mapsto 0,\quad
	a_{3}\mapsto 0,\quad
	a_{4}\mapsto 0,\quad
	a_{5}\mapsto 0,\quad
	a_{6}\mapsto 0,\quad
	a_{7}\mapsto 0,\quad
	a_{8}\mapsto 0,\quad
	a_{9}\mapsto 0,\quad
	a_{10}\mapsto 0,\quad
	a_{11}\mapsto 0.
\end{equation}
When~\cref{AntiCosmologicalRules,AllAZero} are applied to the model in~\cref{fullAction}, the resulting partial particle spectrum is shown in~\cref{ParticleSpectrographMaxwellTheory}. In the first instance, the analysis is performed without considering the massless sector at all, i.e. by only elucidating the massive poles. It is clear from~\cref{ParticleSpectrographMaxwellTheory} that there will be multiple massive d.o.f, because these denominators contain various roots. Out of necessity, we will eliminate the~$3^-$ and two~$2^-$ d.o.f by sending their square masses to infinite magnitude, and as a means of simplifying the model we do the same for the~$1^+$ mode. Neither of the~$0^+$ or~$0^-$ modes contains a massive pole; the~$1^-$ mode contains a single massive pole, but the expression for its mass is still slightly cumbersome. The relevant masses are made to diverge by imposing the further replacements
\begin{equation}\label{FurtherRules}
	c_{6}\mapsto 0,\quad c_{8}\mapsto -c_{7},\quad c_{12}\mapsto c_{7}+c_{9}.
\end{equation}
The complete particle spectrograph resulting from the sequential imposition of~\cref{AntiCosmologicalRules,AllAZero,FurtherRules} on~\cref{fullAction} is shown in~\cref{ParticleSpectrographMaxwellNoHigherSpinTheory}. This full analysis (with the massless sector included this time) confirms the massive~$1^-$ mode, along with four massless polarisations whose presence was not previously obvious. Two of these polarisations may be associated with the graviton, since their no-ghost conditions depend only on the~$a_{0}$-coupling. The other two polarisations presumably have their origin in the massless limit of some higher-spin mode, whose mass was removed by the conditions in~\cref{AllAZero}. The expression for the mass of the~$1^-$ mode is now simpler, but it turns out to have a cumbersome pole residue which still hinders the unitarity analysis: to remove this d.o.f entirely we once again take the limit of infinite mass with the final condition
\begin{equation}\label{InfiniteMassVector}
	c_{15}\mapsto -c_{14}.
\end{equation}
The complete particle spectrograph resulting from the sequential imposition of~\cref{AntiCosmologicalRules,AllAZero,FurtherRules,InfiniteMassVector} on~\cref{fullAction} is shown in~\cref{ParticleSpectrographMaxwellNoHigherSpinNo1m}. The massless spectra are the same as in~\cref{ParticleSpectrographMaxwellNoHigherSpinTheory}, though the extra massless pole residue is somwhat simplified. The total conditions for unitarity can be readily computed as
\begin{equation}\label{MaxwellUnitary}
	a_{0}<0<c_{13}.
\end{equation}
As explained in~\cref{EvenHubble}, the model leading to~\cref{MaxwellUnitary} can be understood as the square of the homothetic curvature.

\begin{figure*}[ht]
\includegraphics[height=0.9\textwidth,angle=90]{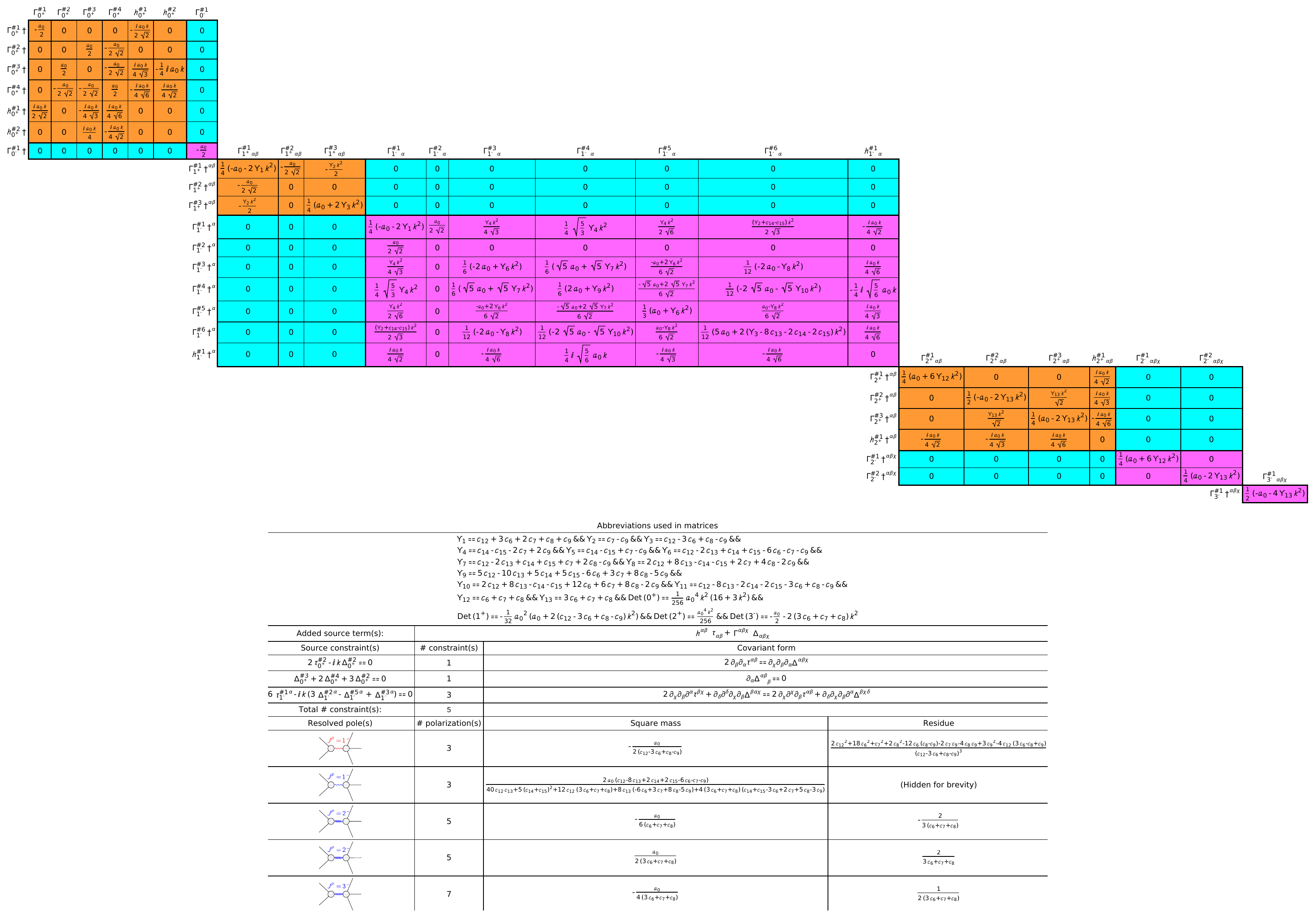}
	\caption{\label{ParticleSpectrographMaxwellTheory} Partial particle spectrograph of~\cref{fullAction} after the sequential imposition of~\cref{AntiCosmologicalRules,AllAZero}. For brevity, the analysis is halted once the massive spectrum has been computed; the massless spectrum is not considered. All quantities are defined in~\cref{FieldKinematicsMetricPerturbation,FieldKinematicsConnection,ParticleSpectrographKScreeningNoA2m1p1mA1mB}. See~\cite{Barker:2024juc} for further notational details. This is a vector graphic: all details are visible under magnification.}
\end{figure*}
\begin{figure*}[ht]
\includegraphics[height=0.9\textwidth,angle=90]{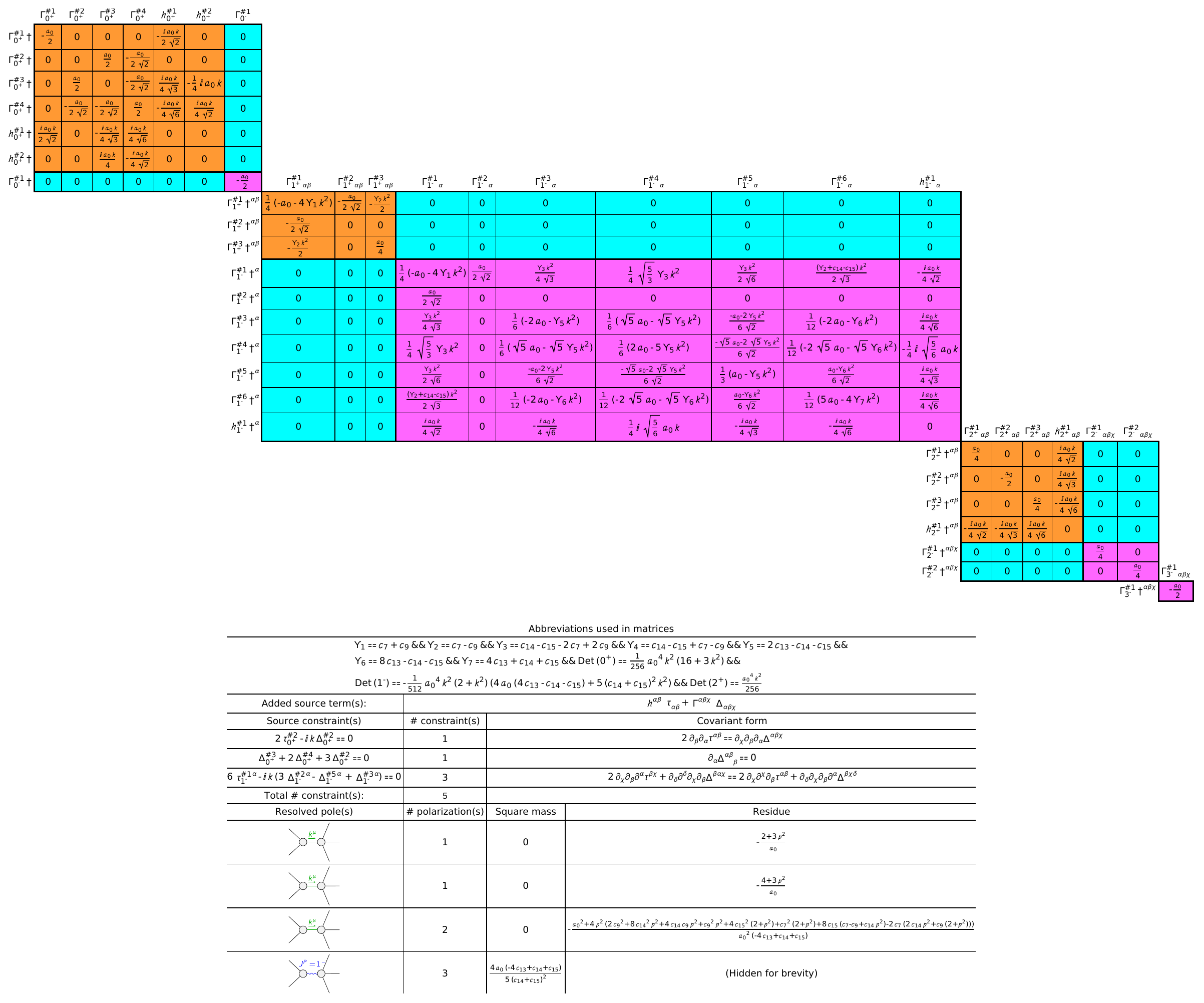}
	\caption{\label{ParticleSpectrographMaxwellNoHigherSpinTheory} Complete particle spectrograph of~\cref{fullAction} after the sequential imposition of~\cref{AntiCosmologicalRules,AllAZero,FurtherRules}, to be compared with~\cref{ParticleSpectrographMaxwellTheory}. All quantities are defined in~\cref{FieldKinematicsMetricPerturbation,FieldKinematicsConnection,ParticleSpectrographKScreeningNoA2m1p1mA1mB}. See~\cite{Barker:2024juc} for further notational details. This is a vector graphic: all details are visible under magnification.}
\end{figure*}
\begin{figure*}[ht]
\includegraphics[height=0.9\textwidth,angle=90]{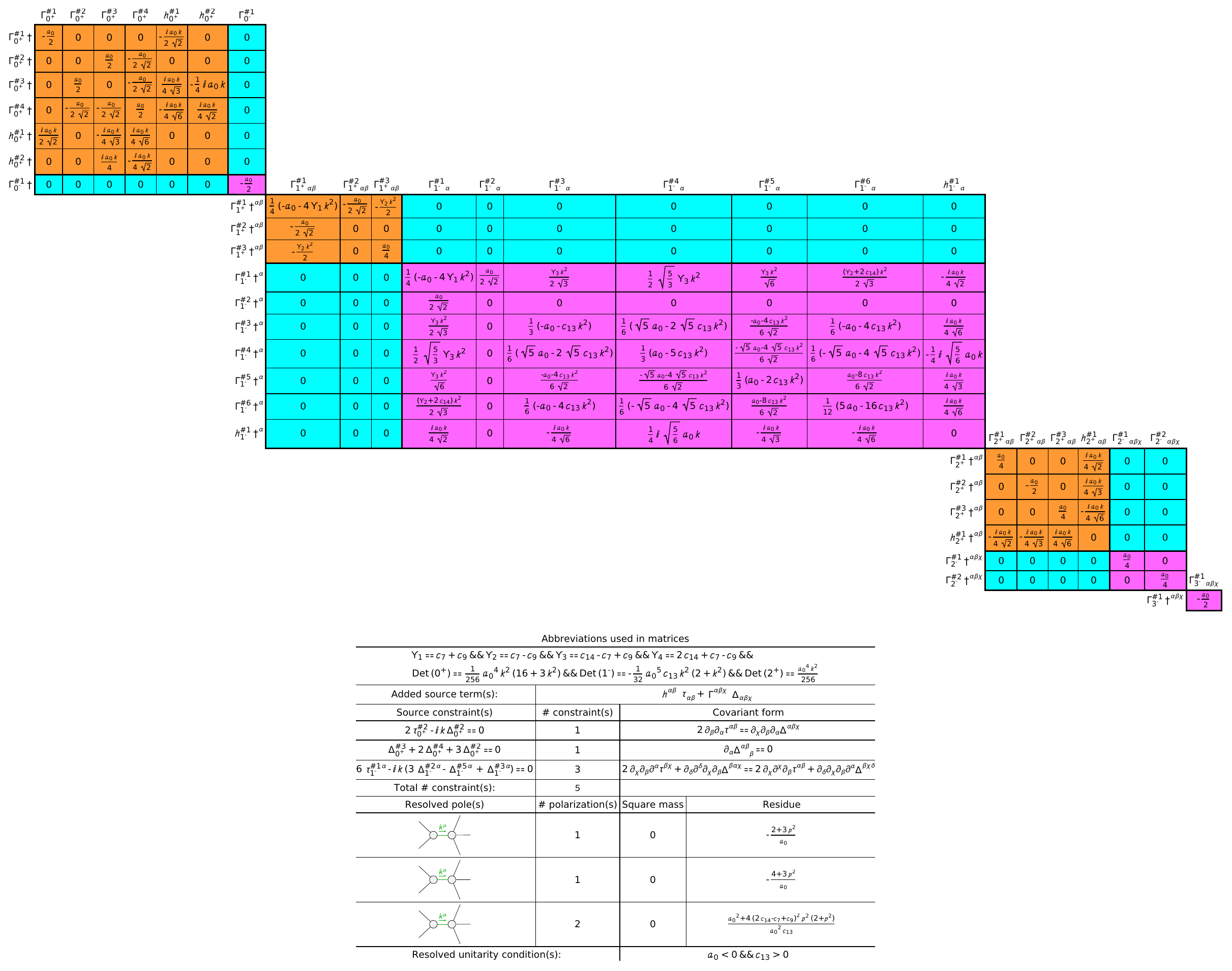}
	\caption{\label{ParticleSpectrographMaxwellNoHigherSpinNo1m} Complete particle spectrograph of~\cref{fullAction} after the sequential imposition of~\cref{AntiCosmologicalRules,AllAZero,FurtherRules,InfiniteMassVector}, to be compared with~\cref{ParticleSpectrographMaxwellNoHigherSpinTheory}. All quantities are defined in~\cref{FieldKinematicsMetricPerturbation,FieldKinematicsConnection,ParticleSpectrographKScreeningNoA2m1p1mA1mB}. See~\cite{Barker:2024juc} for further notational details. This is a vector graphic: all details are visible under magnification.}
\end{figure*}
\end{widetext}

\end{document}